\documentclass[11pt]{article}

\newcommand\spacingset[1]{\renewcommand{\baselinestretch}%
	{#1}\small\normalsize}

\newenvironment{set11}
{%
	\clearpage
	\let\orignewcommand\newcommand
	\let\newcommand\renewcommand
	\makeatletter
	\input{size11.clo}%
	\makeatother
	\let\newcommand\orignewcommand
}
{%
	\clearpage
}

\RequirePackage{amsthm, amsmath,amsfonts,amssymb}
\RequirePackage[authoryear]{natbib}
\numberwithin{equation}{section}

\usepackage{latexsym,multirow}
\usepackage{amssymb,amsmath, bm}
\usepackage{graphicx}
\usepackage{booktabs}
\usepackage{caption}
\usepackage{subcaption}
\usepackage{enumerate}
\usepackage{comment}
\usepackage[normalem]{ulem}
\usepackage{appendix}
\usepackage{setspace}
\usepackage{natbib}

\usepackage{dcolumn}
\newcolumntype{.}{D{.}{.}{-1}}
\newcolumntype{d}[1]{D{.}{.}{#1}}
\theoremstyle{plain}

\newtheorem{proposition}{Proposition}
\newtheorem{theorem}{Theorem}

\newtheorem{appprop}{Proposition}[section]

\newtheorem{appresult}{Result}[section]
\newtheorem{applemma}{Lemma}[section]

\newcommand{\ind}{\mbox{$\perp\!\!\!\perp$}}

\newcommand{\argmax}{\operatornamewithlimits{argmax}}
\newcommand{\argmin}{\operatornamewithlimits{argmin}}

\newcommand\E{\mathbb{E}}
\newcommand\cT{\mathcal{T}}
\newcommand\bT{\bm{T}}
\newcommand\bt{\bm{t}}
\newcommand\bC{\bm{C}}
\newcommand\bF{\bm{F}}
\newcommand\bX{\bm{X}}
\newcommand\bbeta{\bm{\beta}}
\newcommand\bphi{\bm{\phi}}
\newcommand\bzero{\bm{0}}

\usepackage{algpseudocode}
\usepackage{algorithm}

\algtext*{EndFor}

\usepackage{xcolor}
\usepackage{colortbl}
\usepackage{hyperref}

\usepackage[compact]{titlesec}

\allowdisplaybreaks

\newcommand{\blind}{0}
\begin{document}

\newcommand{\tit}{Estimating Heterogeneous Causal Effects of
	High-Dimensional Treatments:\\ Application to Conjoint Analysis}
%
%

\pdfbookmark[1]{Title Page}{Title Page}

\spacingset{1.1}

\if0\blind

{\title{{\bf \tit}\thanks{The methods described in this paper can be
			implemented using open-source software \texttt{FactorHet}
			available at
			\url{https://CRAN.R-project.org/package=FactorHet}. We thank Jelena Bradic, Ray Duch, Tom Robinson, Teppei Yamamoto, and participants at the 2021 Joint Statistical Meetings, the University of North Carolina Chapel Hill Methods and Design Workshop, the Bocconi Institute for Data Science and Analytics Seminar, and the 2022 American Political Science Association Annual Meeting for helpful feedback on this paper. We also thank two anonymous reviewers from the Maguro Peer Pre-Review Program at Harvard's Institute for Quantitative Social Science. This research was done using services provided by the OSG Consortium (\url{https://doi.org/10.21231/906P-4D78}), which is supported by the National Science Foundation awards \#2030508 and \#2323298. Imai thanks the Alfred P. Sloan Foundation (2020--13946) for partial support. Pashley was partially supported by the NSF Graduate Research Fellowship (DGE1745303). Any opinion, findings, and conclusions or recommendations expressed in this material are those of the authors and do not necessarily reflect the views of the National Science Foundation.}}
	
	\author{Max Goplerud\thanks{Assistant Professor, Department of Government, University of Texas at Austin. 158 W 21st Street, Austin, TX 78712. Email: \href{mailto:mgoplerud@austin.utexas.edu}{mgoplerud@austin.utexas.edu}. URL: \href{https://www.mgoplerud.com}{https://mgoplerud.com}}
		\and
		Kosuke
		Imai\thanks{Professor, Department of Government and Department of
			Statistics, Harvard University.  1737 Cambridge Street,
			Institute for Quantitative Social Science, Cambridge MA 02138.
			Email: \href{mailto:imai@harvard.edu}{imai@harvard.edu} URL:
			\href{https://imai.fas.harvard.edu}{https://imai.fas.harvard.edu}}
		\and Nicole E. Pashley\thanks{Assistant Professor, Department of Statistics, Rutgers University.  501 Hill Center, 
			110 Frelinghuysen Road, Piscataway, NJ 08854.
			Email: \href{mailto:nicole.pashley@rutgers.edu}{nicole.pashley@rutgers.edu}}}

	\date{
		First draft: August 15, 2022 \\
		This draft: \today
	}
	
	\maketitle
	
}\fi

\if1\blind
\title{\bf \tit}

\maketitle
\fi

\setcounter{page}{0}
\thispagestyle{empty}
\vspace{-1em}
\begin{abstract}
	
	Estimation of heterogeneous treatment effects is an active area of research.  Most of the existing methods, however, focus on estimating the conditional average treatment effects of a single, binary treatment given a set of pre-treatment covariates.  In this paper, we propose a method to estimate the heterogeneous causal effects of high-dimensional treatments, which poses unique challenges in terms of estimation and interpretation.  The proposed approach finds maximally heterogeneous groups and uses a Bayesian mixture of regularized logistic regressions to identify groups of units who exhibit similar patterns of treatment effects.  By directly modeling group membership with covariates, the proposed methodology allows one to explore the unit characteristics that are associated with different patterns of treatment effects.  Our motivating application is conjoint analysis, which is a popular type of survey experiment in social science and marketing research and is based on a high-dimensional factorial design.  We apply the proposed methodology to the conjoint data, where survey respondents are asked to select one of two immigrant profiles with randomly selected attributes.  We find that a group of respondents with a relatively high degree of prejudice appears to discriminate against immigrants from non-European countries like Iraq. An open-source software package is available for implementing the proposed methodology.

	
\end{abstract}

\section{Introduction}
\spacingset{1.375}

Over the past decade, a number of researchers have exploited modern
machine learning algorithms and proposed new methods to estimate
heterogeneous treatment effects using experimental data.  They include
tree-based methods
\citep[e.g.,][]{imai:stra:11,athe:imbe:16,wage:athe:18,hahn:murr:carv:20},
regularized regressions \citep[e.g.,][]{imai:ratk:13, tian:etal:14,
  kunz:etal:19}, ensemble methods \citep[e.g.,][]{vand:rose:11,
  grim:mess:west:17}, and frameworks that allow for the use of generic
machine learning methods \citep[e.g.,][]{cher:etal:19,imai:li:24}.
This methodological development, however, has largely been confined to
settings with a single, binary treatment variable; some exceptions
include a time-varying treatment \citep[e.g.,][]{almirall2014time},
and a relatively small number of treatments
\citep[e.g.,][]{imai:ratk:13}.

In this paper, we estimate the heterogeneous effects of a {\it
  high-dimensional} treatment by analyzing the data from conjoint
experiments, in which the number of possible treatment combinations
exceeds the sample size.  While the high dimensionality in treatment
effect heterogeneity problems typically comes from the number of
covariates or moderators, conjoint experiments provide an additional
difficulty due to high dimensionality of treatment.  We address the
methodological challenge of effectively summarizing the complex
patterns of heterogeneous treatment effects that are induced by the
interactions among the treatments themselves as well as the
interactions between the treatments and unit characteristics.

\paragraph*{Methodological contributions.}
We consider a common setting where researchers wish to use a small
number of groups to summarize heterogeneous treatment effects and
characterize these groups using several pre-treatment covariates
\citep[e.g.,][]{cher:etal:19,imai:li:24}.  We show that once
researchers select the number of groups to be used for summarizing
heterogeneous treatment effects, finding the maximally heterogeneous
groups in terms of potential outcomes is equivalent to maximizing the
likelihood function based on the latent group membership. Furthermore,
modeling the conditional probability of an individual's latent group
membership using the moderators of interest yields maximally
heterogeneous groups that are predicted well by these moderators.

A primary methodological challenge with high-dimensional treatments is characterizing both the interactions among a large number of treatment variables and their relationships with moderating covariates. Our methodology addresses this by finding maximally heterogeneous groups while characterizing the relationship between group membership and unit characteristics. Thus, it is possible to understand the types of units that are likely to exhibit similar treatment effect patterns.

Since optimizing over the latent group membership is difficult, we marginalize it out, leading to a mixture of experts model \citep[e.g.,][]{gormley2019mixture,gupta1994using}. We also develop estimation strategies by bringing together two previously disconnected literatures, one on mixture models and the other on sparsity-inducing penalties to fuse factor levels. 
 
\paragraph*{Empirical application.}
Conjoint analysis is a popular survey experimental methodology in
social sciences and marketing research
\citep[e.g.,][]{hainmueller2014causal,rao:14}.  Conjoint analysis is a
variant of factorial designs \citep{DasPilRub15} with a large number
of factorial treatments---so large that typically not all possible
treatments are observed.  Under the most commonly used
``forced-choice'' design, respondents are asked to evaluate a pair of
profiles whose attributes are randomly selected based on factorial
variables with several levels.

In the specific experiment we reanalyze, the original authors used a
conjoint analysis to measure immigration preferences by presenting
each survey respondent with several pairs of immigrant profiles with
varying attributes including education, country of origin, and job
experience \citep{hainmueller2015hidden}.  For each pair, the
respondent was asked to choose one profile they prefer.  The authors
then analyzed the resulting response patterns to understand which
immigrant characteristics play a critical role in forming the
immigration preferences of American citizens.

In the methodological literature on factorial designs and conjoint
analysis, researchers have focused on average marginal effects,
which represent the average effect of one factor level relative to
another level of the same factor averaging over the randomization
distribution of the remaining factors \citep{hainmueller2014causal,
  DasPilRub15}.  Many empirical researchers use subgroup analysis to
explore how these marginal effects depend on a small number of
moderating covariates
\citep[e.g.,][]{hainmueller2015hidden,newman2019economic}.

Unfortunately, such an approach often results in low statistical power
and may suffer from multiple testing problems \citep{liu:shir:23}.
More fundamentally, by marginalizing other treatments, researchers may
miss important interactions among treatments.  Although some have
explored the estimation of interaction effects
\citep[e.g.,][]{DasPilRub15,egam:imai:19,de2022improving}, few have
investigated how to estimate heterogeneous treatment effects of
high-dimensional treatments. 

Moreover, there is even less prior research that models how the effects of high-dimensional treatments vary as a function of
moderators.  One exception is \cite{robinsondetect} which uses a
BART-based approach for conjoint experiments, but their heterogeneous effects of interest are different from ours (see
Section~\ref{subsec:cjbart} for comparison).

\paragraph*{Related models.}
To overcome this challenge, we develop a mixture of regularized
logistic regression model under our general methodological framework
of treatment effect heterogeneity with high-dimensional treatments. We
combine and extend two distinct strands of methodological
research. First, a growing literature explores regularization with
high-dimensional factors, and their interactions, by fusing or
grouping levels of factors together
\citep[e.g.,][]{bondell2009anova,post2013factor,stokell2021modelling}. This
methodology is well-suited to factorial experiments because it
provides a natural way of interpreting empirical findings by
identifying a set of factor levels that characterize distinct
treatment effects (e.g., \citealt{egam:imai:19}).

However, since our goal is to identify groups of individuals with
heterogeneous effects, we use a mixture model that finds the maximally
heterogeneous groups (see Section~\ref{subsec:kmeans}). Although the
marketing literature has long applied mixture models to analyzing
heterogeneity in conjoint experiments
\citep[e.g.,][]{gupta1994using,andrews2002empirical}, they focused on
settings with low-dimensional treatments. In the high-dimensional
setting, some combine mixture models with sparsity constraints
\citep[e.g.,][]{khal:chen:07,stadler2010lasso,khalili2010mixture}, but
these constraints are not designed to induce the fusion of factor
levels that is essential in conjoint analysis.

Our model, therefore, synthesizes both of these approaches by using a finite
mixture model with a prior that encourages fusing levels, while respecting
the hierarchical structure---fusing main effects of factors only if
their interactions are also fused \citep{yan2017hierarchical}. For
efficient computation, we develop an Expectation-Maximization (EM)
algorithm \citep{demp:lair:rubi:77} by exploiting the representation
of $\ell_1$ and $\ell_2$ penalties as a mixture of Gaussians
\citep[e.g.,][]{figueiredo2003adaptive,polson2011svm,ratkovic2017sparse,goplerud2021sparsity}.
We derive a tractable algorithm that adapts the latent overlapping
group LASSO developed in sparse modeling to fusion required in
factorial experiments.

The rest of the paper is organized as follows.  In
Section~\ref{sec:con}, we discuss the motivating application, which is
a conjoint analysis of American citizens' preferences regarding
immigrant features.  We also briefly describe a methodological
challenge to be addressed.  In Section~\ref{sec:methods}, we present
our proposed methodology. In Section~\ref{sec:simulation}, we show our
method performs well in a realistic numerical simulation. In
Section~\ref{sec:analysis}, we apply this methodology and reanalyze
the data from the motivating conjoint analysis.
Section~\ref{sec:disc} concludes with a discussion.
The R package \texttt{FactorHet} \citep{FactorHet} can be used to implement
our methodology and \cite{replicationAOAS} provides replication code for our
application and simulations.

\section{Motivating Application: Conjoint Analysis of Immigration
  Preferences}\label{sec:con}

Our motivating application is a conjoint analysis of American
immigration preferences.  In this section, we introduce the
experimental design and discuss the results of previous analyses that
motivate our methodology for estimating heterogeneous treatment
effects.

\subsection{The Experimental Design}

In an influential study, \cite{hainmueller2015hidden} use conjoint
analysis to estimate the effect of immigrant attributes on preferences
for admission to the United States (Data are available at the AJPS
Dataverse \url{https://doi.org/10.7910/DVN/25505}). The authors
conduct an online survey experiment using a sample of 1,407 American
adults.  Each survey respondent assessed five pairs of immigrant
profiles with randomly selected attributes.  For each pair, a
respondent was asked to choose which of the two immigrant profiles
they preferred to admit to the United States.

The attributes of immigrant profiles used in this factorial experiment, with number of levels provided in parentheses,
are gender (2), education (7), employment plans (4), job experience (4), profession (11),
language skills (4), country of origin (10), reasons for applying (3), and prior
trips to the United States (5).  For completeness, these factors and their
levels are reproduced as Table~A1 of the Supplementary Material \citep{aoas_supp}. In total, there exist
over 1.4 million possible profiles, implying more than
$2 \times 10^{12}$ possible comparisons of two profiles that are
possible in the experiment.  It is clear that with 1,407 respondents,
even though each respondent performs five comparisons, not all
possible profiles can be included.  Thus, exploring treatment effect
heterogeneity requires a methodological development that goes beyond
the models used previously in the causal inference literature for
binary treatments.
  
The levels of each factor variable were independently randomized to
yield one immigrant profile.  Randomization was subject to some
restrictions such that profession and education factors result in
sensible pairings (e.g., ruling out doctors with less than two-years
of college education) and immigrants whose reason for applying is
persecution must come from Iraq, Sudan, Somalia, or China.  The
ordering of attributes was also randomized for each respondent.  The
experiment additionally collected data on the respondents, including
demographic information, partisanship, attitudes towards immigration,
and ethnocentrism.  A rating for each immigrant profile was also
recorded, but that metric is not the focus of our analysis.

\subsection{Heterogeneous Treatment Effects}

\cite{hainmueller2015hidden} conducted their primary analysis based on
linear regression model where the unit of analysis is an immigrant
profile (rather than a pair) and the outcome variable is an indicator
for whether a given profile was chosen.  The predictors of the model
include the indicator variable for each immigrant attribute.  The
model also includes the interactions between education and profession,
as well as between country of origin and reasons for applying, to
account for the restricted randomization scheme mentioned above.
Finally, the standard errors are clustered by respondent.

\begin{figure}[t]
\centering 
\includegraphics[width=\textwidth]{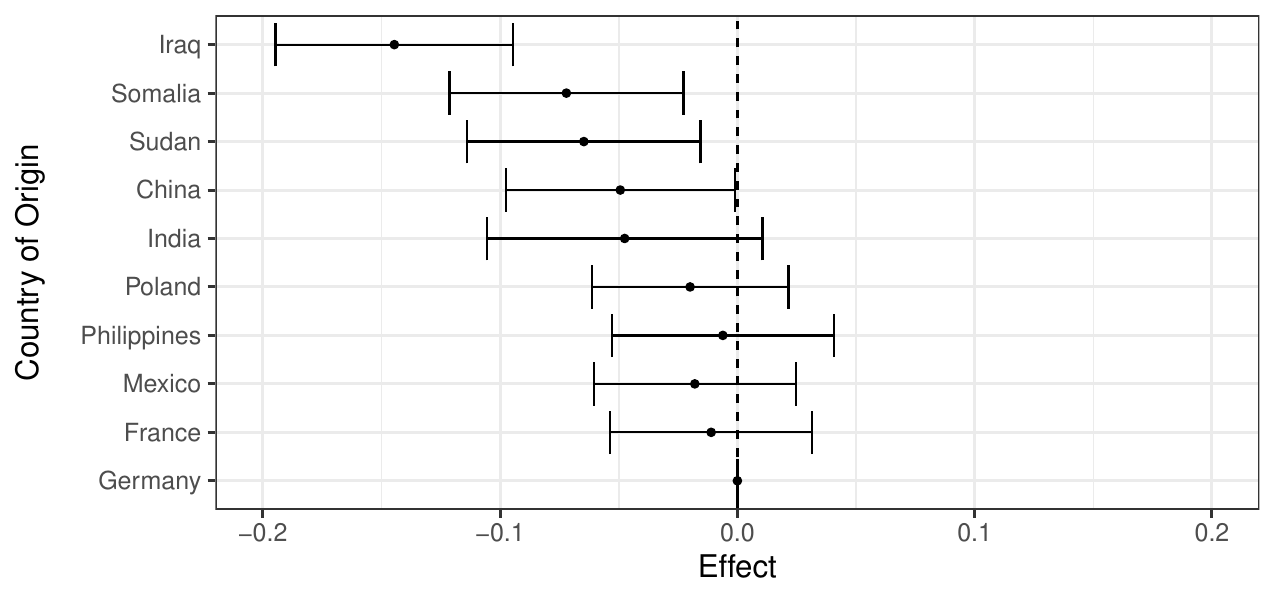}
\caption{Estimated average marginal component effects of
  country of origin where the baseline is Germany, with effect
  estimates as given in
  \cite{hainmueller2015hidden}.} \label{fig:hhy_country_est}
\end{figure}

As formalized in \cite{hainmueller2014causal}, the regression
coefficient represents the average marginal component effect (AMCE) of
each attribute averaging over all the other attributes including those
of the other profile in a given pair.  Fig~\ref{fig:hhy_country_est}
reproduces the estimated overall AMCEs of country of origin where the
baseline category is Germany.  There is little country effect with the
exception of Iraq, which negatively affects the likelihood of being
preferred by a respondent.

\begin{table}[t!]
    \centering\small
    \begin{tabular}{|p{0.35\linewidth}|p{0.65\linewidth}|}
    \hline
    Moderator &Split\\
     \hline
     Education & Any college education or no college education\\
     Ethnocentrism & Median ethnocentrism measure\\
     Political party & Republican or Democrat\\
     Percent of foreign born workers in respondent's industry & High or low \\
     Household income & More or less than \$50,000\\
     Fiscal exposure to immigration & High or low\\
     ZIP code demographics & $<$ 5\% immigrants, $>$ 5\% immigrants (primarily from Latin America), or $>$ 5\% immigrants (primarily not from Latin America) \\
     Race/ethnicity & White or non-white\\
     Hispanic ethnicity & Hispanic or non-hispanic \\
     Ideology & Liberal or conservative\\
     Immigration attitudes & Supports or does not support reducing immigration\\
     Gender & Male or female \\
     Age & Young or old \\
     \hline
    \end{tabular}
     \caption{List of subset analyses performed in \cite{hainmueller2015hidden}, listed by moderator and how it was split to form sub-groups.}\label{tab:subsets}
\end{table}

Beyond the AMCEs, these authors and others including
\citet{newman2019economic} have explored the heterogeneous treatment
effects among respondents by conducting many sub-group analyses based
on a number of respondent characteristics including partisanship and
level of education.  Table~\ref{tab:subsets} shows all of the
sub-group analyses performed by \cite{hainmueller2015hidden} and how
the respondents were broken up into groups.  We find that 13
sub-group analyses were performed (excluding those used for robustness
checks), with results from the first three (education, ethnocentrism,
and political party) presented in the main paper.  Of those three
analyses, the authors find some evidence of heterogeneous effects of
country of origin between subsets that differ on ethnocentrism, but
little evidence of heterogeneity beyond this.  The other 10 analyses
can be found in their appendix, and the authors conclude for that
participants responded similarly, in general, across those sub-groups.

Our goal is to build a methodology that enables one to more
systematically explore heterogeneous treatment effects in conjoint
experiments.  Sub-group analyses like those conducted in the original
analysis can be problematic for several reasons.  First, the analyst
must conduct a separate analysis for each moderator of interest,
leading to multiple testing problem.  Second, typically the moderators
are dichotomized (or broken up into a small number of groups),
requiring the analyst to decide how to split the data.  Third, they are
not amenable to exploration of how multiple moderators might work
together to change outcomes.

To address these issues, one could include the moderators as
covariates within the regression.  However, if the goal is to provide
estimated heterogeneous effects with straightforward interpretations,
regressions with possibly complex interactions are not ideal.  To
estimate heterogeneous effects, we need to not only interact a large
number of treatments, but we will have to further interact all main
and interaction effects of treatments with the moderators.  It is
unclear how to best reduce the dimensionality of both the moderator
and treatment space in a classic regression set up.  It is also
challenging to interpret the interactions from these models to
understand the characteristics of units that lead to different
treatment effect patterns.

In sum, researchers must parsimoniously characterize how a large
number of possible treatment combinations interact with several key
moderators of interest.  The goal is to obtain estimates of
heterogeneous effects and understand how the covariate distributions of
units with different treatment effects differ.  We now turn to our
methodology which is designed to address these challenges and result
in interpretable estimates.

\section{Modeling Heterogeneous Effects of High-dimensional
  Treatments}\label{sec:methods}

We now describe the proposed methodology.  To simplify the exposition,
we focus on a general factorial design. 
This design corresponds to
conjoint analysis with a single task per person, where there is only
one profile assessed rather than a comparison of profiles, and
complete randomization of all combinations of factor levels. Extensions to independent factor randomization and realistic conjoint analyses are
immediate and will be discussed and applied in
Section~\ref{sec:analysis}.

\subsection{Set Up}

Suppose that we have a simple random sample of $N$ units.  Consider a
factorial design with $J$ factors where each factor
$j \in \{1, \cdots, J\}$ has $L_j \ge 2$ levels.  The treatment
variable for unit $i$, denoted by $\bT_i$, is a $J$-dimensional vector
of random variables, each of which represents the assigned level of
the corresponding factor variable.  For example, the $j$th element of
this random vector $T_{ij} \in \{0,1,2,\ldots,L_j-1\}$ represents the
level of factor $j$ which is assigned to unit $i$.

Following \cite{DasPilRub15}, we define the potential outcome for unit
$i$ as $Y_i(\bt)$ where $\bt \in \cT$ represents the realized
treatment with $\cT$ representing the support of the randomization
distribution for $\bT_i$. 
Then, the observed outcome is given by
$Y_i = Y_i(\bT_i)$.  The notation implicitly assumes no interference
between units \citep{Rubin80}.  In this paper, for the sake of
concreteness, we focus on the binary outcome $Y_i \in \{0, 1\}$.
Extensions to non-binary outcomes are straightforward.  Lastly, we
observe a vector of $p_x$ pre-treatment covariates for each unit $i$
and denote it by $\bX_i$.  All together, we observe
$(Y_i, \bT_i, \bX_i)$ for each unit $i$.

To illustrate the notation, consider a simplified version of our
motivating example where each respondent $i$ observes a single
immigrant profile and must decide whether to support that immigrant's
admission or not.  Then, $\bT_i$ is a vector indicating the level
respondent $i$ sees for each of the nine immigrant attributes.  The
outcome variable $Y_i$ is an indicator for whether respondent $i$
chooses to support admission for the immigrant they are presented
with.  Lastly, $\bX_i$ denotes a vector of covariates for respondent
$i$ that we hypothesize might moderate the treatment effect.  In our
application, $\bX_i$ included political party, education, demographics
of their ZIP code, ethnicity, and Hispanic prejudice score (see
Section~\ref{sec:analysis:data_model} for details).

The randomness in our data, $(Y_i, \bT_i, \bX_i)$ comes from two
sources: random sampling of units into the study and random assignment
of units to treatments.  For simplicity, we assume units are sampled
via simple random sampling (though our method can incorporate sampling
weights).  The randomization of treatment assignment implies
$\{Y_i(\bt)\}_{\bt \in \cT} \ \ind \ \bT_i$ for each $i$ where the
exact mode of randomization will determine the distribution of
$\bT_i$.  In many conjoint experiments, researchers independently and
uniformly randomize each factor.  However, in some cases including our
application, researchers may exclude certain unrealistic combinations
of factor levels (e.g., doctor without a college degree), leading to
the dependence between factors.  In all cases, researchers have
complete knowledge of the randomization distribution of the factorial
treatment variables.

Based on random sampling and random treatment assignment alone, we can
conduct valid inference for marginal treatment effects of interest
using simple regression or difference-in-means estimator
\citep[see][]{hainmueller2014causal}.  If we wish to explore treatment
effect heterogeneity across treatments and covariates, however, a
model-based approach is useful.  We next introduce our model, which
will allow us to explore heterogeneous effects in a principled manner
while also handling the high-dimensional nature of the data.

\subsection{General Framework}\label{subsec:kmeans}

The most basic causal quantity of interest is the AMCE, which is
defined for any given factor $j$ as
\begin{equation}
  \delta_j (l, l^\prime) \ = \ \E[Y_i(T_{ij}=l, \bT_{i,-j}) - Y_i(T_{ij}
  = l^\prime, \bT_{i,-j})], \label{eq:AMCE}
\end{equation}
where $l \ne l^\prime \in \cT_j$ with $\cT_j$ representing the support
of the randomization distribution for $T_j$.  The expectation in
Equation~\eqref{eq:AMCE} is taken over the distribution of the other
factors $\bT_{i,-j}$ as well as the random sampling of units from the
population.  Thus, the AMCE averages over two sources of causal
heterogeneity---heterogeneity across treatment combinations and
across units.  Different treatment combinations may have distinct
impacts on units with varying characteristics.  Our goal is to model
these potentially complex heterogeneous treatment effects using an
interpretable model.

We propose to model heterogeneous treatment effects based on $K$
distinct treatment effect patterns where $K \ge 2$ is chosen by a
researcher, based on their desired granularity of heterogeneity.  This
approach, which is based on a fixed number of subgroups to
characterize treatment effect heterogeneity, is commonly used by
empirical researchers.  Others have studied various methodological
aspects of this approach albeit in the context of binary treatment
\citep[][]{cher:etal:19,imai:li:24}.

Our goal is to summarize the treatment effect heterogeneity by
dividing the population into $K$ subpopulations and characterizing
these groups based on a set of pre-treatment covariates or
``moderators'' denoted by $\bX_i$.  In particular, we would like to
construct $K$ groups such that across-group treatment effect
heterogeneity is maximized while minimizing the within-group
heterogeneity. Since the treatments of interest are high-dimensional,
we focus on finding maximally heterogeneous groups in terms of average
potential outcomes rather than their contrasts.  We can then estimate
any treatment effects of interest within each group.

Let $Z_i \in \{1, \cdots, K\}$ denote the latent group membership of
unit $i$ and $\mathcal{Z}=\{Z_i\}_{i=1}^n$.  We use
$\zeta_k(\bm{t}) = \E[Y_i(\bm{t}) \mid Z_i = k]$ to represent the
average potential outcome under treatment $\bm{t}$ for group $k$.
Under the randomization of $T_i$, define the estimated within-group
average outcome under treatment $t$ for group $k$ and the estimated
overall average outcome as
$\hat{\zeta}_k(\bm{t}; \mathcal{Z}) = \sum_{i=1}^N I\{Z_i = k,
\bm{T}_i = \bm{t}\} Y_i/\sum_{i=1}^N I\{Z_i = k, \bm{T}_i = \bm{t}\}$
and
$\widehat{\overline{Y}}(\bm{t}) = \sum_{i=1}^N I\{\bm{T}_i = \bm{t}\}
Y_i/\sum_{i=1}^N I\{\bm{T}_i = \bm{t}\}$, respectively.

Given the number of groups $K$ selected by researchers, we show how to
find maximally heterogeneous groups in terms of potential outcomes.
The following proposition establishes that maximizing the
Kullback-Leibler (KL) divergence of potential outcomes between groups
is equivalent to maximizing the log-likelihood over groups and their
centroids.  We emphasize that this equivalence result does not assume
the existence of a ``correct'' number of groups.
\begin{proposition}[Finding maximally heterogeneous
  groups] \label{prop:kmeans} Maximally heterogeneous groups in the
  terms of the KL divergence of potential outcomes can be found by
  maximizing the log-likelihood function over the group membership and
  the centroids of groups,
  \begin{equation}
  \begin{aligned}
    & \argmax_{\mathcal{Z}} \left\{\sum_{k=1}^K \sum_{i=1}^N \mathbf{1}\{Z_i = k\} \mathrm{KL}\left(\hat{\zeta}_k(\bm{T}_i; \mathcal{Z}) \Vert 
      \widehat{\overline{Y}}(\bm{T}_i)\right) \right\} \\
 = \ & \argmax_{\mathcal{Z}} \sum_{k=1}^K \sup_{\zeta_k}
\sum_{i=1}^N  \mathbf{1}\{Z_i = k \} \left[Y_{i} \log
       \zeta_k(\bm{T}_i)+ (1-Y_i)\log \{1- \zeta_k(\bm{T}_i)\}\right] 
   \end{aligned} \label{eq:equivalence}
   \end{equation}
where $Y_{i}$ is binary, the KL divergence of two Bernoulli
distributions with means $\mu_1$ and $\mu_2$ is given by
$\mathrm{KL}\left(\mu_1 \Vert \mu_2\right) \ = \ \mu_1 \log
\mu_1/\mu_2 + (1-\mu_1)\log (1-\mu_1)/(1-\mu_2)$, and
$\{\hat{\zeta}_k(\bt; \mathcal{Z})\}_{k=1}^K$ denotes the maximizers of
the right hand side of Equation~\eqref{eq:equivalence} given
$\mathcal{Z}$.
\end{proposition}
Section~C of the Supplementary Material provides a proof of a more general result
for the natural exponential family distributions \citep[see][for a
similar result in the Gaussian case]{chi2016kmeans}. The
log-likelihood formulation is equivalent to the classification maximum
likelihood approach in mixture modeling
\citep{mclachlan1982classification}.

We now extend the above equivalence result to the settings in which
we further model the group membership $Z_i$ using a set of moderators
$\bX_i$, i.e.,
$\pi_k(\bm{x}) =\mathrm{Pr}(Z_i = k\mid \bX_i = \bm{x})$ for
$k=1,2,\ldots,K$.  Such a model helps characterize and understand the
types of units that comprise each group.  The next proposition shows
that maximizing the log-likelihood function of this extended model is
equivalent to finding $K$ maximally heterogeneous groups such that the
group memberships are predicted well by the moderators.
\begin{proposition}[Finding maximally heterogeneous groups with
  moderators] \label{prop:moe} Suppose that we extend the setting of
  Proposition~\ref{prop:kmeans} and additionally model the conditional
  probability of each individual's group membership given categorical
  moderators $\{\pi_k(\bX_i)\}_{k=1}^K$.  Then, maximally
  heterogeneous groups in terms of the KL divergence of potential
  outcomes with the entropy of group membership probabilities as a
  penalty term can be found by maximizing the log-likelihood function
  of the extended model,
  \begin{equation}
\begin{aligned}
& \argmax_{\mathcal{Z}} \left\{\sum_{k=1}^K \sum_{i=1}^N \mathbf{1}\{Z_i = k\} \mathrm{KL}\left(\hat{\zeta}_k(\bm{T}_i; \mathcal{Z}) \Vert 
\widehat{\overline{Y}}(\bm{T}_i)\right) - \sum_{i=1}^N H(\{\hat{\pi}_k(\bm{X}_i; \mathcal{Z})\}_{k=1}^K) \right\}\\
= \ & \argmax_{\mathcal{Z}} \sum_{k=1}^K \sup_{\zeta_k, \pi_k}
\sum_{i=1}^N  \mathbf{1}\{Z_i = k \} \left[Y_{i} \log
\zeta_k(\bm{T}_i)+ (1-Y_i)\log \{1- \zeta_k(\bm{T}_i)\} + \log \pi_k(\bX_i)\right] 
\end{aligned} \label{eq:moe}
\end{equation}
where $H(\{p_k\}_{k=1}^K) = -\sum_{k=1}^K p_k \log p_k$ (by
convention, if $p_k = 0$, then $p_k \log p_k = 0$) is the entropy, and
$\hat{\pi}_k(\bm{x}; \mathcal{Z}) = \sum_{i=1}^N \bm{1}\{Z_i = k,
\bX_i = \bm{x}\}/\sum_{i=1}^N \bm{1}\{\bX_i = \bm{x}\}$ and
$\hat\zeta_k(\bt; \mathcal{Z})$ are the maximizers of the
log-likelihood function of the right hand side of
Equation~\eqref{eq:moe} given $\mathcal{Z}$.
\end{proposition}
Proof is given in Section~D of the Supplementary Material.  Since the entropy
$H(\{\hat{\pi}_k(\bm{x}; \mathcal{Z})\}_{k=1}^K)$ is maximized when
$\hat{\pi}_k(\bm{x}) = 1/K$, Proposition~\ref{prop:moe} shows that
adding a group membership model based on moderators encourages finding
groups whose memberships are well predicted by the moderators.

Direct optimization of
Equations~\eqref{eq:equivalence}~and~\eqref{eq:moe} over $\mathcal{Z}$
has been studied under the name of ``classification maximum
likelihood'' in the literature on mixture models
\citep{mclachlan1982classification}. For completeness,
Section~G.3 of the Supplementary Material provides an estimation algorithm
for this approach, which modifies the proposed algorithm described in
Section~\ref{sec:est_inf}.  Unfortunately, the classification maximum
likelihood approach suffers from the incidental parameter problem
because the cardinality of $\mathcal{Z}$ increases with the sample
size $N$, leading to an asymptotic bias and inconsistency
\citep{bryant1978asymptotic}.

To address this problem, a dominant approach in the literature is
Bayesian, treating the right hand side of Equation~\eqref{eq:moe} as a
log-posterior that consists of a log-likelihood and a log-prior over
$\mathcal{Z}$, i.e., $\mathrm{Pr}(Z_i = k \mid \bX_i) = \pi_k(\bX_i)$.
By marginalizing out $\mathcal{Z}$, we avoid the incidental parameter
problem, yielding the objective function known as a mixture maximum
likelihood \citep{mclachlan1982classification}.

The model is called ``mixture-of-experts'' when $\pi_k$ depends on
$\bX_i$ \citep{gormley2019mixture} with the following objective
function,
\begin{equation}
\label{eq:profile_moe}
\{\hat{\zeta}_k, \hat{\pi}_k\}_{k=1}^K = \argmax_{\{\zeta_k, \pi_k\}_{k=1}^K} \sum_{i=1}^N \log\left[\sum_{k=1}^K \pi_k(\bX_i) \zeta_k(\bT_i)^{Y_i} \{1 - \zeta_k(\bT_i)\}^{1-Y_i}\right].
\end{equation}
While this setup no longer appears to provide a direct
characterization of the optimal groups,
Proposition~\ref{prop:equivmoe} shows that a mixture-of-experts model
finds maximally heterogeneous groups as in Proposition~\ref{prop:moe}
but with an additional penalty that encourages less extreme posterior
probabilities of group memberships.
\begin{proposition}[Finding maximally heterogeneous groups with a
  mixture of experts] \label{prop:equivmoe} Maximizing the likelihood
  function under a mixture-of-experts model is equivalent to finding
  maximally heterogeneous groups as in Proposition~\ref{prop:moe} with
  an additional penalty.  That is, the following equality holds for
  any $\mathcal{Z}$,
\begin{equation*}
  \begin{aligned}
    & \argmax_{\zeta, \pi} \sum_{i=1}^N \log\left[\sum_{k=1}^K \pi_k(\bX_i)
      \zeta_k(\bT_i)^{Y_i} \{1 - \zeta_k(\bT_i)\}^{1-Y_i}\right] \\
    = \ & \argmax_{\zeta, \pi} \sum_{i=1}^N 
    \sum_{k=1}^K  \mathbf{1}\{Z_i = k \} \left[Y_{i} \log
      \zeta_k(\bm{T}_i)+ (1-Y_i)\log \{1- \zeta_k(\bm{T}_i)\} \right.\\
      & \left. \hspace{1in} + \log
      \pi_k(\bX_i) - \log \tilde\pi_{k}(\bX_i, Y_i, \bT_i; \{\zeta_{k^\prime},
        \pi_{k^\prime}\}_{k^\prime=1}^K) \right], 
  \end{aligned}
\end{equation*}
      where \begin{equation*}\label{eq:obscomplete}
	\begin{aligned}
	\tilde\pi_{k}(\bX_i, Y_i, \bT_i; \{\zeta_{k^\prime},
        \pi_{k^\prime}\}_{k^\prime=1}^K) \ = \ & \mathrm{Pr}(Z_i = k
        \mid Y_i, \bT_i, \bX_i, \{\zeta_{k^\prime},
        \pi_{k^\prime}\}_{k^\prime=1}^K) \\
       \ = \ & \frac{\pi_k(\bX_i)\zeta_k(\bT_i)^{Y_i}\{1-\zeta_k\}^{1-Y_i}}{\sum_{k'=1}^K \pi_{k'}(\bX_i)\zeta_{k'}(\bT_i)^{Y_i}\{1-\zeta_{k'}\}^{1-Y_i}}
	\end{aligned}
      \end{equation*}
      is the posterior membership probability for group $k$.
\end{proposition}
Proof of the proposition directly follows from a well-known identity
\citep[e.g.,][]{celeux2019mixture}, and hence is omitted.  The
equality in Proposition~\ref{prop:equivmoe} holds for any group
membership $\mathcal{Z}$, including its maximum-a-posteriori (MAP)
estimate, i.e., \\
$\hat{Z}_i = \argmax_{k} \tilde\pi_{k}(\bX_i, Y_i, \bT_i;
\{\hat\zeta_{k^\prime}, \hat\pi_{k^\prime}\}_{k^\prime=1}^K) $. Thus, our proposed model can be seen as finding maximally
heterogeneous groups while imposing a penalty that encourages finding
groups that are well predicted by the moderators $\bX_i$ but with less
extreme group membership probabilities based on the data.

All together, our results provide a justification for using a
mixture-of-experts model for heterogeneous effect estimation under the
settings with high-dimensional treatments. We emphasize that a primary
motivation for the use of Bayesian approach is to resolve the
incidental parameter problem with classification maximum
likelihood. Importantly, the results above do not assume a specific
data generating process.  Instead, we have shown that given the number
of groups and appropriate prior distributions, researchers can find
maximally heterogeneous groups by fitting a mixture-of-experts model.

\subsection{Model Specification}\label{sec:regmodel}

Since $\cT$ is high-dimensional, many treatment combinations are
unobserved with a typical sample size.  Thus, nonparametric estimation
is not applicable. We, therefore, model $\zeta_k(\bm{t})$ using a
regularized logistic regression where an ANOVA-style sum-to-zero
constraint is imposed separately for each factor to facilitate merging
of different levels within each factor.  This modeling strategy
identifies a relatively small number of treatment combinations while
avoiding the specification of a baseline level for each factor
\citep{egam:imai:19}. The interpretation of $\zeta_k(\bm{t})$ under
this model is still the average of potential outcome under treatment
$\bm{t}$ in group $k$. Note that we do not assume homogeneity of
outcomes or effects within each group.

We use a multinomial logistic regression for $\pi_k(\bm{x})$:
\begin{equation}
  \zeta_{k}(\bT_i) \
  = \ \frac{\exp(\psi_{k}(\bT_i))}{1+\exp(\psi_{k}(\bT_i))}, \quad
  \text{and} \quad
  \pi_{k}(\bX_i) = \frac{\exp(\bX_i^\top \bphi_k)}{\sum_{k'=1}^K
    \exp(\bX_i^\top \bphi_{k'})}, \label{eq:basic_models}
\end{equation}
where $\bphi_1 = \bzero$ for identification. For $\psi_{k}(\bT_i)$, we
assume an additive model and include both main effects and two-way
interaction effects with a common intercept $\mu$ shared across all
groups,
\begin{align*}
\psi_{k}(\bT_i)& \ =\ \mu + \sum_{j=1}^J \sum_{l =0}^{L_j-1} 
\mathbf{1}\{T_{ij} = l\} \beta^j_{kl} + \sum_{j=1}^{J-1} \sum_{j' >
	j} \sum_{l=0}^{L_j-1} \sum_{l' = 0}^{L_{j'}-1} \mathbf{1}\{T_{ij} = l,
T_{ij'} = l'\} \beta^{jj'}_{kll'}\\
&\ =\ \mu + \tilde{\bm{T}}_i^\top\bbeta_k, 
\end{align*}
for each $k=1,2,\ldots,K$ where $\tilde{\bm{T}}_i$ is the vector of
indicators, $\mathbf{1}\{T_{ij} = l\}$ and
$\mathbf{1}\{T_{ij} = l, T_{ij'} = l'\}$, and $\bbeta_k$ is a stacked
column vector containing all coefficients for group $k$.  Inclusion of
higher-order interactions is straightforward (see Section~E of the
Supplementary Material) and hence is omitted in the main paper for notational simplicity.

For identification, we use the following ANOVA-type sum-to-zero
constraints,
\begin{equation}
\sum_{l =0}^{L_j-1} \beta^j_{kl} = 0, \quad \text{and} \quad
\sum_{l=0}^{L_{j}-1} \beta^{jj'}_{kll'} = \sum_{l'=0}^{L_{j'}-1} \beta^{jj'}_{kll'} = 0, \label{eq:sum-to-zero}
\end{equation}
for $j,j'=1,2,\ldots,J$ with $j' > j$.  We
write them compactly as,
\begin{equation}
\bC^\top \bbeta_k \ = \ \bm{0}, \label{eq:constraint}
\end{equation}
where each row of $\bC^\top \bbeta_k$ corresponds to one of the
constraints given in Equation~\eqref{eq:sum-to-zero}.

\subsection{Sparsity-inducing Prior}

Given the high dimensionality of this model, we use a
sparsity-inducing prior.  In our application, we have a total of 315
$\beta$ coefficients for each group.  In factorial experiments, it is
desirable to regularize the model such that certain levels of each
factor are fused together when their main effects and all interactions
are similar \citep{post2013factor,egam:imai:19}.  For example, we
would like to fuse levels $l_1$ and $l_2$ of factor $j$ if
$\beta^j_{l_1} \approx \beta^j_{l_2}$ and
$\beta^{jj'}_{l_1 l'} \approx \beta^{jj'}_{l_2 l'} $for all other
factors $j'$ and all of its levels $l'$.

We encourage such fusion by applying a structured sparsity approach of
\citet{goplerud2021sparsity} that generalizes the group and
overlapping group LASSO
\citep[e.g.,][]{yuan:lin:06,yan2017hierarchical} while allowing
positive semi-definite penalty matrices.  For computational
tractability, we use $\ell_2$ norm instead of the $\ell_\infty$ norm,
which is used in GASH-ANOVA \citep{post2013factor}.
An additional benefit of the use of regularization is that it gives us some protection 
against finding spurious relations \citep[see][]{gelman2012we}. 

For illustration, consider a simple example with one group and two
factors---factor one has three levels and factor two has two levels.
In this case, our penalty contains four terms,
\begin{equation*}
\begin{split}
& \sqrt{(\beta^1_0 - \beta^1_1)^2 +  (\beta^{12}_{00} -
    \beta^{12}_{10})^2 + (\beta^{12}_{01} -
    \beta^{12}_{11})^2}\\
  ~+~ &\sqrt{(\beta^1_0 - \beta^1_2)^2 +
    (\beta^{12}_{00} - \beta^{12}_{20})^2 +
    (\beta^{12}_{01}-\beta^{12}_{21})^2}\\
~+~ & \sqrt{(\beta^1_1 - \beta^1_2)^2 + (\beta^{12}_{10} - \beta^{12}_{20})^2 + (\beta^{12}_{11} - \beta^{12}_{21})^2}\\
~+~ & \sqrt{(\beta^2_0 - \beta^2_1)^2 +  (\beta^{12}_{00}-\beta^{12}_{01})^2 + (\beta^{12}_{10} - \beta^{12}_{11})^2 + (\beta^{12}_{20}-\beta^{12}_{21})^2} .
\end{split}
\end{equation*}
The first three terms encourages the pairwise fusion of the levels of
factor one whereas the fourth encourages the fusion of the two levels
of factor two. For compact notation, the penalty can also be written
using the sum of Euclidean norms of quadratic forms,
\begin{equation*}
 || \bm{\beta}^\top \bm{F}_1 \bm{\beta} ||_2 +  \ ||
 \bm{\beta}^\top \bm{F}_2 \bm{\beta} ||_2 + \ ||
 \bm{\beta}^\top \bm{F}_3 \bm{\beta} ||_2 + \ ||
 \bm{\beta}^\top \bm{F}_4 \bm{\beta} ||_2,  
\end{equation*}
where $\bF_1, \bF_2, \bF_3$ are appropriate positive semi-definite
matrices to encourage the fusion of the pairs of levels in factor one
and $\bF_4$ encourages the fusion of the two levels in factor two,
and \\
$\bbeta = [\beta^1_0\ \beta^1_1\ \beta^1_2\ \beta^2_0\ \beta^2_1\
\beta^{12}_{00}\ \beta^{12}_{10}\ \beta^{12}_{20}\ \beta^{12}_{01}\
\beta^{12}_{11}\ \beta^{12}_{21}]^\top$.  Note that the sum-to-zero
constraints make this type of fusion of factors together sensible for
sparsity.

We generalize this formulation to an arbitrary number of factors and
factor levels. For each factor that contains $L_j$ levels, we have
$\binom{L_j}{2}$ penalty matrices to encourage pairwise
fusion. Imposing additional constraints is a simple extension; 
for example, for ordered factors, one might use penalties that 
penalize the differences between adjacent levels (e.g. $l$ and $l+1$). 
Let $G = \sum_{j=1}^J \binom{L_j}{2}$ represent
the total number of penalty matrices.  For $g=1,2,\ldots,G$, we use
$\bF_g$ to denote a penalty matrix such that
$\sqrt{\bbeta^\top\bF_{g}\bbeta}$ is equivalent to the $\ell_2$ norm
on the vector of differences between all main effects and interactions
containing a main effect. We note that $\{\bm{F}_g\}_{g=1}^G$ is not
directly chosen but rather are determined by factors in the experiment
($J$, $L_j$, whether $j$ is ordered or unordered) and the included
interactions (as well as the use of ``latent overlapping groups''; see
Section~H.4 of the Supplementary Material. 

We interpret this penalty as a prior under our Bayesian framework
described in Section~\ref{subsec:kmeans},
\begin{equation}
\label{eq:prior_beta}
p\left(\bbeta_k \mid \{\bm{\phi}_k\}_{k=2}^K\right) \ \propto \ \left(\lambda {\bar{\pi}_k^\gamma}\right)^m \exp\left(-\lambda \bar{\pi}_k^\gamma \sum_{g=1}^{G} \sqrt{\bbeta_k^\top\bF_{g} \bbeta_k}\right), 
\end{equation}
where $\bar{\pi}_k = \sum_{i=1}^N \pi_k(\bX_i)/N$ and
$m = \text{rank}\left([\bF_{1}, \cdots, \bF_{G}]\right)$. We follow
existing work in allowing the penalty on the treatment effects
$\bm{\beta}_k$ to be scaled by the group-membership size
$\bar{\pi}_k$ when $\gamma = 1$
\citep{khal:chen:07,stadler2010lasso}.
On the other hand, when $\gamma = 0$ the $\bar{\pi}_k$ disappears, implying no use of the $\bX_i$ in the prior.
 We note that the prior on
$p(\bm{\beta} \mid \{\bm{\phi}_k\}_{k=2}^K)$ is guaranteed to be
proper when all pairwise fusions are encouraged by
$\{\bm{F}_g\}_{g=1}^G$, although in other circumstances it may be
improper
\citep{goplerud2021sparsity}.  Section~F of the Supplementary Material
provides additional details. Following \cite{zahid2013ridge}, we use a
normal prior distribution for the coefficients for the moderators.

The resulting regularization is invariant to the choice of baseline
group $\bphi_1 = \bm{0}$, which is the first row of the $K \times p_x$
coefficient matrix $\bphi$.  The prior distribution is given by
\begin{equation}
\label{eq:prior_phi}
p(\{\bm{\phi}_k\}_{k=2}^K) \ \propto \ \exp\left(-\frac{\sigma^2_\phi}{2}\sum_{l=1}^{p_x} [\bphi_{2l}, \cdots, \bphi_{Kl}]^\top \bm{\Sigma}_\phi [\bphi_{2l}, \cdots, \bphi_{Kl}]\right), 
\end{equation}
where $\bm{\Sigma}_\phi$ is a $(K-1) \times (K-1)$ matrix with
$[\bm{\Sigma}_\phi]_{kk'} = (K-1)/K$ if $k = k'$ and $[\bm{\Sigma}_\phi]_{kk'} =-1/K$
otherwise. We set $\sigma^2_\phi$ to $1/4$ for a relatively diffuse prior.

As noted in a recent survey, ``ensuring generic identifiability for
general [mixture of expert] models remains a challenging issue''
\cite[p. 294]{gormley2019mixture}.  Although mixtures with a Bernoulli
outcome variable are generally unidentifiable, several aspects of our
methodology are expected to alleviate the identifiability
problem. First, a typical conjoint analysis has repeated observations
per unit $i$ \citep{grun2008multinomial}.  Second, our model is a
mixture of experts rather than a mixture model
\citep{jiang1999identifiability}.  Third, our treatment variables,
which act as covariates in a mixture of experts, are randomized and
hence uncorrelated with one another. Lastly, our model regularizes the
coefficients through an informative prior.  While a formal
identifiability analysis of our model is beyond the scope of this
paper, the simulation analysis (Section~\ref{sec:simulation}) shows
that our model can accurately recover the coefficients in a realistic
setting. It is also possible to use a bootstrap-based procedure to
examine the identifiability issue in a specific setting
\citep{grun2008multinomial}.

\subsection{Estimation and Inference}\label{sec:est_inf}

We fit our model by finding a maximum of the log-posterior using an
extension of the Expectation-Maximization (EM;
\citealt{demp:lair:rubi:77}) algorithm known as the Alternating
Expectation-Conditional Maximization (AECM; \citealt{meng:vand:97})
algorithm. Equation~\eqref{eq:aecm_obs} defines our (observed)
log-posterior using the terms defined in
Equations~\eqref{eq:moe},~\eqref{eq:prior_beta},
and~\eqref{eq:prior_phi}, where we collect all model parameters as
$\bm{\theta}$:
\begin{equation}
\label{eq:aecm_obs}
\begin{split}\log p\left(\bm{\theta} \mid \{Y_i, \bX_i, \bT_i\}_{i=1}^N\right) \ = \ &\sum_{i=1}^N \log\left[\sum_{k=1}^K \pi_k(\bX_i) \zeta_k(\bT_i)^{Y_i} \{1-\zeta_k(\bT_i)\}^{1-Y_i}\right] + \\ &\sum_{k=1}^K \log p(\bm{\beta}_k \mid
\{\bm{\phi}\}_{k=2}^K) + \log p(\{\bm{\phi}_k\}_{k=2}^K) + \text{const}.
\end{split}
\end{equation}

For now, we assume the value of regularization parameter $\lambda$ is
fixed, although we discuss this issue in
Section~\ref{sec:additional}. The linear constraints on $\bm{\beta}_k$
given in Equation~\eqref{eq:constraint} still hold but are suppressed for notational simplicity.

Section~G of the Supplementary Material provides a full derivation of our
AECM algorithm; each iteration involves two cycles where the data
augmentation scheme enables iterative updating of the treatment effect
parameters $\bm{\beta}$ and moderators $\bm{\phi}$. After augmenting
with missing data, the update for $\bm{\beta}$ can be done using ridge
regression; Section~G.1 addresses the linear
constraints imposed by $\bC^T\bm{\beta}_k = \bm{0}$. The update for
$\bm{\phi}$ can be performed using a modified version of a multinomial
logistic regression based on a standard optimizer (e.g., L-BFGS) (see
Section~G.2).

\subsection{Additional Considerations}
\label{sec:additional}

Since fitting the proposed model is computationally expensive, we use the Bayesian Information Criteria (BIC), rather than cross validation, to
select the value of the regularization parameter $\lambda$
\citep{khal:chen:07,khalili2010mixture,chamroukhi2019regularized}. Section~G.4
of the Supplementary Material
presents our degrees-of-freedom estimator and explains how we tune
$\lambda$ using Bayesian model-based
optimization. Section~G.5 discusses
additional details of our EM algorithm including initialization and
techniques to accelerate convergence.

We extend the above model and estimation algorithm to accommodate
common features of conjoint analysis: (1) repeated observations for
each individual respondent (Section~H.1 of the Supplementary Material),
(2) a forced choice conjoint design
(Section~H.2), and (3) standardization
weights for factors with different numbers of levels $L_j$
(Section~H.3). Lastly, our
experience suggests that the proposed penalty function, which consists
of overlapping groups, often finds highly sparse
solutions.  Section~H.4 details the integration of the
latent overlapping group formulation of \cite{yan2017hierarchical}
into our framework to address this issue.

Once the model parameters are estimated, we can compute quantities of
interest such as the AMCEs, defined in Equation~\eqref{eq:AMCE}.  We
do this separately for each group, such that
$\delta_{jk} (l, l^\prime)$ is the AMCE for factor $j$, changing from
level $l^{\prime}$ to $l$ in group $k$.  Our estimator is the average
of the estimated difference in predicted responses when changing from
level $l^{\prime}$ to $l$ of factor $j$, where the average is taken
over the empirical distribution of the assignment on the other
factors.  This estimation is described in more detail in
Section~I of the Supplementary Material under various settings.  We can
use the empirical distribution here because treatment is randomly
assigned.

To quantify the uncertainty of the parameter estimates, we rely on a
quadratic approximation to the log posterior distribution. To ensure
its differentiability, we follow a standard approach in the
regularized regression literature \citep[e.g.,][]{fan2001variable} and
fuse pairwise factor levels that are sufficiently close
together. Section~J of the Supplementary Material describes this process,
deriving the Hessian of the log-posterior using \cite{loui:82}'s
method and then using the delta method for inference on other of
quantities of interest, e.g., the AMCE.

Finally, in principle, our framework does not assume a ``correct''
data generating process. The choice of number of groups $K$ should
depend on the desired granularity of discovered heterogeneity, with
more groups leading to finer levels of heterogeneity.  Similarly, the
choice of moderators should reflect the researcher's substantive
interests.  Section~K.3 of the Supplementary Material shows performance of our method
across different values of $K$ and different specifications of the
moderators when the true data generating process is a mixture
model. As expected, the bias of AMCE is not affected by changing the
specification of these parameters. However, there are some impacts on
the estimation of conditional effects in terms of precision.

Common data-driven approaches for choosing $K$ include
use of an information criterion such as the BIC; however, while we
find that these approaches work well under simulation settings (see
Section~K.3.1 for demonstration), they can
perform poorly in practice (see Section~L),
especially when the component densities are mis-specified or not
especially well separated \citep{celeux2019mixture}. Thus, even if a
data-driven heuristic is used as a guide for choosing $K$, we suggest
comparing different $K$ as illustrated in Section~\ref{sec:analysis}.

\section{Simulations}\label{sec:simulation}

We explore the performance of our method using a simple but realistic
simulation study.  Specifically, we consider the case of a conjoint
experiment with ten factors ($J = 10$) each with three levels
($L_j = 3$). To evaluate the performance of the proposed method, we
consider two different settings; in the first, we assume there are
1,000 respondents who each perform five comparison tasks. In the
second, we assume a larger experiment with 2,000 respondents who each
perform ten tasks.

In all cases, we assume that the data generating process follows a
mixture of experts model with three groups ($K=3$).  We calibrate the
true $\bm{\beta}_k$ such that the implied average marginal component
effects (AMCE) are comparable in magnitude to the empirical effects
presented in Section~\ref{sec:analysis}. We use a set of five
correlated continuous moderators and an intercept to again mimic a
realistic empirical setting and choose $\{\bm{\phi}_k\}_{k=2}^3$ to
relatively clearly separate respondents into different groups.
Section~K of the Supplementary Material presents complete description of
the simulation settings and the true parameter values used for the
$\bm{\beta}_k$ and marginal effects.

For each sample size, we independently generate 1,000 simulated data sets by drawing $N$ observations of moderators, randomly assigning a group membership to each observation based on the implied probabilities given their moderators, and generating the observed treatment
profiles completely at random. We fit our model to the data with
$K = 3$ and examine the average marginal component effects in each
group with respect to the first baseline level.

\begin{figure}[!t]
  \centering 
  \includegraphics[width=\textwidth]{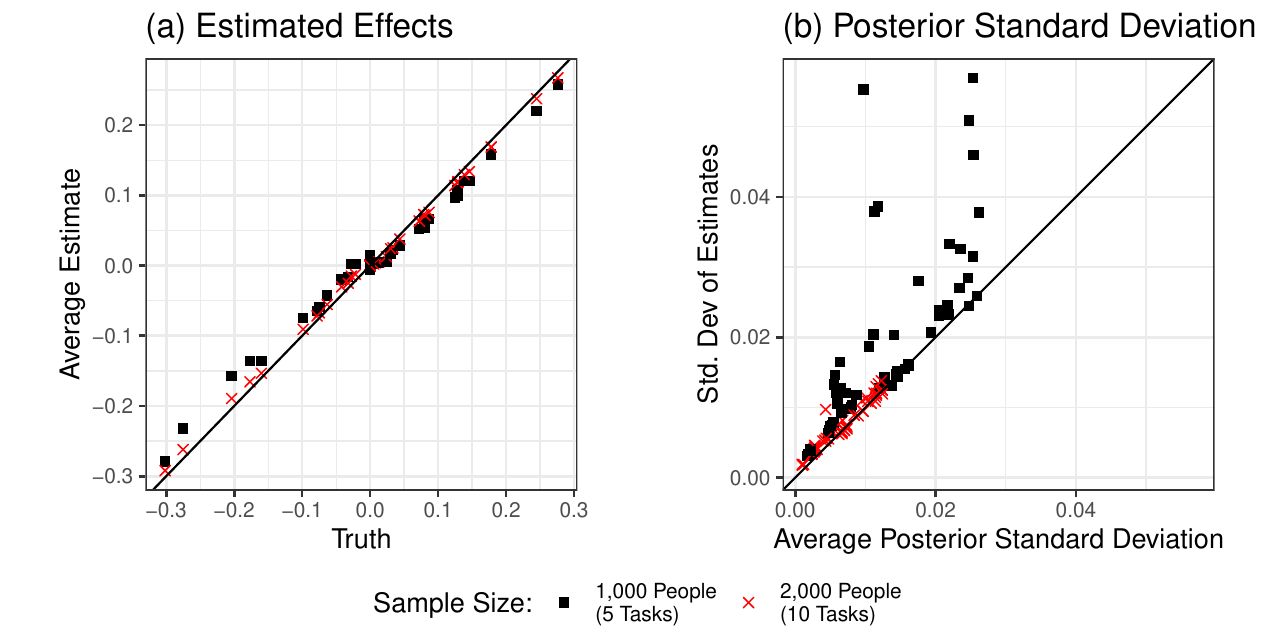}
  \caption{The empirical performance of the proposed estimator on
    simulated data. The black squares indicate the effects estimated
    for each group with the smaller sample size (1,000 people
    completing 5 tasks); the red crosses indicate effects estimated
    with the larger sample size (2,000 people completing 10 tasks).}
  \label{fig:simulation}
\end{figure}

Fig~\ref{fig:simulation} summarizes our results (see
Section~K.2 for the results regarding
the estimated coefficients $\bm{\beta}_k$). The left panel illustrates
a high correlation between the estimated effects and their true values
($\rho = 0.995$ for smaller sample size; $\rho = 0.999$ for larger
sample size). While the performance overall is reasonably strong, we
see that even when the dataset is large there is some degree of
attenuation bias due to shrinkage.

The right panel shows the frequentist evaluation of our Bayesian
posterior standard deviations.  We compare the average posterior
standard deviation against the standard deviation of the estimated
effects across the 1,000 Monte Carlo simulations.  The average
posterior standard deviations are noticeably smaller than the standard
deviation of the estimates when the sample size is small.  For the
large sample size, however, our approximate Bayesian posterior
standard deviations in this simulated example are roughly the same
magnitude of the standard deviation of the sampling distribution of
the estimator.

Even though our method's frequentist coverage is somewhat below the
nominal level in small samples, this undercoverage appears to be
primarily attributable to the shrinkage bias in our regularized
estimation rather than the large sample discrepancy between our
posterior standard deviations and the corresponding standard deviation
of sampling distribution.

Section~K.2 explores one way to
address the limitations of the default estimator by exploring sample
splitting and refitting the model given the estimated sparsity pattern
(i.e., which levels are fused together) and moderator effects
($\{\bm{\phi}_k\}_{k=2}^K$) on half of the data. This results in
smaller bias and improved coverage at both sample sizes.

Section~K.3 explores how when the true data generating
process is a mixture model, the ``wrong'' choice of $K$, e.g.
$K \in \{1,2,4\}$, as well as not using moderators (i.e., $\bX_i = 1$)
or using moderators in a different specification than the true model
impacts our results. In both settings, there is limited impact in
terms of bias in terms of estimating the AMCE, although both types of
misspecification incur a penalty in terms of root mean-squared error.

\section{Empirical Analysis}\label{sec:analysis}

In this section, we apply our methodology to the immigration conjoint
data introduced in Section~\ref{sec:con}.  We find evidence of effect
heterogeneity for immigrant choice based on respondent
characteristics.  In particular, the immigrant's country of origin
plays a greater role in forming the immigration preference of
respondents with increased prejudice, as measured by a Hispanic
prejudice score.  Outside of this group, which accounts for about one
third of the respondents, the country of origin factor plays a much
smaller role.

\subsection{Data and Model}\label{sec:analysis:data_model}

Following the original analysis, our model includes indicator variables
for each factor and interactions between country and reason of
application factors as well as those between education and job factors
in order to account for the restricted randomization.  We additionally
include interactions between country and job as well as those between
country and education, in accordance with the skill premium theory of
\cite{newman2019economic}.  This theory hypothesizes that prejudiced
individuals prefer highly skilled immigrants only for certain
immigrant countries.  This results in a total of 41 AMCEs and 222
average marginal interaction effects (AMIEs) for each group. 

For modeling group membership, we include the respondents' political
party, education, demographics of their ZIP code (we follow the
original analysis and include the variables indicating whether
respondents' ZIP code had few immigrants, meaning $<5\%$, and for
those from ZIPs with more than $5\%$ foreign-born, whether the
majority were from Latin America), ethnicity, and Hispanic prejudice
score.  The Hispanic prejudice score was used by
\cite{newman2019economic}, though we negate it to make lower values
correspond to lower prejudice for easier interpretation.  The score is
based on a standardized (and negated) feeling thermometer for
Hispanics.  The score ranges from $-1.61$ to $2.11$ for our sample,
where higher scores indicate higher levels of prejudice.

We remove respondents who are themselves Hispanic since the Hispanic prejudice score was not measured for these
respondents.  After removing entries with missing data, we have a
sample of 1,069 respondents. Most respondents evaluated five pairs of
profiles, though five respondents have fewer responses in the data set
used.  The total number of observations is 5,337 pairs of profiles.
We do not incorporate the survey weights into our analysis to better
demonstrate our methods though it is possible to include them.

The original experiment was conducted using the forced choice design,
in which a respondent chooses one profile out of a pair of immigrant
profiles.  We follow \cite{egam:imai:19} and model the choice as a
function of differences in treatments as follows,
\begin{align*}
\psi_k(\bT_i^L, \bT_i^R) \ = \ &  \mu + \sum_{j=1}^J \sum_{l \in L_j} \beta^j_{kl}\left(\mathbf{1}\left\{T^L_{ij}=l\right\}-\mathbf{1}\left\{T^R_{ij}=l\right\}\right)\\
& + \sum_{j=1}^{J-1} \sum_{j' > j} \mathbf{1}\left\{\mathcal{I}(j, j')\right\}\sum_{l \in L_j} \sum_{l' \in L_{j'}} \beta^{jj'}_{kll'}\left(\mathbf{1}\left\{T^L_{ij}=l, T^L_{ij'}=l'\right\}-\mathbf{1}\left\{T^R_{ij}=l, T^R_{ij'}=l'\right\}\right),
\end{align*}
where $\bT_i^L$ and $\bT_i^R$ represent the factors for the left and
right profiles and $\mathcal{I}(j, j') = 1$ if an interaction between
$j$ and $j'$ is include in the model. The outcome variable $Y_i$ is
equal to 1 if the left profile is selected and is equal to 0 if the
right profile is chosen.

To account for randomization restrictions, we include interactions
between country of origin and reason for applying as well as between
job and education.  To test relevant theories, we include additional
interactions between country of origin and job as well as country of
origin and education. These interaction effects proved
to be very small in magnitude (see Section~L of the Supplementary Material). Thus, we do not explore higher
order interactions given the commonly adopted principle of hierarchy
and sparsity \citep{wu2021experiments}, which implies that lower-order
effects are expected to be more significant than higher-order effects
and we should expect an even smaller number of nonzero effects. 
With this linear predictor formulation, the estimation
and inference proceed as explained in Section~\ref{sec:methods}.

We conduct two analyses, one with two groups and the other with three
groups.  These two models perform equally well in terms of
out-of-sample classification, a data-driven measure that can be used
to choose the number of groups.  Using more than three groups does not
give improved substantive insights and provides little improvement in
model performance.  As noted previously, each analysis optimizes the
BIC to calibrate the amount of regularization and employs
standardization weights to account for factors with different number
of levels (see Sections~G.4~and~H.3 of the Supplementary Material,
respectively, for details). We treat education and job experience as
ordered factors and only penalize the differences between adjacent
levels.

We report our findings using only the full data estimates, i.e.,
without the sample splitting explored in Section~K.2.
Initial experiments found
that the results were somewhat sensitive to specific folds chosen, and
thus we report only the full data results in the main
text. Section~L illustrates the distribution
of estimates across twenty different sample splits.

\subsection{Estimated Heterogeneity}

We focus on the AMCE for each factor as the quantity of interest and
separately estimate it for each group.  Under our model for the forced
choice design, the AMCE of level $l$ versus level $l^\prime$ of factor
$j$ within group $k$ can be written as,
\begin{align*}
\delta_{jk}(l, l^\prime) & \ = \ \frac{1}{2}\E \left[
                         \left\{\Pr\left(Y_i = 1 \mid Z_i=k, T^L_{ij}=l,
                         \bT^L_{i,-j}, \bT^R_{i}\right) -
                         \Pr\left(Y_i = 1 \mid Z_i=k, T^L_{ij}=l^\prime,
                         \bT^L_{i,-j}, \bT^R_{i}\right)
                         \right\}\right. \\
  & \hspace{.5in} \left.
                         +\left\{\Pr\left(Y_i = 0 \mid Z_i=k, T^R_{ij}=l,
                         \bT^R_{i,-j}, \bT^L_{i}\right) -
                         \Pr\left(Y_i = 0 \mid Z_i=k, T^R_{ij}=l^\prime,
                         \bT^R_{i,-j}, \bT^L_{i}\right)
    \right\}\right]. 
\end{align*}
The expectation is over the population of respondents and the
distribution of the factors not involved in this AMCE.  That is, we
compute the AMCE separately for the left and right profiles and then
average them to obtain the overall AMCE.  We estimate this quantity
using the fitted model and averaging over the empirical distribution
of the factorial treatments.

\begin{figure}[t!]
\centering 
\includegraphics[width=\textwidth]{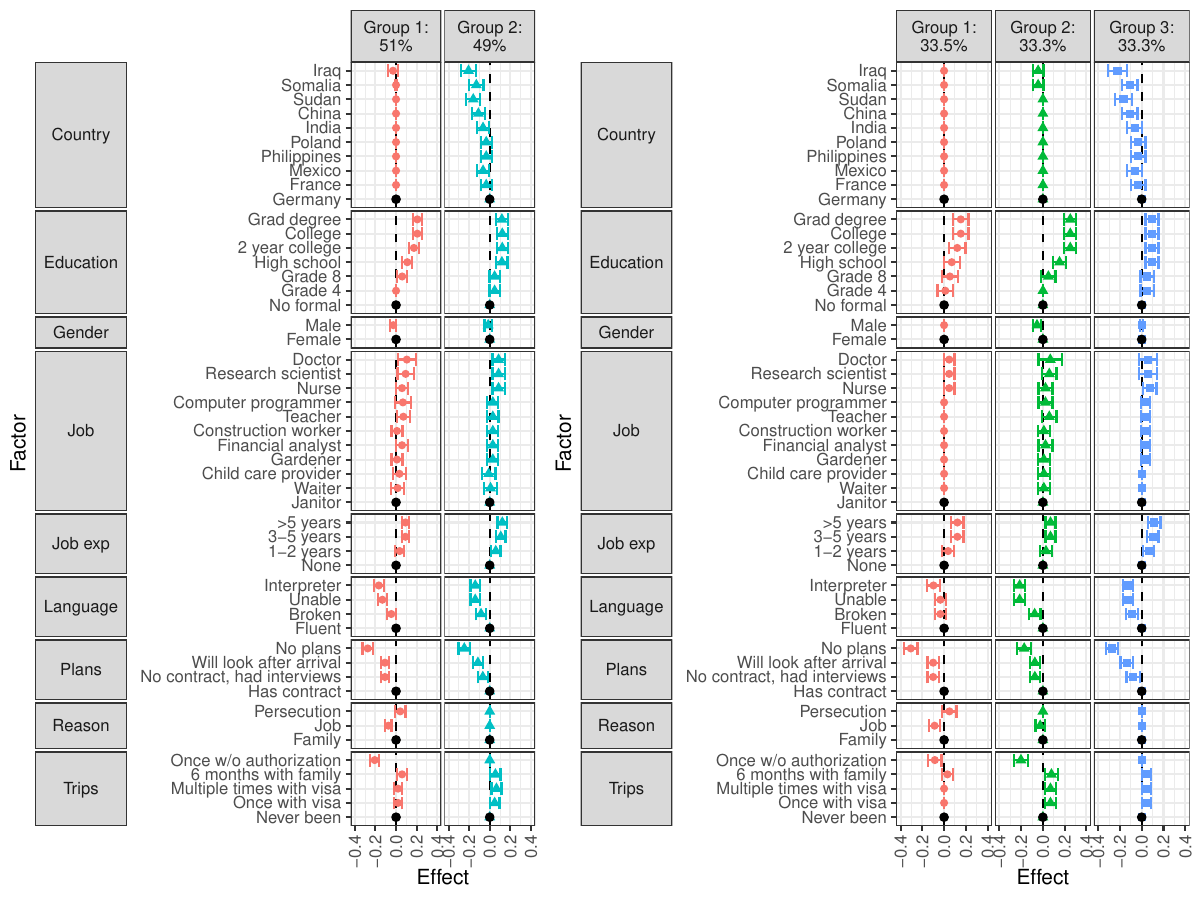}
\caption{Estimated average marginal component effects using a
  two-group (left) and three-group (right) analysis. The point
  estimates and 95\% Bayesian credible intervals are shown. A solid
  circle represents the baseline level of each factor.  Numbers after
  colons give average posterior predictive probabilities for each
  group.} \label{fig:ame}
\end{figure}

Fig~\ref{fig:ame} presents the estimated AMCEs and their 95\% Bayesian
credible intervals for the two-group and three-group analyses in the
left and right panels, respectively. Group~2 in the two-group analysis
and Group~3 in the three-group analysis display stronger impacts of
country of origin than the other groups.  The respondents in these
groups give the most preference to immigrants from Germany and the
least preference to immigrants from Iraq (followed by Sudan).  The
significant negative effects of Iraq in Group~2 of the two-group
analysis and Group~3 of the three-group analysis are consistent with
the significant negative effect for Iraq found by
\cite{hainmueller2015hidden}.  The patterns we observe for the other
factors are also similar for these two groups in the two analyses.

Across all groups, respondents prefer educated and experienced
immigrants who already have contracts (over those who have no
contracts or plans).  Respondents also prefer immigrants who have
better language skills, although this feature matters less for
respondents in Group~1 of the three group analyses.

For both analyses, the respondents in Group~1 do not care much about
immigrant's country of origin.  Instead, they place a greater emphasis
on education and reason for immigration when compared to those in the
other groups.  While the differences between Groups~1~and~2 in the
three-group analysis are generally substantively small, those in Group~2 appear
to place more emphasis on education and prior entry without legal
authorization.  Those in Group~1, on the other hand, give a slight
benefit to immigrants whose reason for immigration is persecution.

\begin{figure}[t!]
\centering 
\includegraphics[width=\textwidth]{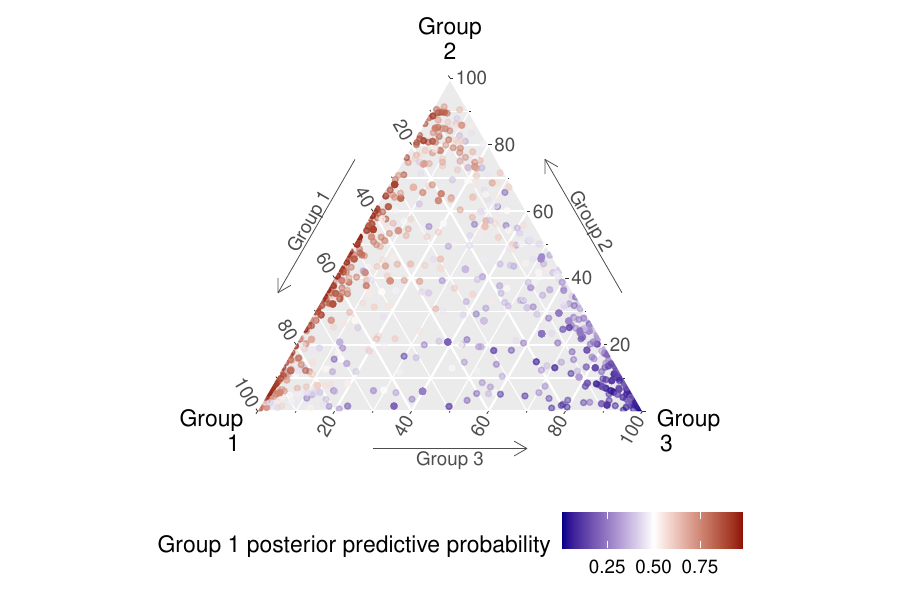}
\caption{Ternary plot of the joint posterior predictive probability of
  belonging to each group in the three-group analysis (three axes)
  where the color of each dot represents the posterior predictive
  probability of belonging to Group 1 under the two-group analysis.
} \label{fig:ternary}
\end{figure}

Indeed, for the three-group analysis, Groups~1~and~2 together
correspond roughly to Group~1 of the two-group analysis.  In fact,
about 81\% of the respondents who belong to Group~1 of the two-group
analysis are the members of either Group~1~or~2 in the three-group
analysis, using a weighted average of their estimated group membership
posterior predictive probabilities.

Fig~\ref{fig:ternary} visualizes these posterior predictive
probabilities of group membership under the three-group analysis with
each dot colored by the posterior predictive probability of belonging
to Group~1 under the two-group analysis.  According to this ternary
plot, those observations that are likely to be part of Group~1 under
the two-group analysis (i.e., red dots) are likely to be split between
Groups~2~and~3 under the three group analysis.  In contrast, those who
have a high probability of belonging to Group~2 under the two-group
analysis (i.e., blue dots) tend to be part of Group~3.  

Fig~\ref{fig:ame} shows fusion of various factor levels due to
regularization.  The levels being fused appear sensible.  For example,
``doctor'' and ``research scientist,'' both occupations requiring high
levels of education, are consistently fused together.  For education,
use of the ordinal structure ensures only adjacent levels can be
fused. We see sensible cut points for fusion; in the two group
analysis, Group~1 differentiates individuals who have at least a
college degree and Group~2 differentiates individuals who have at
least a high school degree.

The comparison of AMCEs across subgroups can be misleading as they
depend on the choice of baseline category
\citep{leeper2020measuring}. Section~L of the Supplementary Material presents an alternative quantity that avoids issues of baseline dependency (marginal means; \citealt{leeper2020measuring}). The results are generally similar to AMCEs shown above.

\begin{figure}[t!]
\centering 
\includegraphics[width=\textwidth]{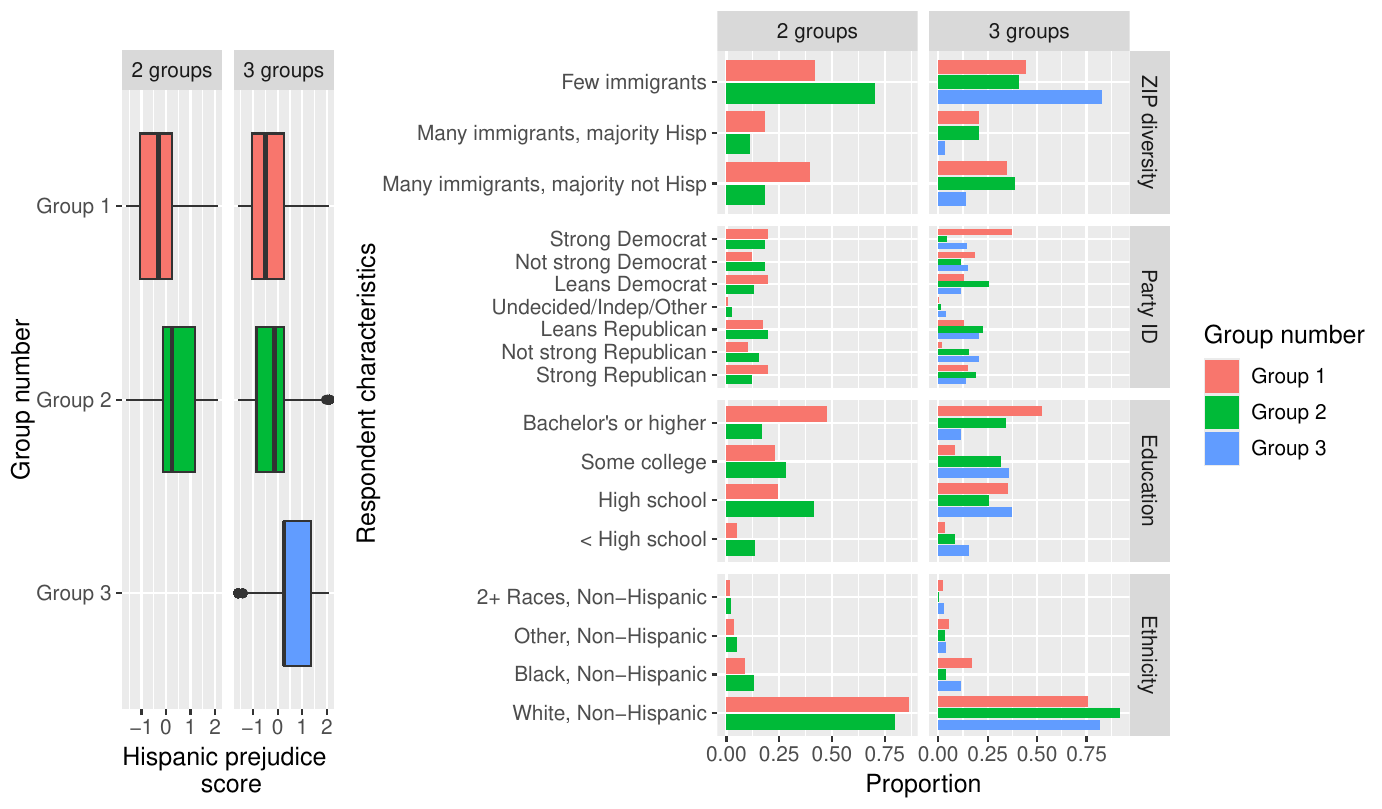}
\caption{Distribution of respondent characteristics for each group.
  Left set of plots shows weighted box plots of the Hispanic prejudice
  moderator within each group over the posterior predictive
  distribution using a two-group (left) and three-group (right)
  analysis.  Right set of plots shows the distribution of categorical
  moderators within each group over the posterior predictive
  distribution using a two-group (left) and three-group (right)
  analysis. 
} \label{fig:mod2}
\end{figure}

\subsection{Group Membership}

Who belongs to each group?  The left panel of Fig~\ref{fig:mod2} shows
the distribution of Hispanic prejudice score for each group weighted
by the corresponding posterior predictive group membership probability
for each individual respondent.  The plot shows that for the two-group
analysis, those with high prejudice score are more likely to be part
of Group~2. For the three-group analyses, those with high prejudice
are more likely to be in Group~3. This is consistent with the finding
above that the respondents in those groups put more emphasis on
immigrant's country of origin.

The right panel of the figure shows the distribution of other
respondent characteristics.  In general, Group~2 in the two-group
analysis and Group~3 in the three-group analysis consist of those who
live in ZIP codes with few immigrants and have lower educational
achievements.  For the three-group analysis, those in Group~2 tend to
be Republicans, whereas those in Group~1 are more likely to be
Democrats.  This is consistent with the finding of a larger penalty
for entry without legal authorization in Group~2.  Group~3 contains a
mix of political ideologies, though it has more respondents who
identify as Undecided/Independent/Other or not strong Republican than
the other two groups.

Which respondent characteristics are predictive of the group
membership?  In addition to the covariate distribution for each group
shown in Fig~\ref{fig:mod2}, we can also find how important each
moderator is in predicting group membership, conditional on all other
moderators.  We examine how the predicted probabilities of group
memberships change across respondents with different characteristics.
Specifically, we estimate
\begin{equation}
\label{eq:mfx_moderator}
\mathbb{E}\left[\pi_k(X_{ij} = x_1, \bm{X}_{i,-j}) - \pi_{k}(X_{ij} =
  x_0, \bm{X}_{i,-j})\right]
\end{equation}
where $x_0$ and $x_1$ are different values of covariate of interest
$X_{ij}$.  If $X_{ij}$ is a categorical variable, we set $x_0$ to the
baseline level and $x_1$ to the level indicated on the vertical axis.
If $X_{ij}$ is a continuous variable as in the case of the Hispanic
prejudice score, then $x_0$ and $x_1$ represent the 25th and 75th
percentile values.  The solid arrows represent whether the
corresponding 95\% Bayesian credible interval covers zero or
not. Section~L of the Supplementary Material shows the effect of changing a
moderator on the absolute value of the changes in predicted
probabilities of group membership. In some cases, changing a moderator
shows a small average change but a larger average of absolute changes.

\begin{figure}[!t]
\centering
\includegraphics[width=\textwidth]{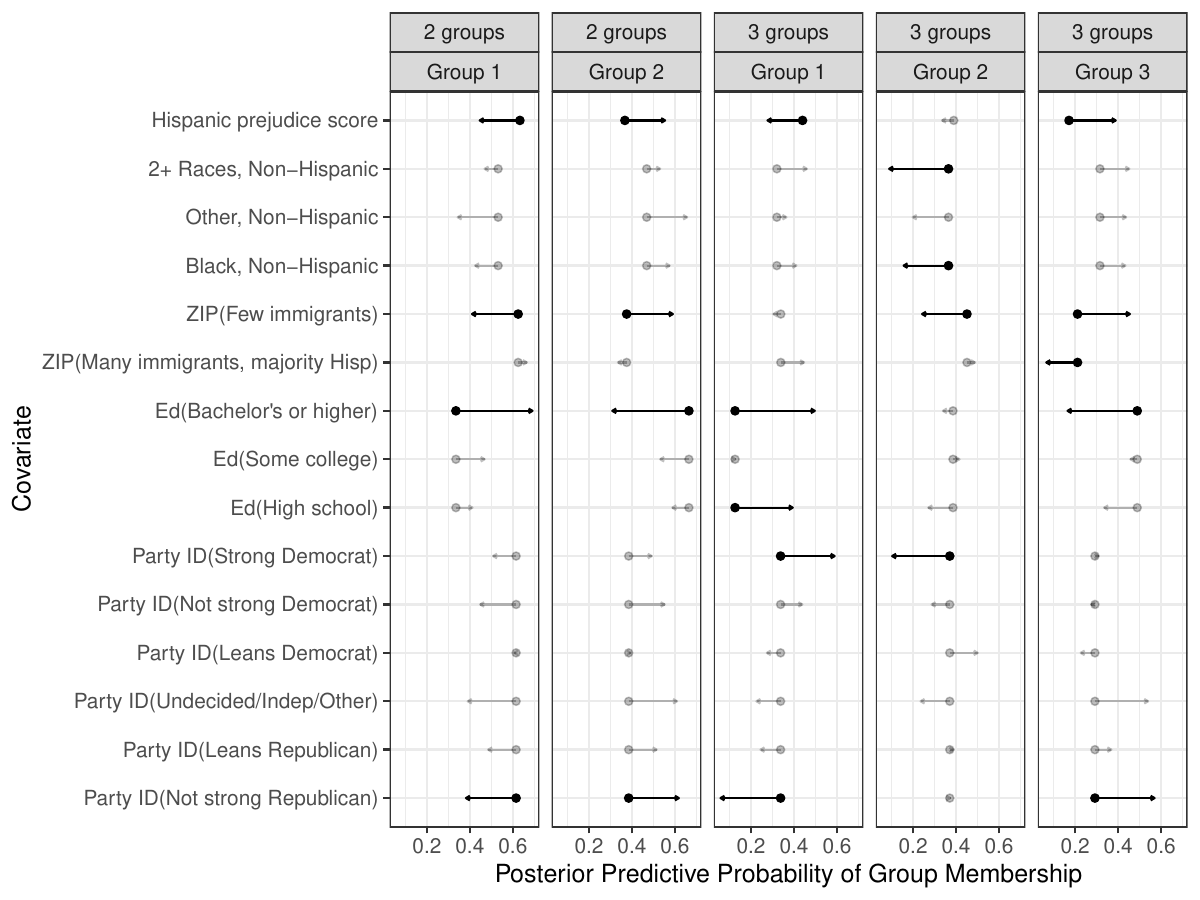}
\caption{The impact of moderator values on likelihood of being
  assigned to groups, for two-group (left two plots) and
  three-group (right three plots) analysis. 
 Dark arrows indicate that there is a significant effect of the moderator on group membership, i.e., that the corresponding quantity defined in Equation~\eqref{eq:mfx_moderator} is statistically significant.} \label{fig:mod}
\end{figure}

Consistent with the earlier findings, Fig~\ref{fig:mod} shows that
those with high Hispanic prejudice scores tend to be part of Group~2
in the two-group analysis and Group~3 in the three-group analysis even
after controlling for other moderators.  These respondents are also
less likely to be members of Group~1 in both analyses. Party ID also plays a statistically significant role (indicated by dark arrow). Controlling for other factors, in the three-group
analysis, not strong Republicans tend to be part of Group~3 and more strong Democrats belonging to Group~1.  On average, respondents in Group~1 have higher education in both analyses. 

Finally, we estimate the average marginal interaction effects (AMIEs)
between two factors \citep{egam:imai:19}, which can be computed by
subtracting the two AMCEs from the average effect of changing the two
factors of interest at the same time.  Thus, the AMIE represents the
additional effect of changing the two factors beyond the sum of the
average effects of changing one of the factors alone.  Formally, we
can define the AMIE of changing factors $j$ and $j'$ from levels $l_j$
and $l_{j'}$ to levels $l_j^\prime$ and $l_{j'}^\prime$, respectively,
as follows, 
\begin{equation*}
  \E[Y_i(T_{ij} = l_j, T_{ij'} = l_{j'}, \bT_{i,-j,-j'}) - Y_i(T_{ij}
  = l_j^\prime, T_{ij'} = l_{j'}^\prime, \bT_{i,-j,-j'})] -
  \delta_j(l_j, l_j^\prime) - \delta_{j'}(l_{j'}, l_{j'}^\prime).
\end{equation*}

All of the AMIE effects found are quite small, so we do not present
those results here.  According to the skill-premium theory of
\cite{newman2019economic}, we expect to find an interaction between
job and country or education and country, in at least some groups.
Unfortunately, our analysis does not find support for this hypothesis.

\subsection{Comparison to an alternative method}\label{subsec:cjbart}

While there exist few methods to estimate heterogeneous effects of
high-dimensional treatments, an exception is \cite{robinsondetect},
who develop a BART-based method for analyzing heterogeneity in
conjoint experiments.  The primary goal of their method is the
estimation of the conditional average marginal effects (CAMCE) for
each individual given their covariate values.

While our method is motivated by a different goal---finding an
interpretable set of groups with distinctive treatment effects---our
method can also produce estimates of the CAMCE for any set of
covariates. The two methods can be compared in this task by examining
CAMCE. Formally, under our model, the CAMCE for factor $j$ comparing
levels $l$ and $l'$ for covariates $\bX_i$ is a weighted average of
the group-specific AMCEs, denoted by $\delta_{jk}(l,l')$.
\begin{equation}
\mathrm{CAMCE}_{j}(l, l'; \bX_i) = \sum_{k=1}^K \delta_{jk}(l,l')
\pi_k(\bX_i) \label{eq:camce}
\end{equation}
By plugging in our estimates $\hat{\pi}_k(\bX_i)$ and
$\hat{\delta}_{jk}(l,l')$, we can estimate the CAMCE.

Section~B of the Supplementary Material compares the estimated CAMCE
obtained from our method and \cite{robinsondetect}'s (\texttt{cjbart})
using the same moderators and treatments. Our method discovers a
considerable degree of heterogeneity in the CAMCEs whereas
\texttt{cjbart} shows limited treatment effect variation for most
countries. Under our model, the estimated heterogeneous effects are
more strongly associated with predictors than \texttt{cjbart}; for
example, our method finds a clear association, on average, between the
estimated CAMCE and prejudice or party identification whereas
\texttt{cjbart} does not.

\section{Concluding Remarks}\label{sec:disc}

We have shown that a Bayesian mixture of regularized logistic
regressions can be effectively used to estimate heterogeneous
treatment effects of high-dimensional treatments.  The proposed
approach finds maximally heterogeneous groups and yields interpretable
results, illuminating how different sets of treatments have
heterogeneous impacts on distinct groups of units.  We apply our
methodology to conjoint analysis, which is a popular survey
experiment.  Our analysis shows that individuals with high prejudice
score tend to discriminate against immigrants from certain
non-European countries.  These individuals tend to be less educated
and live in areas with few immigrants.  Future research should
consider the derivation of optimal treatment rules in this setting as
well as the empirical evaluation of such rules.  Another important
research agenda is the estimation of heterogeneous effects of
high-dimensional treatments in observational studies.

\bibliographystyle{pa}
\bibliography{het_ref,my,imai}

@Article{aoas_supp,
  author = 	 {Goplerud, Max and Imai, Kosuke and Pashley, Nicole E.}, 
  title = 	 {Supplementary Material for ``Estimating Heterogeneous Causal Effects of High-Dimensional Treatments: Application to Conjoint Analysis''},
  journal = 	 {Annals of Applied Statistics},
  year = 	 {in-press},
  OPTkey = 	 {},
  OPTvolume = 	 {},
  OPTnumber = 	 {},
  OPTpages = 	 {},
  OPTmonth = 	 {},
  OPTnote = 	 {},
  OPTannote = 	 {}
}

@TechReport{cher:etal:19,
  author = 	 {Chernozhukov, Victor and Demirer, Mert and Duflo,
                  Esther and {Fernandez-Val}, Ivan},
  title = 	 {Generic Machine Learning Inference on Heterogeneous
                  Treatment Effects in Randomized Experiments},
  institution =  {arXiv:1712.04802},
  year = 	 {2019},
  OPTkey = 	 {},
  OPTtype = 	 {},
  OPTnumber = 	 {},
  OPTaddress = 	 {},
  OPTmonth = 	 {},
  OPTnote = 	 {},
  OPTannote = 	 {}
}

@article{hainmueller2014causal,
  title={Causal Inference in Conjoint Analysis: Understanding Multidimensional Choices via Stated Preferenc<e Experiments},
  author={Hainmueller, Jens and Hopkins, Daniel J and Yamamoto, Teppei},
  journal={Political Analysis},
  volume={22},
  number={1},
  pages={1--30},
  year={2014},
  publisher={Cambridge University Press}
}

@article{de2022improving,
  title={Improving the External Validity of Conjoint Analysis: The Essential Role of Profile Distribution},
  author={De la Cuesta, Brandon and Egami, Naoki and Imai, Kosuke},
  year={2022},
  journal={Political Analysis},
  volume={30},
  number={1},
  month={January},
  pages={19--45}
}

@article{hainmueller2015hidden,
  title={The Hidden American Immigration Consensus: A Conjoint Analysis of Attitudes toward Immigrants},
  author={Hainmueller, Jens and Hopkins, Daniel J},
  journal={American Journal of Political Science},
  volume={59},
  number={3},
  pages={529--548},
  year={2015},
  publisher={Wiley Online Library}
}

@article{newman2019economic,
  title={Economic Reasoning with Racial Hue: Is the Immigration Consensus Purely Race Neutral?},
  author={Newman, Benjamin J and Malhotra, Neil},
  journal={The Journal of Politics},
  volume={81},
  number={1},
  pages={153--166},
  year={2019},
  publisher={University of Chicago Press Chicago, IL}
}

@article{DasPilRub15,
  title =        {Causal Inference from {$2^K$} Factorial Designs by
                  using Potential Outcomes},
  volume =       77,
  language =     {English},
  number =       4,
  journal =      {Journal of the Royal Statistical Society. Series B
                  (Statistical Methodology)},
  author =       {Dasgupta, Tirthankar and Pillai, Natesh S. and
                  Rubin, Donald B.},
  month =        sep,
  year =         2015,
  pages =        {727--753}
}

@article{Rubin80,
  author =       {Donald B. Rubin},
  journal =      {Journal of the American Statistical Association},
  number =       371,
  pages =        {591--593},
  title =        {Randomization Analysis of Experimental Data: The
                  Fisher Randomization Test Comment},
  volume =       75,
  year =         1980
}

@article{bondell2009anova,
	year  = {2009},
	volume = {65},
	number = {1},
	pages = {169--177},
	author = {Howard D. Bondell and Brian J. Reich},
	title = {Simultaneous Factor Selection and Collapsing Levels in {ANOVA}},
	journal = {Biometrics}
}

@article{post2013factor,
	title={Factor Selection and Structural Identification in the Interaction ANOVA Model},
	author={Post, Justin B. and Bondell, Howard D.},
	journal={Biometrics},
	volume={69},
	number={1},
	pages={70--79},
	year={2013}
}

@article{yan2017hierarchical,
	title={Hierarchical Sparse Modeling: A Choice of Two Group Lasso Formulations},
	author={Yan, Xiaohan and Bien, Jacob},
	journal={Statistical Science},
	volume={32},
	number={4},
	pages={531--560},
	year={2017}
}

@article{stadler2010lasso,
	title={$\ell$-1-Penalization for Mixture Regression Models},
	author={St{\"a}dler, Nicolas and B{\"u}hlmann, Peter and Van De Geer, Sara},
	journal={Test},
	volume={19},
	number={2},
	pages={209--256},
	year={2010},
	publisher={Springer}
}

@article{ratkovic2017sparse,
	title = {Sparse Estimation and Uncertainty with Application to Subgroup Analysis},
	author = {Ratkovic, Marc and Tingley, Dustin},
	volume = {25},
	number = {1},
	pages = {1--40},
	year = {2017},
	journal = {Political Analysis}
}

@article{polson2013polyagamma,
	journal = {Journal of the American Statistical Association},
	volume = {108},
	number = {504},
	pages = {1339--1349},
	title = {Bayesian Inference for Logistic Models Using P{\'o}lya–Gamma Latent Variables},
	author = {Polson, Nicholas G. and Scott, James G. and Windle, Jesse},
	year = {2013}
}

@article{figueiredo2003adaptive,
	journal = {IEEE Transactions on Pattern Analysis and Machine Intelligence},
	title = {Adaptive Sparseness for Supervised Learning},
	author = {M\'{a}rio A.T. Figueiredo},
	year = {2003},
	volume = {25},
	number = {9},
	pages = {1150--1159}
}

@article{polson2011svm,
	title = {Data Augmentation for Support Vector Machines},
	author = {Polson, Nicholas G. and Scott, Steve L.},
	journal = {Bayesian Analysis},
	volume = {6},
	number = {1},
	pages = {1--24},
	year = {2011}
}

@book{lawson1974linear,
	title = {Solving Least Squares Problems},
	author = {Lawson, Charles L. and Hanson, Richard J.},
	publisher = {Prentice-Hall},
	location = {Englewood Cliffs, NJ},
	year = {1974}
}

@article{goplerud2021sparsity,
	title = {Modelling Heterogeneity Using Bayesian Structured Sparsity},
	author = {Goplerud, Max},
	journal = {Working paper available at \url{https://arxiv.org/pdf/2103.15919.pdf}},
	year = {2021}
}

@article{zahid2013ridge,
	year = {2013},
	volume = {28},
	number = {3},
	pages = {1017--1034},
	author = {Zahid, Faisal Maqbool  and Tutz, Gerhard},
	title = {Ridge Estimation for Multinomial Logit Models with Symmetric Side Constraints},
	journal = {Computational Statistics}
}

@article{murphy2020init,
	year = {2020},
	volume = {14},
	number = {2},
	pages = {293--325},
	author = {Keefe Murphy and Thomas Brendan Murphy},
	title = {Gaussian Parsimonious Clustering Models with Covariates and a Noise Component},
	journal = {Advances in Data Analysis and Classification}
}

@article{varadhan2008simple,
	title={Simple and Globally Convergent Methods for Accelerating the Convergence of Any {EM} Algorithm},
	author={Varadhan, Ravi and Roland, Christophe},
	journal={Scandinavian Journal of Statistics},
	volume={35},
	number={2},
	pages={335--353},
	year={2008},
	publisher={Wiley Online Library}
}

@article{chamroukhi2019regularized,
	title={Regularized Maximum Likelihood Estimation and Feature Selection in Mixtures-of-Experts Models},
	author={Chamroukhi, Faicel and Huynh, Bao-Tuyen},
	journal={Journal de la Soci{\'e}t{\'e} Fran{\c{c}}aise de Statistique},
	volume={160},
	number={1},
	pages={57--85},
	year={2019}
}

@article{khalili2010mixture,
	title={New Estimation and Feature Selection Methods in Mixture-of-Experts Models},
	author={Khalili, Abbas},
	journal={Canadian Journal of Statistics},
	volume={38},
	number={4},
	pages={519--539},
	year={2010}
}

@article{fan2001variable,
	title={Variable Selection via Nonconcave Penalized Likelihood and Its Oracle Properties},
	author={Fan, Jianqing and Li, Runze},
	journal={Journal of the American Statistical Association},
	volume={96},
	number={456},
	pages={1348--1360},
	year={2001},
	publisher={Taylor \& Francis}
}

@article{oelker2017uniform,
	title={A Uniform Framework for the Combination of Penalties in Generalized Structured Models},
	author={Oelker, Margret-Ruth and Tutz, Gerhard},
	journal={Advances in Data Analysis and Classification},
	volume={11},
	number={1},
	pages={97--120},
	year={2017},
	publisher={Springer}
}

@article{celeux1992classification,
	title={A Classification {EM} Algorithm for Clustering and Two Stochastic Versions},
	author={Celeux, Gilles and Govaert, G{\'e}rard},
	journal={Computational Statistics \& Data Analysis},
	volume={14},
	number={3},
	pages={315--332},
	year={1992},
	publisher={Elsevier}
}

@article{bischl2018mlrmbo,
	title={{mlrMBO}: A Modular Framework for Model-Based Optimization of Expensive Black-Box Functions}, 
	author={Bernd Bischl and Jakob Richter and Jakob Bossek and Daniel Horn and Janek Thomas and Michel Lang},
	year={2018},
	journal={Working paper available at \url{https://arxiv.org/pdf/1703.03373.pdf}}
}

@article{stokell2021modelling,
	title={Modelling High-Dimensional Categorical Data Using Nonconvex Fusion Penalties}, 
	author={Benjamin G. Stokell and Rajen D. Shah and Ryan J. Tibshirani},
	year={2021},
	journal={Journal of the Royal Statistical Society: Series B (Statistical Methodology)},
	volume = {83},
	number = {3},
	pages = {579--611}
}

@article{zou2006adaptive,
	title = {The Adaptive LASSO and Its Oracle Properties},
	author = {Zou, Hui},
	volume = {101},
	issue = {476},
	year = {2006},
	pages = {1418--1429},
	journal = {Journal of the American Statistical Association}
}

@article{lim2015learning,
	title={Learning Interactions via Hierarchical Group-Lasso Regularization},
	author={Lim, Michael and Hastie, Trevor},
	journal={Journal of Computational and Graphical Statistics},
	volume={24},
	number={3},
	pages={627--654},
	year={2015},
	publisher={Taylor \& Francis}
}

@article{gupta1994using,
	title={On using Demographic Variables to Determine Segment Membership in Logit Mixture Models},
	author={Gupta, Sachin and Chintagunta, Pradeep K.},
	journal={Journal of Marketing Research},
	volume={31},
	number={1},
	pages={128--136},
	year={1994}
}

@article{andrews2002empirical,
	title={An Empirical Comparison of Logit Choice Models with Discrete versus Continuous Representations of Heterogeneity},
	author={Andrews, Rick L and Ainslie, Andrew and Currim, Imran S},
	journal={Journal of Marketing Research},
	volume={39},
	number={4},
	pages={479--487},
	year={2002}
}

@article{leeper2020measuring,
  title={Measuring Subgroup Preferences in Conjoint Experiments},
  author={Leeper, Thomas J and Hobolt, Sara B and Tilley, James},
  journal={Political Analysis},
  volume={28},
  number={2},
  pages={207--221},
  year={2020},
  publisher={Cambridge University Press}
}

@article{grun2008multinomial,
	year = {2008},
	volume = {25},
	number = {2},
	pages = {225--247},
	author = {Bettina Gr\"{u}n and Friedrich Leisch},
	title = {Identifiability of Finite Mixtures of Multinomial Logit Models with Varying and Fixed Effects},
	journal = {Journal of Classification}
}

@article{jiang1999identifiability,
	title={On the Identifiability of Mixtures-of-Experts},
	author={Jiang, Wenxin and Tanner, Martin A},
	journal={Neural Networks},
	volume={12},
	number={9},
	pages={1253--1258},
	year={1999},
	publisher={Elsevier}
}

@incollection{celeux2019mixture,
	title={Model Selection for Mixture Models -- Perspectives and Strategies},
	author={Celeux, Gilles and Fr{\"u}hwirth-Schnatter, Sylvia and Robert, Christian P.},
	booktitle={Handbook of Mixture Analysis},
	editor={Fr{\"u}hwirth-Schnatter, Sylvia and Celeux, Gilles and Robert, Christian P.},
	pages={118--154},
	year={2019},
	publisher={Chapman and Hall/CRC}
}

@incollection{gormley2019mixture,
	title={Mixture of Experts Models},
	author={Gormley, Isobel Claire and Fr{\"u}hwirth-Schnatter, Sylvia},
	booktitle={Handbook of Mixture Analysis},
	editor={Fr{\"u}hwirth-Schnatter, Sylvia and Celeux, Gilles and Robert, Christian P.},
	pages={271--307},
	year={2019},
	publisher={Chapman and Hall/CRC}
}

@article{almirall2014time,
  title={Time-Varying Effect Moderation Using the Structural Nested Mean Model: Estimation Using Inverse-Weighted Regression with Residuals},
  author={Almirall, Daniel and Griffin, Beth Ann and McCaffrey, Daniel F and Ramchand, Rajeev and Yuen, Robert A and Murphy, Susan A},
  journal={Statistics in Medicine},
  volume={33},
  number={20},
  pages={3466--3487},
  year={2014},
  publisher={Wiley Online Library}
}

@article{robinsondetect,
  title={How to Detect Heterogeneity in Conjoint Experiments},
  author={Robinson, Thomas S and Duch, Raymond M},
  year={2024},
  volume = {86},
  number = {2},
  journal = {The Journal of Politics},
  pages = {412--427}
}

@incollection{mclachlan1982classification,
	title={The Classification and Mixture Maximum Likelihood Approaches to Cluster Analysis},
	author={McLachlan, Geoffrey J.},
	booktitle={Classification, Pattern Recognition and Reduction of Dimensionality},
	editor={Krishnaiah, Paruchuri R. and Kanal, Laveen N.},
	volume={2},
	pages={199-208},
	year={1982},
	location={Amsterdam},
	publisher={North-Holland}
}

@article{bryant1978asymptotic,
	title={Asymptotic Behaviour of Classification Maximum Likelihood Estimates},
	author={Bryant, Peter and Williamson, John A.},
	journal={Biometrika},
	volume={65},
	number={2},
	pages={273--281},
	year={1978}
}

@Manual{cjbart,
    title = {cjbart: Heterogeneous Effects Analysis of Conjoint Experiments},
    author = {Thomas Robinson and Raymond Duch},
    year = {2023},
    note = {R package version 0.3.2},
    url = {https://CRAN.R-project.org/package=cjbart},
  }

@article{chi2016kmeans,
	year = {2016},
	volume = {70},
	number = {1},
	pages = {91--99},
	author = {Jocelyn T. Chi and Eric C. Chi and Richard G. Baraniuk},
	title = {k-{POD}: A Method for k-Means Clustering of Missing Data},
	journal = {The American Statistician}
}

@article{hastie1987closer,
	title={A Closer Look at the Deviance},
	author={Hastie, Trevor},
	journal={The American Statistician},
	volume={41},
	number={1},
	pages={16--20},
	year={1987}
}

@book{everitt2011cluster,
	title={Cluster Analysis},
	author={Everitt, Brian S. and Landau, Sabine and Leese, Morven and Stahl, Daniel},
	year={2011},
	edition = {5th Edition},
	publisher={John Wiley \& Sons}
}

@book{wu2021experiments,
  title={Experiments: {P}lanning, Analysis, and Optimization},
  author={Wu, CF Jeff and Hamada, Michael S},
  year={2021},
  edition = {3rd},
  publisher={John Wiley \& Sons}
}

@article{gelman2012we,
  title={Why We (Usually) Don't Have to Worry About Multiple Comparisons},
  author={Gelman, Andrew and Hill, Jennifer and Yajima, Masanao},
  journal={Journal of Research on Educational Effectiveness},
  volume={5},
  number={2},
  pages={189--211},
  year={2012},
  publisher={Taylor \& Francis}
}

@article{kang2007dr,
	title = {Demystifying Double Robustness: A Comparison of Alternative Strategies for Estimating a Population Mean from Incomplete Data},
	volume = {22},
	number = {4},
	journal = {Statistical Science},
	publisher = {Institute of Mathematical Statistics},
	author = {Kang,  Joseph D. Y. and Schafer,  Joseph L.},
	year = {2007},
	pages = {523--539}

}

@article{shi2023forward,
  title={Forward screening and post-screening inference in factorial designs},
  author={Shi, Lei and Wang, Jingshen and Ding, Peng},
  journal={arXiv preprint arXiv:2301.12045},
  year={2023}
}

@Manual{FactorHet,
    title = {FactorHet: Estimate Heterogeneous Effects in Factorial Experiments Using
Grouping and Sparsity},
    author = {Max Goplerud and Nicole E. Pashley and Kosuke Imai},
    year = {2025},
    note = {R package version 1.0.0},
    doi = {10.32614/CRAN.package.FactorHet},
    url = {https://CRAN.R-project.org/package=FactorHet},
  }

@misc{replicationAOAS,
	author = {Goplerud, Max and Imai, Kosuke and Pashley, Nicole E.},
	publisher = {Harvard Dataverse},
	title = {{Replication Data for: ``Estimating Heterogeneous Causal Effects of High-Dimensional Treatments: Application to Conjoint Analysis''}},
	year = {2025},
	doi = {10.7910/DVN/YAHPEH},
	url = {https://doi.org/10.7910/DVN/YAHPEH}
}

@article{egam:imai:19,
  author = 	 {Egami, Naoki and Imai, Kosuke},
  title = 	 {Causal Interaction in Factorial Experiments: Application to Conjoint Analysis},
  journal =  {Journal of the American Statistical Association},
  year = 	 {2019},
  volume = 	 {114},
  number = 	 {526},
  pages = {529--540},
  month = 	 {June},
  OPTnote = 	 {},
  OPTannote = 	 {}
}

@Article{imai:li:24,
  author = 	 {Imai, Kosuke and Li, Michael Lingzhi},
  title = 	 {Statistical Inference for Heterogeneous Treatment Effects Discovered by Generic Machine Learning in Randomized Experiments},
  journal = 	 {Journal of Business \& Economic Statistics},
  year = 	 {2025},
  volume = {43},
  number = {1},
  OPTkey = 	 {},
  OPTvolume = 	 {},
  OPTnumber = 	 {},
  pages = 	 {256--268},
  OPTmonth = 	 {},
  OPTannote = 	 {}
}

@article{imai:ratk:13,
	Author = {Imai, Kosuke and Ratkovic, Marc},
	Journal = {Annals of Applied Statistics},
	Title = {Estimating Treatment Effect Heterogeneity in Randomized Program Evaluation},
	Year = {2013},
        Volume = {7},
        Number = {1},
        Month = {March},
        Pages = {443--470}}

@article{imai:stra:11,
	Author = {Imai, Kosuke and Strauss, Aaron},
	Journal = {Political Analysis},
	Month = {Winter},
	Number = {1},
	Pages = {1--19},
	Title = {Estimation of Heterogeneous Treatment Effects from Randomized Experiments, with Application to the Optimal Planning of the Get-out-the-vote Campaign},
	Volume = {19},
	Year = {2011}}

@article{liu:shir:23,
title={Multiple Hypothesis Testing in Conjoint Analysis},
volume={31},
number={3},
journal={Political Analysis},
author={Liu, Guoer and Shiraito, Yuki},
year={2023},
pages={380--395}
}

@Book{rao:14,
  author = 	 {Rao, Vithala R.},
  ALTeditor = 	 {},
  title = 	 {Applied Conjoint Analysis},
  publisher = 	 {Springer},
  year = 	 {2014},
  OPTkey = 	 {},
  OPTvolume = 	 {},
  OPTnumber = 	 {},
  OPTseries = 	 {},
  address = 	 {Berlin Heidelberg},
  OPTedition = 	 {},
  OPTmonth = 	 {},
  OPTnote = 	 {},
  OPTannote = 	 {}
}

@article{tian:etal:14,
author = {Tian, Lu and Slizadeh, Ash A. and Gentles, Andrew J. and Tibshirani, Robert},
title = {A Simple Method for Estimating Interactions Between a Treatment and a Large Number of Covariates},
journal = {Journal of the American Statistical Association},
volume = {109},
number = {508},
pages = {1517--1532},
year  = {2014}
}

@Article{athe:imbe:16,
  author = 	 {Athey, Susan and Imbens, Guido},
  title = 	 {Recursive Partitioning for Heterogeneous Causal Effects},
  journal = 	 {Proceedings of the National Academy of Sciences},
  year = 	 {2016},
  OPTkey = 	 {},
  volume = 	 {113},
  number = 	 {27},
  pages = 	 {7353--7360},
  OPTmonth = 	 {},
  OPTnote = 	 {},
  OPTannote = 	 {}
}

@article{demp:lair:rubi:77,
	Author = {Dempster, Arthur P. and Laird, Nan M. and Rubin, Donald B.},
	Journal = {Journal of the Royal Statistical Society, {S}eries {B}, {M}ethodological},
	Number = 1,
	Pages = {1--22},
	Title = {Maximum Likelihood from Incomplete Data Via the {EM} Algorithm},
	Volume = 39,
	Year = 1977}

@Article{grim:mess:west:17,
  author = 	 {Grimmer, Justin and Messing, Solomon and Westwood, Sean J.},
  title = 	 {Estimating Heterogeneous Treatment Effects and the Effects of Heterogeneous Treatments with Ensemble Methods},
  journal = 	 {Political Analysis},
  year = 	 {2017},
  OPTkey = 	 {},
  volume = 	 {25},
  OPTnumber = 	 {},
  pages = 	 {413--434},
  OPTmonth = 	 {},
  OPTnote = 	 {},
  OPTannote = 	 {}
}

@article{hahn:murr:carv:20,
  title={Bayesian Regression Tree Models for Causal Inference: Regularization, Confounding, and Heterogeneous effects},
  author={Hahn, P. Richard and Murray, Jared S. and Carvalho, Carlos M.},
  journal={Bayesian Analysis},
  year={2020},
  volume={15},
  number={3},
  pages={965--1056},
  publisher={International Society for Bayesian Analysis}
}

@article{khal:chen:07,
	Author = {Khalili, Abbas and Chen, Jiahua},
	Journal = {Journal of the American Statistical Association},
	Month = {September},
	Number = 479,
	Pages = {1025--1038},
	Title = {Variable Selection in Finite Mixture of Regression Models},
	Volume = 102,
	Year = 2007}

@article{kunz:etal:19,
	volume={116},
	number={10},
	pages={4156--4165},
	year={2019},
  	author = 	 {K\"{u}nzel, S\"{o}ren R. and Sekhon, Jasjeet S. and Bickel, Peter J. and Yu, Bin},
  	title = {Metalearners for Estimating Heterogeneous Treatment Effects using Machine Learning},
	journal = {Proceedings of the National Academy of Sciences},
}

@article{loui:82,
	Author = {Louis, Thomas A.},
	Journal = {Journal of the Royal Statistical Society, {S}eries {B}, {M}ethodological},
	Pages = {226--233},
	Title = {Finding the Observed Information Matrix When Using the {EM} Algorithm},
	Volume = 44,
	Year = 1982}

@article{meng:vand:97,
	Author = {Meng, Xiao-Li and van Dyk, David A.},
	Journal = {Journal of the Royal Statistical Society, {S}eries {B}, {M}ethodological},
	Pages = {511--567},
	Title = {The {EM} Algorithm -- an Old Folk Song Sung to a Fast New Tune (with Discussion)},
	Volume = 59,
	Year = 1997}

@Book{vand:rose:11,
  author = 	 {{van der Laan}, Mark J. and Rose, Sheri},
  ALTeditor = 	 {},
  title = 	 {Targeted Learning: Causal Inference for Observational and
Experimental Data},
  publisher = 	 {Springer},
  year = 	 {2011},
  OPTkey = 	 {},
  OPTvolume = 	 {},
  OPTnumber = 	 {},
  OPTseries = 	 {},
  OPTaddress = 	 {},
  OPTedition = 	 {},
  OPTmonth = 	 {},
  OPTnote = 	 {},
  OPTannote = 	 {}
}

@Article{wage:athe:18,
  author = 	 {Wager, Stefan and Athey, Susan},
  title = 	 {Estimation and Inference of Heterogeneous Treatment Effects using Random Forests},
  journal = 	 {Journal of the American Statistical Association},
  year = 	 {2018},
  OPTkey = 	 {},
  volume = 	 {113},
  number = 	 {523},
  pages = 	 {1228--1242},
  OPTmonth = 	 {},
  OPTnote = 	 {},
  OPTannote = 	 {}
}

@article {yuan:lin:06,
author = {Yuan, Ming and Lin, Yi},
title = {Model Selection and Estimation in Regression with Grouped Variables},
journal = {Journal of the Royal Statistical Society: Series B (Statistical Methodology)},
volume = {68},
number = {1},
pages = {49--67},
year = {2006},
}

\newpage
\begin{appendices}
	\begin{set11}
	\def\spacingset#1{\renewcommand{\baselinestretch}%
		{#1}\small\normalsize} \spacingset{1}
	\renewcommand\appendixname{Supplementary Material}
	
			\setlength\paperwidth{614.295pt}
	\setlength\paperheight{794.96999pt}
	\setlength\textwidth{469.75502pt}
	\setlength\textheight{650.43001pt}
	\setlength\oddsidemargin{0.0pt}
	\setlength\evensidemargin{0.0pt}
	\setlength\topmargin{-37.0pt}
	\setlength\headheight{12.0pt}
	\setlength\headsep{25.0pt}
	\setlength\topskip{11.0pt}
	\setlength\footskip{30.0pt}
	\setlength\marginparwidth{59.0pt}
	\setlength\marginparsep{10.0pt}
	\setlength\columnsep{10.0pt}
	\setlength\hoffset{0.0pt}
	\setlength\voffset{0.0pt}
	\setlength\mag{1000}
	
	\def\spacingset#1{\renewcommand{\baselinestretch}%
		{#1}\small\normalsize} \spacingset{1}
	\renewcommand\appendixname{Supplementary Material}
	
\def\spacingset#1{\renewcommand{\baselinestretch}%
	{#1}\small\normalsize} \spacingset{1}

\appendix
\setcounter{page}{1}
\begin{center}
{\bf \Large  Supplementary Material for\\``Estimating Heterogeneous Causal Effects of
  High-Dimensional Treatments:  Application to Conjoint Analysis''}

\vspace{0.5cm}
{ \large Max Goplerud \hspace{.5in} Kosuke Imai \hspace{.5in} Nicole E. Pashley}
\end{center}

\setcounter{equation}{0}
\setcounter{figure}{0}
\setcounter{section}{0}
\setcounter{table}{0}
\setcounter{proposition}{0}

\setcounter{appprop}{0}
\setcounter{appthm}{0}
\setcounter{appresult}{0}
\setcounter{applemma}{0}
\renewcommand {\theequation} {A\arabic{equation}}
\renewcommand {\thefigure} {A\arabic{figure}}
\renewcommand {\thealgorithm} {A\arabic{algorithm}}
\renewcommand {\thetable} {A\arabic{table}}

\medskip 

\section{The Details of the Immigration Conjoint Experiment}
\label{app:conjoint}

\begin{table}[ht!]
	\centering\small
	\begin{tabular}{|p{0.25\linewidth}|p{0.12\linewidth}| p{0.58\linewidth}|}
		\hline
		Attribute &\# of Levels & Levels\\
		\hline
		Education & 7 & No formal education; Equivalent to completing fourth grade in the U.S.; Equivalent to completing eighth grade in the U.S.; Equivalent to completing high school in the U.S.; Equivalent to completing two years at college in the U.S.; Equivalent to completing a college degree in the U.S.; Equivalent to completing a graduate degree in the U.S.\\
		Gender & 2 & Female; Male\\
		Country of origin & 10 & Germany; France; Mexico; Philippines; Poland; India; China; Sudan; Somalia; Iraq\\
		Language & 4 & During admission interview, this applicant spoke fluent English; During admission interview, this applicant spoke broken English; During admission interview, this applicant tried to speak English but was unable; During admission interview, this applicant spoke through an interpreter\\
		Reason for Application & 3 &  Reunite with family members already in U.S.; Seek better job in U.S.; Escape political/religious persecution\\
		Profession & 11 & Gardener; Waiter; Nurse; Teacher; Child care provider; Janitor; Construction worker; Financial analyst; Research scientist; Doctor; Computer programmer\\
		Job experience & 4 & No job training or prior experience; One to two years; Three to five years\\
		Employment Plans & 4 & Has a contract with a U.S. employer; Does not have a contract with a U.S. employer, but has done job interviews; Will look for work after arriving in the U.S.; Has no plans to look for work at this time\\
		Prior Trips to the U.S. & 5 & Never been to the U.S.; Entered the U.S. once before on a tourist visa; Entered the U.S. once before without legal authorization; Has visited the U.S. many times before on tourist visas; Spent six months with family members in the U.S.\\
		\hline
	\end{tabular}\caption{Table 1 in \cite{hainmueller2015hidden}. All attributes for immigrants and their levels.}\label{tab:attributes}
\end{table}

\section{Additional Results for Comparison with \texttt{cjbart}}
\label{append:cjbart_graphs}

We compare the performance of our method with that of
\cite{robinsondetect} whose method is implemented using an open-source
software package, \texttt{cjbart} \citep{cjbart}.  We use the same set
of moderators and factors considered in our earlier
analyses. Figure~\ref{fig:cjbart} compares the estimated CAMCEs for
country with Germany set as the reference category, calculated across
all individual covariate vectors in the sample. Our method discovers a
considerable degree of heterogeneity in the CAMCEs whereas
\texttt{cjbart} shows limited treatment effect variation for
most countries.  Under our model, the estimated heterogeneous effects
are more strongly associated with predictors than \texttt{cjbart}.
For example, our method finds a clear association, on average, between
the estimated CAMCEs and prejudice or party identification, whereas
\texttt{cjbart} does not.

\begin{figure}[!t]
	\centering
	\includegraphics[scale = 0.8]{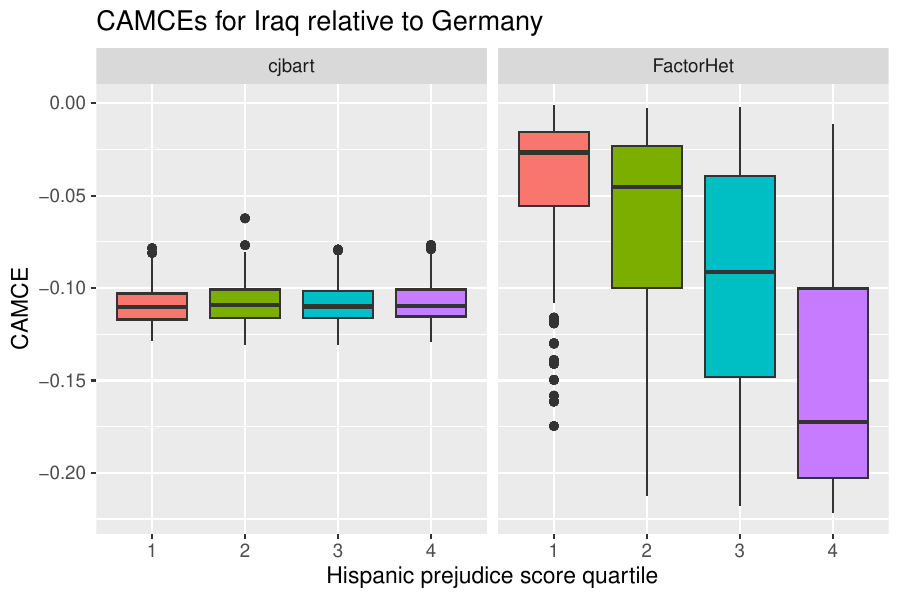}
	\includegraphics[scale = 0.8]{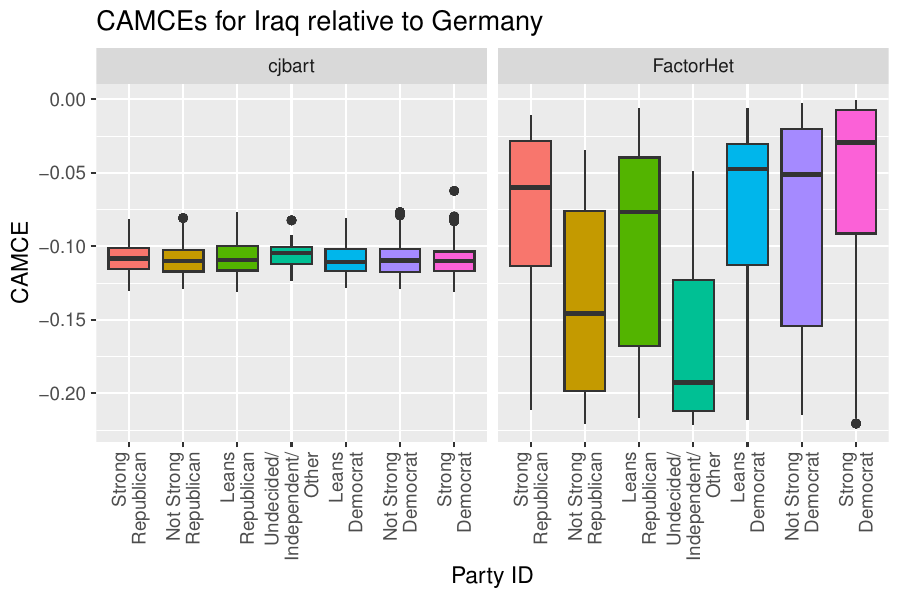}
	\caption{Comparison of discovered heterogeneous effects (the
		conditional average marginal component effects or CAMCEs) between
		the proposed method and the BART-based method \texttt{cjbart}.  In both 
		plots the  y-axis corresponds to values estimated, either by our
		method (right) or by \texttt{cjbart} \citep{robinsondetect} (left).
		The plots show the estimated effect of Iraq as compared to the
		baseline of Germany.  In the top figure, the x-axis and color
		corresponds to the categories of individuals based on the quartile
		of their Hispanic prejudice score.  In the bottom figure, the x-axis
		and color corresponds to party ID.} \label{fig:cjbart}
\end{figure}

Figure~\ref{fig:cjbart_supp} shows the distribution of CAMCEs for all countries. To simplify the visualization, we subset party ID to strong Republicans, strong Democrats, and Independent/other.

\begin{figure}[p]
	\centering \spacingset{1}
	\includegraphics{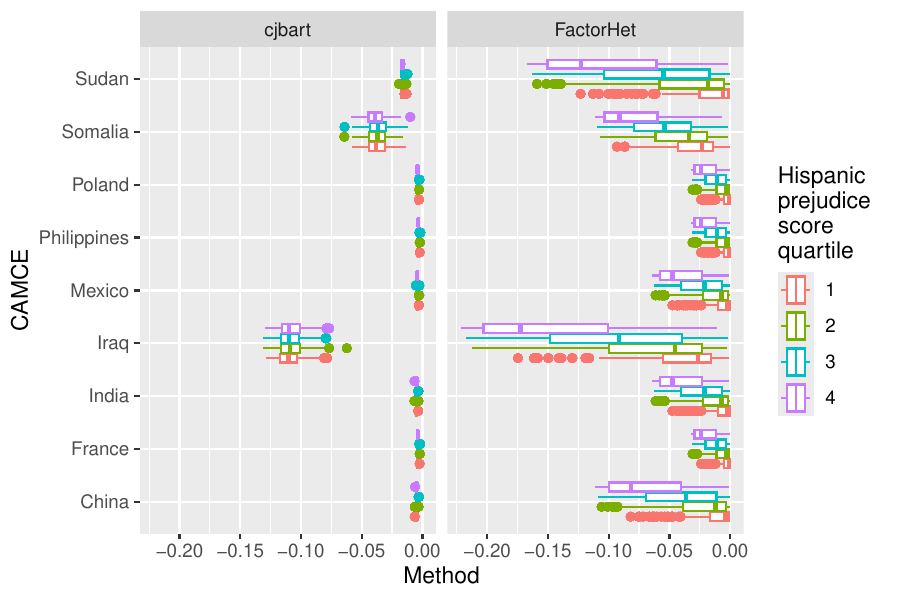}
	\includegraphics{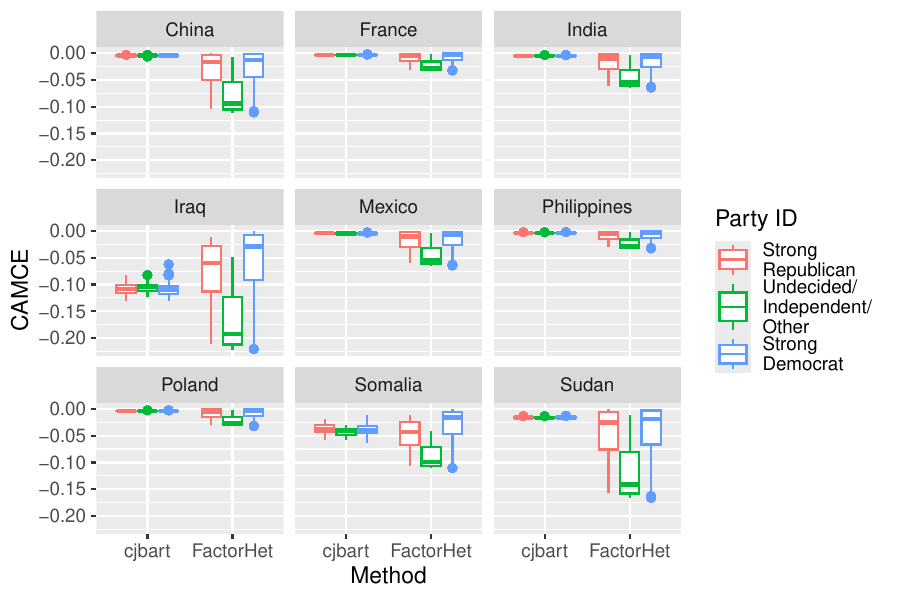}
	\caption{In both plots, the y-axis corresponds to the estimated values,
		either based on our method (right) or based on the method of
		\cite{robinsondetect} (left), for the effect of a given country
		relative to the baseline of Germany.  In the top figure, we color
		code based on quartile for the Hispanic prejudice score.  In the
		bottom figure, we reduce the sample to those who identify as
		``Strong Republican'', ``Strong Democrat'', or ``Independent/Other''
		and color code by party ID.} \label{fig:cjbart_supp}
\end{figure}

\section{Proof of Proposition~\ref{prop:kmeans}}\label{app:kmeans}

To prove Proposition~\ref{prop:kmeans}, we provide a more general
result for one-parameter exponential family distributions, which
include the specific Bernoulli result in the main text as a special
case. We consider a random variable $Y$ that is assumed to follow a
single-parameter exponential family distribution with canonical
parameter $\theta$ and $\mu = d\psi(\theta)/d\theta = \psi'(\theta)$.
Since $\mu$ is monotone in $\theta$, we index the density $f$ using
$\mu$:
$$f_\mu(y) = c(y) \exp\left(y \theta - \psi(\theta)\right).$$
The maximum likelihood estimate of the mean $\hat{\mu}$ given $N$ observations $\{y_i\}_{i=1}^{N}$ from $Y$ is the sample average, $\frac{1}{N} \sum_{i=1}^{N} y_i = \hat{\mu}$, and the corresponding estimate of the canonical parameter is $\hat{\theta}$.

Proposition~\ref{prop:kmeans_general} states that maximally heterogeneous groups in terms of Kullback-Leibler (KL) divergence of potential outcomes is equivalent to maximizing the log-likelihood over groups and their centroids for any choice of single parameter exponential family $f$.

\begin{appprop}\label{prop:kmeans_general} \spacingset{1}
	
	Assume a partition of $N$ observations, indexed by
	$i \in \{1, \cdots, N\}$, into $K$ groups whose memberships
	$Z_i \in \{1, \cdots, K\}$ are denoted by $\mathcal{Z}$. Define the
	estimated within-group average outcome under treatment $\bm{t}$ for
	group $k$ and the estimated overall average outcome as
	$\hat{\zeta}_k(\bm{t}; \mathcal{Z}) = \sum_{i=1}^N I\{Z_i = k,
	\bm{T}_i = \bm{t}\} Y_i/\sum_{i=1}^N I\{Z_i = k, \bm{T}_i =
	\bm{t}\}$ and
	$\widehat{\overline{Y}}(\bm{t}) = \sum_{i=1}^N I\{\bm{T}_i =
	\bm{t}\} Y_i/\sum_{i=1}^N I\{\bm{T}_i = \bm{t}\}$, respectively.
	
	Then, maximally heterogeneous groups in the terms of the
	Kullback-Leibler (KL) divergence of potential outcomes can be found by maximizing the
	log-likelihood function over the group membership and the centroids
	of groups, i.e.,
	\begin{align*}
	& \argmax_{\mathcal{Z}} \left\{\sum_{k=1}^K \sum_{i=1}^N \mathbf{1}\{Z_i = k\} \mathrm{KL}\left(\hat{\zeta}_k(\bm{T}_i; \mathcal{Z}) \Vert 
	\widehat{\overline{Y}}(\bm{T}_i)\right) \right\} = \argmax_{\mathcal{Z}} \sum_{k=1}^K \sup_{\zeta_k}
	\sum_{i=1}^N  \mathbf{1}\{Z_i = k \} \log f_{\zeta_k}(Y_i)
	\end{align*}
	where $\mathrm{KL}(\mu_1, \mu_2)$ indicates the KL divergence
	between two single-parameter exponential family distributions with
	means $\mu_1$ and $\mu_2$ is defined as \citep{hastie1987closer}:
	$$\mathrm{KL}(\mu_1, \mu_2) = \mathbb{E}_{f_{\mu_1}(Y)}\left[\log f_{\mu_1}(Y) - \log f_{\mu_2}(Y)\right] = \left(\theta_1 - \theta_2\right)\mu_1 - \left[\psi(\theta_1) - \psi(\theta_2)\right].$$
\end{appprop}

To prove this proposition, we use Lemma~\ref{res:partition_deviance} which decomposes the total deviance of
the observed data into the between and within components as in $k$-means \citep[ch. 5]{everitt2011cluster}.  This generalizes the standard Gaussian result \citep[see][]{chi2016kmeans}.

\begin{applemma}[Deviance Decomposition for Exponential
	Family] \label{res:partition_deviance} \spacingset{1} 
	
	Define the deviance for a single observation $y$ as follows:
	$$D(y, \mu) = 2 \left[\log f_y(y) - \log f_\mu(y)\right]$$
	and the total deviance of the observed data when evaluated at the maximum
	likelihood estimate for each treatment $\bm{t}$---the sample average
	$\widehat{\overline{Y}}(\bm{t})$ given randomization of $\bT_i$---as
	follows
	\begin{align*}
	D_{\mathrm{Total}} = \sum_{i=1}^N \sum_{\bm{t} \in \cT} D\left(Y_i(\bm{t}), \widehat{\overline{Y}}(\bm{t})\right) \bm{1}\{\bm{T}_i = \bm{t}\} = \sum_{i=1}^N D\left(Y_i, \widehat{\overline{Y}}(\bm{T}_i)\right),
	\end{align*}
	where
	$\widehat{\overline{Y}}(\bm{t}) = \sum_{i=1}^N \bm{1}\{\bm{T}_i =
	\bm{t}\} Y_i/\sum_{i=1}^N \bm{1}\{\bm{T}_i = \bm{t}\}$.  Then, for any
	partition $\mathcal{Z}$ of the observations into $K$ groups,
	$D_{\mathrm{Total}}$ can be decomposed as follows:
	\begin{align*}
	D_{\mathrm{Total}} =\underbrace{\sum_{k=1}^K \sum_{i=1}^N \bm{1}\{Z_i = k\} \cdot 2~\mathrm{KL} \left(\hat{\zeta}_k(\bm{T}_i; \mathcal{Z}), \widehat{\overline{Y}}(\bm{T}_i)\right)}_{=D_{\mathrm{Between}}} + \underbrace{\sum_{k=1}^K \sum_{i=1}^N \bm{1}\{Z_i = k\} D\left(Y_i(\bm{T}_i), \hat{\zeta}_k(\bm{T}_i; \mathcal{Z})\right)}_{=D_\mathrm{Within}} 
	\end{align*}
	where 
	$\hat{\zeta}_k(\bm{t}; \mathcal{Z}) = \sum_{i=1}^N \bm{1}\{Z_i = k,
	\bm{T}_i = \bm{t}\} Y_i/N_k(\bm{t}; \mathcal{Z})$ and
	$N_k(\bm{t}; \mathcal{Z}) = \sum_{i=1}^N \bm{1}\{Z_i = k, \bm{T}_i =
	\bm{t}\}$.
\end{applemma}

\begin{proof}
	Define $\hat{\bar{\theta}}(\bm{t})$, $\hat{\theta}_k(\bm{t}; \mathcal{Z})$ and $\theta_i(\bm{t})$ as the canonical parameters associated with, respectively, means $\widehat{\overline{Y}}(\bm{t})$, $\hat{\zeta}_k(\bm{t}; \mathcal{Z})$, $Y_i$ where $\theta_i(\bm{t})$ is used to define a saturated model for $Y_i(\bm{t})$. The result is proved below by re-arranging $D_{\mathrm{Total}}$. 
	\begin{align*}
	D_{\mathrm{Total}} = &\sum_{i=1}^N \sum_{\bm{t} \in \cT} \bm{1}\{\bm{T}_i = \bm{t}\} \cdot 2 \left[ \left(\theta_i(\bm{t}) - \hat{\bar{\theta}}(\bm{t})\right) Y_i(\bm{t}) - \left(\psi(\theta_i(\bm{t})) - \psi(\hat{\bar{\theta}}(\bm{t}))\right)\right]  \\
	= & \sum_{k=1}^K \sum_{i=1}^N \sum_{\bm{t} \in \cT} \bm{1}\{Z_i = k, \bm{T}_i = \bm{t}\} \cdot 2 \left(\theta_i(\bm{t}) - \hat{\bar{\theta}}(\bm{t}) +  \hat{\theta}_k(\bm{t}; \mathcal{Z}) - \hat{\theta}_k(\bm{t}; \mathcal{Z})\right) Y_i(\bm{t}) \\
	&  - \sum_{k=1}^K \sum_{i=1}^N \sum_{\bm{t} \in \cT}\bm{1}\{Z_i = k, \bm{T}_i = \bm{t}\}  \cdot 2\left(\psi(\theta_i(\bm{t})) - \psi(\hat{\bar{\theta}}(\bm{t})) + \psi(\hat{\theta}_k(\bm{t}; \mathcal{Z})) - \psi(\hat{\theta}_k(\bm{t}; \mathcal{Z})) \right) \\
	=& \sum_{k=1}^K \sum_{i=1}^N \sum_{\bm{t} \in \cT} \bm{1}\{Z_i = k, \bm{T}_i = \bm{t}\} \cdot 2 \hat{\zeta}_k(\bm{t}; \mathcal{Z})\left[ \left(\hat{\theta}_k(\bm{t}; \mathcal{Z}) - \hat{\bar{\theta}}(\bm{t})\right) - \left(\psi(\hat{\theta}_k(\bm{t}; \mathcal{Z})) -\psi(\hat{\bar{\theta}}(\bm{t})) \right)\right]  \\ 
	& + \sum_{k=1}^K \sum_{i=1}^N  \sum_{\bm{t} \in \cT} \bm{1}\{Z_i = k, \bm{T}_i = \bm{t}\} D\left(Y_i(\bm{t}), \hat{\zeta}_k(\bm{t}; \mathcal{Z})\right) \\
	= & \sum_{k=1}^K \sum_{i=1}^N \bm{1}\{Z_i = k\} \cdot 2 ~\mathrm{KL}\left(\hat{\zeta}_k(\bm{T}_i; \mathcal{Z}),
	\widehat{\overline{Y}}(\bm{T}_i)\right) + \sum_{k=1}^K \sum_{i=1}^N \bm{1}\{Z_i = k\} D\left(Y_i(\bm{T}_i), \hat{\zeta}_k(\bm{T}_i; \mathcal{Z})\right)
	\end{align*}
	where the simplification of $D_{\mathrm{Between}}$ follows from noting
	that
	$\sum_{i=1}^N Y_i(\bm{t}) \bm{1}\{\bm{T}_i = \bm{t}, Z_i = k\} =
	N_k(\bm{t}; \mathcal{Z}) \hat{\zeta}_k(\bm{t}) = \sum_{i=1}^N
	\bm{1}\{\bm{T}_i = \bm{t}, Z_i = k\} \hat{\zeta}_k(\bm{t})$ by
	definition.
\end{proof}

\paragraph*{Proof of Proposition~\ref{prop:kmeans_general}.}
Given Lemma~\ref{res:partition_deviance}, maximizing
$D_{\mathrm{Between}}$ over $\mathcal{Z}$ is equivalent to minimizing
$D_\mathrm{Within}$ over $\mathcal{Z}$. Then, $D_{\mathrm{Between}}$
can be divided by two to obtain the left-hand side of the
proposition. The right-hand side of the proposition is derived as
follows. Minimizing the deviance is equivalent to maximizing the
log-likelihood, i.e.,
\begin{equation*}
\argmin_{\mathcal{Z}} D_{\mathrm{Within}} = \argmax_{\mathcal{Z}} \left\{\sum_{k=1}^K \sum_{i=1}^N  \bm{1}\{Z_i = k\} \log f_{\hat{\zeta}_k(\bm{T}_i; \mathcal{Z})}(Y_i(\bm{T}_i)) \right\}.
\end{equation*}
This can be written as a two-level optimization problem, noting that
$Y_i = Y_i(\bT_i)$ by the consistency assumption and that for fixed
$\mathcal{Z}$, the maximum likelihood estimate of $\zeta_k(\bm{t})$ is
$\hat{\zeta}_k(\bm{t}; \mathcal{Z})$, i.e., the within-group observed
average.
\begin{align*}
\argmin_{\mathcal{Z}} D_{\mathrm{Within}} &=\argmax_{\mathcal{Z}} \left\{\sum_{k=1}^K  \sup_{\zeta_k}  \sum_{i=1}^N   \bm{1}\{Z_i = k\}\log f_{\zeta_k(\bT_i)}(Y_i)\right\}
\end{align*}
\qed

Finally, Proposition~\ref{prop:kmeans} in the main text uses the
Bernoulli likelihood for $f$ and is shown below.
\begin{equation*}
\argmin_{\mathcal{Z}} D_{\mathrm{Within}} = \argmax_{\mathcal{Z}} \left\{\sum_{k=1}^K  \sup_{\{\zeta_k\}}  \sum_{i=1}^N  \bm{1}\{Z_i = k\} \left[Y_i \log \zeta_k(\bT_i) + \{1 - Y_i\}\log\{1 - \zeta_k(\bT_i)\}\right]\right\}
\end{equation*}

\section{Proof of  Proposition~\ref{prop:moe}}
\label{app:moe}

As before, we prove a more general result using the one-parameter
exponential family distributions.
\begin{appprop}[Finding maximally heterogeneous groups with
	moderators] \label{prop:moe_general} \spacingset{1} Suppose we extend the setting of Proposition~\ref{prop:kmeans_general} and additionally model the conditional probability of each individual's group membership given categorical moderators $\{\pi_k(\bX_i)\}_{k=1}^K$. Then, maximally heterogeneous groups in terms of KL divergence of potential outcomes with the entropy of group membership probabilities as a penalty term can be found by maximizing the log-likelihood function of the extended model,
	\begin{align*}
	& \argmax_{\mathcal{Z}} \left\{\sum_{k=1}^K \sum_{i=1}^N \mathbf{1}\{Z_i = k\} \mathrm{KL}\left(\hat{\zeta}_k(\bm{T}_i; \mathcal{Z}) \Vert 
	\widehat{\overline{Y}}(\bm{T}_i)\right) - \sum_{i=1}^N H(\{\hat{\pi}_k(\bm{X}_i; \mathcal{Z})\}_{k=1}^K) \right\} \nonumber\\
	= \ & \argmax_{\mathcal{Z}} \sum_{k=1}^K \sup_{\zeta_k, \pi_k}
	\sum_{i=1}^N  \mathbf{1}\{Z_i = k \} \left[\log f_{\zeta_k}(Y_i) + \log \pi_k(\bX_i)\right]
	\end{align*}
	where $H(\{p_k\}_{k=1}^K) = -\sum_{k=1}^K p_k \log p_k$ (by convention, if $p_k = 0$, then $p_k \log p_k = 0$) is the entropy, and $\hat{\pi}_k(\bm{x}; \mathcal{Z}) = \sum_{i=1}^N \bm{1}\{Z_i = k, \bX_i = \bm{x}\}/\sum_{i=1}^N \bm{1}\{\bX_i = \bm{x}\}$ and $\hat\zeta_k(\bt; \mathcal{Z})$ are the maximizers of the log-likelihood function of the right hand side of
	the above equation given $\mathcal{Z}$.
\end{appprop}

To prove this proposition, we use Lemma~\ref{res:entropy}.
\begin{applemma}[Entropy of Groups with Respect to Moderators]\label{res:entropy}
	Define the set of observed categorical moderator values as
	$\mathcal{X}$ with
	$N(\bm{x}) = \sum_{i=1}^N \bm{1}\{\bX_i = \bm{x}\}$ and
	$N_k(\bm{x}; \mathcal{Z}) = \sum_{i=1}^N \bm{1}\{Z_i = k, \bX_i =
	\bm{x}\}$. Given $\mathcal{Z}$, the entropy of group membership
	probabilities given moderators, weighted by the frequency of the
	moderators, is defined as follows:
	$$H(\mathcal{Z}) = \sum_{\bm{x} \in \mathcal{X}} N(\bm{x}) H(\{\hat{\pi}_k(\bm{x}; \mathcal{Z})\}_{k=1}^K).$$
	Then, $H(\mathcal{Z})$ can be expressed in the
	following two equivalent ways:
	$$H(\mathcal{Z}) = \sum_{i=1}^N H(\{\hat{\pi}_k(\bX_i; \mathcal{Z})\}_{k=1}^K) = -\sum_{i=1}^N \sum_{k=1}^K \bm{1}\{Z_i = k\} \log \hat{\pi}_k(\bX_i; \mathcal{Z}).$$
\end{applemma}

\begin{proof}
	The first expression follows by noting that the summation merely
	counts the number of times each $\bm{x}$ appears. The second
	expression is derived below by re-arranging $H(\mathcal{Z})$,
	\begin{align*}
	-H(\mathcal{Z}) &= \sum_{\bm{x} \in \mathcal{X}} \sum_{k=1}^K N(\bm{x})\hat{\pi}_k(\bm{x}; \mathcal{Z}) \log\hat{\pi}_k(\bm{x}; \mathcal{Z}) = \sum_{i=1}^N \sum_{k=1}^K \bm{1}\{Z_i = k\} \log\hat{\pi}_k(\bX_i; \mathcal{Z}),
	\end{align*}
	where the last equality follows because
	$N(\bm{x})\hat{\pi}_k(\bm{x}; \mathcal{Z}) = N_k(\bm{x}; \mathcal{Z})$
	and it counts the number of times each combination of $(k, \bm{x})$
	appears. \qed

	Next, to prove Proposition~\ref{prop:moe_general}, we note that for
	any $\mathcal{Z}$, the KL divergence is equal to the log-likelihood
	evaluated at the maximum likelihood estimates plus a constant that
	does not depend on $\mathcal{Z}$ (see the definition of
	$D_{\mathrm{Total}}$):
	
	$$\sum_{k=1}^K \sum_{i=1}^N \bm{1}\{Z_i = k\} \mathrm{KL}\left(\hat{\zeta}_k(\bm{T}_i; \mathcal{Z}) \Vert 
	\widehat{\overline{Y}}(\bm{T}_i)\right) \ = \ \sum_{k=1}^K
	\sum_{i=1}^N \bm{1}\{Z_i = k\} \log_{\hat{\zeta}_k(\bT_i;
		\mathcal{Z})}(Y_i(\bT_i)) + \mathrm{const.}$$
	
	Adding the negative of group-moderator entropy $H(\mathcal{Z})$ to
	both sides and taking the maximum over $\mathcal{Z}$ gives the
	left-hand side of Proposition~\ref{prop:moe_general}. The equivalent
	right-hand side, using Lemma~\ref{res:entropy} can be expressed as:
	
	\begin{equation*}
	\argmax_{\mathcal{Z}} \left\{\sum_{k=1}^K \sum_{i=1}^N  \bm{1}\{Z_i = k\}\left[ \log f_{\hat{\zeta}_k(\bm{T}_i; \mathcal{Z})}(Y_i(\bm{T}_i)) + \log \hat{\pi}_k(\bX_i; \mathcal{Z})\right]\right\}.
	\end{equation*}
	
	As in the proof of Proposition~\ref{prop:kmeans_general}, observing
	that $Y_i = Y_i(\bT_i)$ by the consistency assumption and writing the
	above equation as two-level optimization problem over $\zeta_k$ and
	$\pi_k$ establishes Proposition~\ref{prop:moe_general}. This follows
	by noting that for a fixed $\mathcal{Z}$, the maximum likelihood
	estimate of $\pi_k(\bm{x})$ is $\hat{\pi}_k(\bm{x})$ and the estimate
	of $\zeta_k(\bm{t})$ as $\hat{\zeta}_k(\bm{t})$ is unchanged as the
	optimization problem is separable.  In addition, using the Bernoulli
	likelihood for $f$ gives Proposition~\ref{prop:moe} in the main text.
\end{proof}

\section{Inclusion of Higher Order Interactions}
\label{sec:high_order_int}

Here we illustrate how the model and regularization penalties in Section~\ref{sec:regmodel} can be extended to include higher order interactions in a straightforward manner.
We show below the model including all higher order interactions, and including only a subset is direct.

Let $\mathcal{J} = \{1,\dots,J\}$ be the set of $J$ factors and let $\mathcal{T}$ be the set of all possible assignments on the $\mathcal{J}$ factors.
Then our model for $\psi_{k}(\bT_i)$ with all interactions among factors is
\begin{align*}
\psi_{k}(\bT_i)& \ =\ \mu + \sum_{j=1}^J \sum_{l =0}^{L_j-1} 
\mathbf{1}\{T_{ij} = l\} \beta^j_{kl} + \sum_{j=1}^{J-1} \sum_{j' >
	j} \sum_{l=0}^{L_j-1} \sum_{l' = 0}^{L_{j'}-1} \mathbf{1}\{T_{ij} = l,
T_{ij'} = l'\} \beta^{jj'}_{kll'}\\
&\qquad \ +\ \cdots +  \sum_{\bm{t} \in \mathcal{T}} \mathbf{1}\{\bm{T}_{i} = \bm{t}\} \beta^{12\cdots K}_{k\bm{t}} \\
&\ =\ \mu + \tilde{\bm{T}}_i^\top\bbeta_k.
\end{align*}
In the above formulation, $\beta^{12\cdots K}_{k\bm{t}}$ is the $K$-way interaction coefficient in cluster $k$ for assignment $\bm{t}$.

Let $\mathcal{T}_{-j}$ be the set of all possible assignments on the $\mathcal{J}$ factors except for factor $j$.
With some slight notation abuse by letting $\beta^{12\cdots K}_{kl\bm{t}_{-j}}$ be the $K$-way interaction coefficient in cluster $k$ for assignment $l$ for $j$ and $\bm{t}_{j}$ for the other $J-1$ factors, the ANOVA-type sum-to-zero constraints extend as follows:
\begin{equation}
\sum_{l =0}^{L_j-1} \beta^j_{kl} = 0, 
\sum_{l=0}^{L_{j}-1} \beta^{jj'}_{kll'} = \sum_{l'=0}^{L_{j'}-1} \beta^{jj'}_{kll'} = 0, \dots,  \sum_{l=0}^{L_{j}-1}  \beta^{12\cdots K}_{kl\bm{t}_{-j}} = 0  \label{eq:appsum-to-zero}
\end{equation}
for $j,j'=1,2,\ldots,J$ with $j' > j$ and for all $\bm{t}_{-j} \in \mathcal{T}_{-j}$.  We
write them compactly as,
\begin{equation}
\bC^\top \bbeta_k \ = \ \bm{0}, \label{eq:appconstraint}
\end{equation}
where each row of $\bC^\top \bbeta_k$ corresponds to one of the
constraints given in Equation~\eqref{eq:appsum-to-zero}.

For the structured sparsity, we have penalties of the form
\begin{align*}
\sum_{j=1}^J \sum_{l_j = 1}^{L_j}\sum_{l_j^\prime >l_j}^{L_j}\sqrt{(\beta^j_{l_j} - \beta^j_{l_j^\prime})^2 + \sum_{j^\prime \neq j}\sum_{l_{j^\prime} = 1}^{L_{j^\prime}}(\beta^{j j^\prime}_{l_jl_{j^\prime}} - \beta^{j{j^\prime}}_{l_j^\prime l_{j^\prime}})^2  +\cdots + \sum_{\bm{t}_{-j} \in \mathcal{T}_{-j}} (\beta^{12\cdots K}_{l_j	\bm{t}_{-j}} -  \beta^{12\cdots K}_{l_j^\prime \bm{t}_{-j}})^2}
\end{align*}
This will have $\sum_{j=1}^JL_j(L_j-1)/2$ terms, $L_j(L_j-1)/2$ terms for the $j$th factor.

For illustration, consider a simple example with one group and three
factors---factor one has three levels, factor two has two levels, and factor three has two levels.
In this case, our penalty contains 5 terms,

\begin{equation*}
\begin{split}
& \sum_{l_1 = 1}^{L_1}\sum_{l_1^\prime >l_1}^{L_1}\sqrt{(\beta^1_{l_1} - \beta^1_{l_1^\prime})^2 + \sum_{l_2 = 1}^{L_2}(\beta^{12}_{l_1l_2} - \beta^{12}_{l_1^\prime l_2})^2 + \sum_{l_3 = 1}^{L_3}(\beta^{13}_{l_1l_3} - \beta^{13}_{l_1^\prime l_3})^2 +  \sum_{l_2 = 1}^{L_2}\sum_{l_3 = 1}^{L_3}(\beta^{123}_{l_1l_2l_3} - \beta^{123}_{l_1^\prime l_2l_3})^2}\\
~+~ & \sum_{l_2 = 1}^{L_2}\sum_{l_2^\prime >l_2}^{L_2}\sqrt{(\beta^1_{l_2} - \beta^2_{l_2^\prime})^2 + \sum_{l_2 = 1}^{L_2}(\beta^{12}_{l_1l_2} - \beta^{12}_{l_1 l_2^\prime})^2 + \sum_{l_3 = 1}^{L_3}(\beta^{23}_{l_2l_3} - \beta^{23}_{l_2^\prime l_3})^2 +  \sum_{l_1 = 1}^{L_1}\sum_{l_3 = 1}^{L_3}(\beta^{123}_{l_1l_2l_3} - \beta^{123}_{l_1l_2^\prime l_3})^2}\\\\
~+~ & \sum_{l_3 = 1}^{L_3}\sum_{l_3^\prime >l_3}^{L_3}\sqrt{(\beta^1_{l_3} - \beta^3_{l_3^\prime})^2 + \sum_{l_3 = 1}^{L_2}(\beta^{13}_{l_1l_3} - \beta^{12}_{l_1 l_3^\prime})^2 + \sum_{l_1 = 1}^{L_3}(\beta^{13}_{l_1l_3} - \beta^{23}_{l_1 l_3^\prime})^2 +  \sum_{l_1 = 1}^{L_1}\sum_{l_2 = 1}^{L_2}(\beta^{123}_{l_1l_2l_3} - \beta^{123}_{l_1l_2 l_3^\prime})^2}\\
\end{split}
\end{equation*}

The first three terms encourages the pairwise fusion of the levels of
factor one whereas the fourth encourages the fusion of the two levels
of factor two and the fifth encourages the fusion of the two levels
of factor three.

Using the sum of Euclidean norms of quadratic forms, we can write the penalty as
\begin{equation*}
|| \bm{\beta}^\top \bm{F}_1 \bm{\beta} ||_2 +  \ ||
\bm{\beta}^\top \bm{F}_2 \bm{\beta} ||_2 + \ ||
\bm{\beta}^\top \bm{F}_3 \bm{\beta} ||_2 + \ ||
\bm{\beta}^\top \bm{F}_4 \bm{\beta} ||_2+ \ ||
\bm{\beta}^\top \bm{F}_5 \bm{\beta} ||_2,  
\end{equation*}
where $\bF_1, \bF_2, \bF_3$ are appropriate positive semi-definite
matrices to encourage the fusion of the pairs of levels in factor one, $\bF_4$ encourages the fusion of the two levels in factor two,
$\bF_5$ encourages the fusion of the two levels in factor three,
and
$\bbeta = [\beta^1_0\ \beta^1_1\ \beta^1_2\ \beta^2_0\ \beta^2_1\
\beta^{12}_{00}\ \beta^{12}_{10}\ \beta^{12}_{20}\ \beta^{12}_{01}\
\beta^{12}_{11}\ \beta^{12}_{21} \cdots \beta^{123}_{211}]^\top$. 

More generally, for a fully interacted model we will have $\sum_{j=1}^JL_j(L_j-1)/2 = G$ terms, 
\begin{equation*}
\sum_{g=1}^{G} || \bm{\beta}^\top \bm{F}_g \bm{\beta} ||_2 .
\end{equation*}

\section{Propriety of the Structured Sparse Prior}
\label{sec:app_ssparse_prior}

The proof of propriety for the structured sparse prior used in our
paper is an application of Theorem~1 established in
\cite{goplerud2021sparsity} and is reproduced here.
\begin{theorem}[\cite{goplerud2021sparsity}] 
	\label{result:goplerud_1}
	Consider the following structured sparse prior on
	$\bm{\beta} \in \mathbb{R}^p$ with regularization strength
	$\lambda > 0$ penalizes $K$ linear constraints $\bm{d}_k$ and
	$L$ quadratic constraints $\bm{F}_\ell$ on the parameters
	where $\bm{F}_\ell$ is symmetric and positive
	semi-definite. The kernel of the prior is shown below.
	$$p(\bm{\beta}) \propto \exp\left(-\lambda \left[\sum_{k=1}^K
	|\bm{d}_k^\top \bm{\beta}| + \sum_{\ell=1}^L \sqrt{
		\bm{\beta}^\top \bm{F}_{\ell} \bm{\beta}}\right] \right)$$ Further
	define $\bm{D}^\top = [d_1, \cdots, d_K]^\top$ and
	$\bar{\bm{D}}^\top = [\bm{D}^\top, \bm{F}_1, \cdots, \bm{F}_L]$.
	Then, for $\lambda >0 $, the prior above is proper if and only if
	$\bar{\bm{D}}$ is full column rank.
\end{theorem}
In our specific case, we note that $K = 0$, $L = G$, and
$\lambda = \lambda \bar{\pi}_k^\gamma$. Prior propriety of
$p(\bm{\beta}_k \mid \{\bm{\phi}_k\}_{k=2}^K, \lambda)$, therefore,
can be determined by empirically investigating whether $\bar{\bm{D}}$,
i.e. the vertically stacked $\bm{F}_\ell$, is full column rank.

It is also possible to analytically show the propriety of the prior
distribution in all cases considered in this paper. We focus on the
case of $K = 1$ and arbitrary $\lambda > 0$ as the result follows
automatically for the case in our paper. 
\begin{appresult}
	\label{thm:proper_prior}
	Assume a structured sparse prior for a factorial or conjoint
	design with $J$ factors each with $L_j$ levels where all
	pairwise interactions are included and levels of each factor
	are encouraged to be fused together (i.e. the model in the
	main text). The kernel of the prior is shown below where
	$\bm{F}_g$ are as defined in the main text. 	
	$$k(\bm{\beta}) = \exp\left(-\lambda \sum_{g=1}^G \sqrt{\bm{\beta}^\top \bm{F}_g \bm{\beta}}\right)$$
	Assume that the linear sum-to-zero constraints
	$\bm{C}^\top \bm{\beta} = \bm{0}$ hold.  Then, the structured
	sparse prior on the unconstrained $\tilde{\bm{\beta}}$ such
	that $\tilde{\bm{\beta}} \in \mathcal{N}(\bm{C}^\top)$ is
	proper. Or, equivalently, the following result holds.
	$$\int_{\bm{\beta}: \bm{C}^\top \bm{\beta} = \bm{0}} k(\bm{\beta}) d\bm{\beta} < \infty.$$
\end{appresult}
\begin{proof}
	Let $\mathcal{B}_{\bm{C}^\top}$ represent a basis for the
	linear constraints $\bm{C}^\top$. The integral for evaluating
	propriety can be written as,
	\begin{equation*}
	\int_{\tilde{\bm{\beta}}} \tilde{k}(\tilde{\bm{\beta}})
	d\tilde{\bm{\beta}}\quad \text{where} \quad \tilde{k}(\tilde{\bm{\beta}}) =
	\exp\left(-\lambda \sum_{g=1}^G
	\sqrt{\tilde{\bm{\beta}}^\top\mathcal{B}_{\bm{C}^\top}^\top \bm{F}_g
		\mathcal{B}_{\bm{C}^\top} \tilde{\bm{\beta}}}\right). 
	\end{equation*}
	Note that $\bm{F}_g$ can be expressed as a sum of $N_g$ outer products
	of $|\bm{\beta}|$-length vectors of the form
	$\bm{l}_i \in \{-1, 0, 1\}$ where $-1$ and $1$ correspond to the two
	terms that are fused together and all other elements are $0$, i.e.,
	$ \bm{F}_g = \sum_{g'=1}^{N_g} \bm{l}_{g'}\bm{l}_{g'}^\top$. Thus, one
	can define a matrix
	$\bm{Q}_g^\top = \left[\bm{l}_1, \cdots, \bm{l}_{N_g}\right]$ such
	that $\bm{Q}_g^\top \bm{Q}_g = \bm{F}_g$, which allows us to rewrite
	$\tilde{k}(\tilde{\bm{\beta}})$ as:
	$$\tilde{k}(\tilde{\bm{\beta}}) = \exp\left(-\lambda \sum_{g=1}^G
	\sqrt{\tilde{\bm{\beta}}^\top
		\left[\mathcal{B}_{\bm{C}^\top}\right]^\top \bm{Q}_g^\top \bm{Q}_g
		\mathcal{B}_{\bm{C}^\top} \tilde{\bm{\beta}}}\right).$$
	
	By applying Theorem~\ref{result:goplerud_1} and noting that the
	null spaces of $\bm{A}^T\bm{A}$ and $\bm{A}$ are identical, the
	integral of $\tilde{k}(\tilde{\bm{\beta}})$ is finite if and only if
	$\bm{Q}\mathcal{B}_{\bm{C}^\top}$ is full column rank, where
	$\bm{Q}^\top = [\bm{Q}^\top_1, \cdots, \bm{Q}^\top_G]$.  We
	demonstrate this fact in two steps.  First, there exists a permutation
	matrix $\bm{P}_{Q}$ such that $\bm{P}_Q \bm{Q} $ has a block diagonal
	structure with $J+1$ diagonal blocks. The first $J$ blocks
	corresponding to the main terms for each factor $j$ and the last block
	corresponds to all interaction terms. The null space of each block is
	spanned by the vector $\bm{1}$ as the corresponding block of
	$\bm{P}_Q\bm{Q}$ is a (transposed) orientated incidence matrix of a
	fully connected graph. Thus, the null space of $\bm{P}_Q \bm{Q}$, and
	hence $\bm{Q}$, is spanned by the $J+1$ columns of a block diagonal
	matrix with $\bm{1}$ on each block.  Second, consider the linear
	constraints $\bm{C}^\top \bm{\beta} = \bm{0}$.  The only vector to
	satisfy this constraint and lie in the null space of $\bm{Q}$ must be
	$\bm{0}$ as, for each block, the only vector proportional to $\bm{1}$
	and satisfying the corresponding sum-to-zero constraints must be
	$\bm{0}$. Thus, $\bm{Q}\mathcal{B}_{\bm{C}^\top}$ is full column rank
	and the prior is proper.
\end{proof}

\section{Derivations for the Basic Model}
\label{sec:app_derivations}

This section derives a number of results for the basic model. It first
restates the main results concerning the elimination of the linear
constraints $\bC^\top \bm{\beta}_k = \bm{0}$. Then, it derives the
Expectation Maximization algorithm, our measure of degrees of freedom,
and some additional computational improvements used to accelerate
estimation.  In the following, we use
$\tilde{\bm{T}}_i$ to denote the corresponding vector of indicators for
whether certain treatments or interactions are present (i.e. stacking
all $\bm{1}\{T_{ij} = l\}$, etc. from Equation~\ref{eq:define_lp}).  In
addition, we use $\psi_{ik}$ to indicate the linear predictor for
observation $i$ and group $k$.

\subsection{Removing the Linear Constraints}
\label{sec:app_derivation_LC}

The inference problem in the main text is presented as an optimization
problem subject to linear constraints on the coefficients
$\bm{\beta}_k$. Inference is noticeably easier if these
are eliminated via a transformation of the problem to a
lower-dimensional one by noting that $\bm{\beta}_k$ must lie in the
null space of the constraint matrix $\bm{C}^\top$ (see, e.g., \citealt[ch. 20]{lawson1974linear}). Define
$\tilde{\bm{\beta}}_k = \left(\mathcal{B}_{\bC^\top}^\top
\mathcal{B}_{\bC^\top}\right)^{-1}
\mathcal{B}_{\bC^\top}^\top\bbeta_k$ where $\mathcal{B}_{\bC^\top}$ is
a basis for the null space of $\bC^\top$. The problem can thus
be solved in terms of the unconstrained
$\tilde{\bm{\beta}}_k \in \mathbb{R}^{p-\mathrm{rank}(\bC^\top)}$ given appropriate adjustment of the treatment design vectors, $\tilde{\tilde{\bm{T}}}_i = \mathcal{B}_{\bC^\top} \tilde{\bm{T}}_i$, penalty matrices, $\tilde{\bF}_g=\mathcal{B}_{\bC^\top}^\top \bF_g \mathcal{B}_{\bC^\top}$, and linear predictor, $\psi_{i,k} = \left[\tilde{\tilde{\bm{T}}}_i\right]^\top
\tilde{\bm{\beta}}_{Z_i} + \mu$. Once the algorithm convergences, the constrained parameters can be recovered by noting $\bm{\beta}_k = \mathcal{B}_{\bC^\top}\tilde{\bm{\beta}}_k$.

Given the similarity of the unconstrained and constrained problems and for notational simplicity, we present all results herein dropping the second ``tilde'' notation on $\tilde{\bT}_i$ and the ``tilde'' on $\bm{\beta}_k$ and note that, once estimated, $\tilde{\bm{\beta}}_k$ is projected back into the original space for the reported coefficients, average marginal component effects, etc. The results of Appendix~\ref{sec:app_uncertainty} on
approximating $\tilde{\bm{\beta}}_k$ as multivariate Gaussian imply that $\bm{\beta}_k$ will have a (singular) multivariate Gaussian distribution.

\subsection{Expectation Maximization Algorithm}
\label{sec:app_derivation_EM}

This section considers inference after removing the linear constraints as discussed in the prior subsection. Algorithm~\ref{alg:main} summarizes our approach to maximizing
Equation~\eqref{eq:aecm_obs}.  Each iteration of our AECM algorithm
involves two cycles where the data augmentation scheme enables
iterative updating of the treatment effect parameters $\bm{\beta}$ and
moderators $\bm{\phi}$. $\bm{\theta}$ collects both sets of parameters.

\begin{algorithm}\spacingset{1.25}
	\caption{AECM Algorithm for Estimating $\bm{\theta}$}
	\label{alg:main}
	\begin{algorithmic}
		\State{\textbf{Set Hyper-Parameters}: $K$ (groups), $\lambda$, $\sigma^2_\phi$, $\gamma$ (prior strength), $\epsilon_1, \epsilon_2$ (convergence criteria), $T$ (number of iterations)}
		
		\State{\textbf{Initialize Parameters}: $\bm{\theta}^{(0)}$, i.e. $\bm{\beta}^{(0)}$ and $\bm{\phi}^{(0)}$; Appendix~\ref{sec:app_derivations_improve} provides details.}
		\For{iteration $t \in \{0, \cdots, T-1\}$}
		\State{\textbf{Cycle 1: Update $\bm{\beta}$}}
		\State{1a.} $E$-Step: Find the conditional distributions of $\{Z_i, \omega_i\}_{i=1}^N$ and $\{\{\tau^2_{gk}\}_{g=1}^G\}_{k=1}^K$ given $\{Y_i, \bX_i, \bT_i\}$ and $\bm{\theta}^{(t)}$ (Eq.~\eqref{eq:aecm_ebeta}). Derive $Q_\beta(\bm{\beta}, \bm{\theta}^{(t)})$ (Eq.~\eqref{eq:aecm_qbeta}).
		\State{1b.} $M$-Step: Set $\bm{\beta}^{(t+1)}$ such that $Q_\beta(\bm{\beta}^{(t+1)}, \bm{\theta}^{(t)}) \geq Q_\beta(\bm{\beta}^{(t)}, \bm{\theta}^{(t)})$ 
		\State{\textbf{Cycle 2: Update $\bm{\phi}$}}
		\State{2a.} $E$-Step: Find $p(Z_i=k \mid Y_i, \bX_i, \bT_i, \bm{\beta}^{(t+1)}, \bm{\phi}^{(t)})$. Derive $Q_\phi(\bm{\phi}, \{\bm{\beta}^{(t+1)}, \bm{\phi}^{(t)}\})$ (Eq.~\eqref{eq:aecm_qphi}). 
		\State{2b.} $M$-Step: Set $\bm{\phi}^{(t+1)}$ such that 
		
		$Q_\phi(\bm{\phi}^{(t+1)}, \{\bm{\beta}^{(t+1)}, \bm{\phi}^{(t)}\}) \geq Q_\phi(\bm{\phi}^{(t)}, \{\bm{\beta}^{(t+1)}, \bm{\phi}^{(t)}\})$
		\State{\textbf{Check Convergence}}
		\State{3.} Stop if $\log p\left(\bm{\theta}^{(t+1)} | \{Y_i, \bX_i, \bT_i\}_{i=1}^N\right) - \log p\left(\bm{\theta}^{(t)} | \{Y_i, \bX_i, \bT_i\}_{i=1}^N\right) < \epsilon_1$ (Eq.~\eqref{eq:aecm_obs}) or \\$||\bm{\theta}^{(t+1)} - \bm{\theta}^{(t)}||_\infty < \epsilon_2$.
		\EndFor
	\end{algorithmic}
\end{algorithm}

\subsubsection{Updating Treatment Effect Parameters}

We begin with the cycle of the AECM algorithm for updating $\{\bm{\beta}_k\}_{k=1}^K$ and $\mu$ given $\{\bm{\phi}_{k=2}^K\}$.  To update $\bm{\beta}, \mu$, our data augmentation strategy requires three
types of missing data.  First, we use the standard group memberships of each unit $i$ for inference in finite mixtures, i.e., $Z_i \in \{1, \cdots, K\}$.  We also include two other types of data augmentation that result in a closed-form update.  We use Polya-Gamma augmentation ($\omega_i$;
\citealt{polson2013polyagamma}) for the logistic likelihood and data augmentation on the sparsity-inducing penalty ($\tau^2_{gk}$; see, e.g., \citealt{figueiredo2003adaptive,polson2011svm,ratkovic2017sparse,goplerud2021sparsity}) yielding
\begin{subequations}
	\label{eq:aecm_ebeta}
	\begin{align}
	&p(Y_i, \omega_{i} \mid Z_i, \bX_i, \bT_i) \ \propto \ \frac{1}{2}\exp\left\{\left(Y_i - \frac{1}{2}\right) \psi_{Z_i}(\bT_i) - \frac{\omega_{i}}{2} \left[\psi_{Z_i}(\bT_i)\right]^2\right\} f_{PG}(\omega_{i} \mid 1, 0),\label{eq:augmented1} \\
	&p(\bm{\beta}_k, \{\tau^2_{gk}\}_{g=1}^G \mid \lambda,
	\{\bm{\phi}_k\}) \ \propto \
	\exp\left\{-\frac{1}{2}\bm{\beta}_k^\top\left(\sum_{g=1}^G
	\frac{\bm{F}_{g}}{\tau^2_{gk}}\right)
	\bm{\beta}_k\right\} \prod_{g=1}^G\tau_{gk}^{-1}
	\exp\left\{-\frac{\left(\lambda \bar{\pi}_k\right)^2}{2} \cdot \tau^2_{gk}\right\}, \label{eq:augmented2}
	\end{align}
\end{subequations}

where $f_{PG}(\cdot \mid b, c)$ represents the Polya-Gamma
distribution with parameters $(b,c)$ and
$Z_i \sim \mathrm{Multinomial}\left(1, \bm{\pi}_i\right)$ with the
$k$th element of $\bm{\pi}$ equal to $\pi_k(\bX_i)$. Note that
$\bm{\beta}$ only enters Equation~\eqref{eq:aecm_ebeta} via a
quadratic form. The first cycle of the AECM algorithm involves, therefore, maximizing the following function with respect to $\bm{\beta}$ given $\bm{\theta}^{(t)}$.
\begin{align}
\label{eq:aecm_qbeta}
Q_\beta\left(\bm{\beta}, \bm{\theta}^{(t)}\right)\ = \ &\sum_{i=1}^N \sum_{k=1}^K \E[\mathbf{1}\{Z_i = k\}]\left\{\left(Y_i -
\frac{1}{2}\right) \psi_{k}(\bT_i) - \E[\omega_i \mid Z_i = k]
\frac{\left[\psi_k(\bT_i)\right]^2}{2}\right\}  \nonumber \\
& + \sum_{k=1}^K -\frac{1}{2}
\bm{\beta}_k^\top \left[\sum_{g=1}^K \bm{F}_g \cdot
\E[1/\tau^2_{gk}]\right] \bm{\beta}_k + \text{const.}
\end{align}
where all expectations are taken over the conditional distribution of
the missing data given the current parameter estimates.  We note that
the $E$-Step involves computing
$p(\{\omega_i, Z_i\}, \{1/\tau^2_{gk}\} \mid \{Y_i, \bX_i, \bT_i\},
\bm{\theta}^{(t)})$ which factorizes into, respectively, a collection
of Polya-Gamma (PG), categorical, and Inverse-Gaussian random variables. Their conditional distributions are shown below,

\begin{subequations}
	\begin{align}
	p(\tau^{-2}_{gk} \mid \bm{\theta}) &\sim \mathrm{InverseGaussian}\left(\frac{\lambda}{\sqrt{\bm{\beta}_k^\top\bm{F}_g\bm{\beta}_k}}, \quad \lambda^2\right), \\
	p(Z_i = k \mid Y_i, \bX_i, \bT_i, \bm{\theta}) &\propto p_{ik}^{Y_i} (1-p_{ik})^{1-Y_i} \pi_{ik}; \quad p_{ik} = \frac{\exp(\psi_{ik})}{1+\exp(\psi_{ik})}, \\
	p(\omega_i \mid Z_i = k, \bX_i, \bT_i, \bm{\theta}) &\sim \mathrm{PG}\left(1, \psi_{ik}\right),
	\end{align}
\end{subequations}
as well as the relevant expectations needed in $Q_\beta(\bm{\beta}, \bm{\theta})$,
\begin{subequations}
	\begin{align}
	& \E\left[\tau^{-2}_{gk}\right] = \frac{\lambda}{\sqrt{\bm{\beta}_k^\top\bm{F}_g\bm{\beta}_k}}, \\
	& \E [z_{ik}] = \E\left[\bm{1}\{Z_i = k\}\right] = \frac{p_{ik}^{Y_i} (1-p_{ik})^{1-Y_i} \pi_{ik}}{\sum_{\ell=1}^K p_{i\ell}^{Y_i} (1-p_{i\ell})^{1-Y_i} \pi_{i\ell}}, \\
	& \E [\omega_{i} \mid Z_i = k] = \frac{1}{2\psi_{ik}} \tanh\left(\frac{\psi_{ik}}{2}\right).
	\end{align}
\end{subequations}

Note that as $\bm{\beta}_k^\top \bm{F}_g \bm{\beta}_k$ approaches
zero, $\E[\tau^{-2}_{gk}]$ approaches infinity. To prevent numerical
instability, we rely on the strategy in \cite{goplerud2021sparsity}
(inspired by \citealt{polson2011svm}) where once it is sufficiently
small, e.g. below $10^{-4}$, and thus the restriction is almost
binding, we ensure that restriction holds in all future iterations. We
do so by adding a quadratic constraint
$\bm{\beta}_k^\top \bm{F}_g \bm{\beta}_k = 0$. This implies that $\bm{\beta}_k$ lies in the null space of $\bm{F}_g$ and
thus with an additional transformation, it can be removed and the
problem be solved in an unconstrained space with a modified design.

To compute the update for $\bm{\beta}$, define $\check{\bm{\beta}}^\top = [\mu, \bm{\beta}_1, \cdots,
\bm{\beta}_K]^\top$. We can create a corresponding design matrix
$\check{\bm{T}} = [\bm{1}_{N}, \bm{I}_K \otimes \bm{T}]$
where $\tilde{\bm{T}}^\top = [\tilde{\bm{T}}_1, \cdots, \tilde{\bm{T}}_N]$ and diagonal weight matrix
$\check{\bm{\Omega}} =
\mathrm{diag}\left(\{\{\E[z_{ik}]\E[\omega_{i}\mid Z_i =
k]\}_{i=1}^N\}_{k=1}^K\right)$. Further, we can create the combined
ridge penalty
$\bm{\mathcal{R}} = \mathrm{blockdiag}\left(\{0,
\{\bm{R}_k\}_{k=1}^K\}\right)$ where
$\bm{R}_k = \sum_g \bm{F}_g \E[\tau^{-2}_{gk}]$ and augmented outcome
$\check{\bm{Y}} = \{\{\E[z_{ik}] (Y_i - 1/2)\}_{i=1}^N\}_{k=1}^K$. The $Q_\beta$ function is thus proportional to the following ridge regression problem and yields the update for the $M$-Step,
\begin{align*}
Q_\beta\left(\bm{\beta};
\bm{\theta}^{(t)}\right) & = \check{\bm{Y}}^\top
\left(\check{\bm{T}}
\check{\bm{\beta}}\right) - \frac{1}{2}
\check{\bm{\beta}}^\top
\check{\bm{T}}^\top \check{\bm{\Omega}}
\check{\bm{T}} \check{\bm{\beta}} -
\frac{1}{2} \check{\bm{\beta}}^\top
\bm{\mathcal{R}} \check{\bm{\beta}}  + \text{const.},\\
\check{\bm{\beta}}^{(t+1)} &= \left(\check{\bm{T}}^\top \check{\bm{\Omega}} \check{\bm{T}} + \bm{\mathcal{R}}\right)^{-1} \check{\bm{T}}^\top \check{\bm{Y}}.
\end{align*}

One could reply on a generalized EM algorithm where $Q_\beta$ is improved versus maximized for computational reasons, e.g. by using a conjugate gradient solver initialized at $\check{\bm{\beta}}^{(t)}$.

\subsubsection{Updating Moderator Parameters}

To update the moderator parameters $\bphi$, we use the second cycle of
the AECM algorithm where only the $Z_i$ are treated as missing
data. The $E$-step involves recomputing the group membership probabilities, i.e., $p(Z_i \mid Y_i, \bX_i, \bT_i, \bm{\beta}^{(t+1)}, \bm{\phi}^{(t)})$, given the
updates in the first cycle. The implied $Q$-function is shown below,
\begin{align}
\label{eq:aecm_qphi}
Q_{\phi}(\bm{\phi}, \{\bm{\beta}^{(t+1)}, \bm{\phi}^{(t)}\}) \ = \
&\sum_{k=1}^K \left[\sum_{i=1}^N \E[\mathbf{1}\{Z_{i} = k\}]\log \pi_{k}(\bX_i)\right] \nonumber \\
& + \sum_{k=1}^K \left[m \gamma \log \bar{\pi}_k - \lambda \bar{\pi}_k^\gamma \sum_{g=1}^{G} \sqrt{\bbeta_k^\top \bF_{g} \bbeta_k}\right] + \log p(\{\bphi_k\}_{k=2}^K),
\end{align}
where $\pi_k(\bX_i)$ and $\bar\pi_k= \sum_{i=1}^N \pi_k(\bX_i)/N$ are
functions of $\bphi_k$.  Note that if $\gamma = 0$, this simplifies to
a multinomial logistic regression with $\{\E[\bm{1}\{Z_i = k\}]\}_{k=1}^K$ as the
outcome. We perform the $M$-Step using a standard optimizer (e.g., L-BFGS) to optimize $Q_{\phi}$ and thus obtain $\bm{\phi}^{(t+1)}$.

\subsection{Classification Maximum Likelihood}
\label{app:classificationml}

If classification maximum likelihood approach is desired, despite
statistical concerns about this procedure's asymptotic bias (e.g.,
\citealt{bryant1978asymptotic}), it can be easily implemented by
adapting the preceeding EM algorithm. \cite{celeux1992classification}
propose the ``classification EM'' algorithm in the spirit of how
$k$-means classification is commonly implemented.

The adjustment proceeds as follows
\citep[p. 319]{celeux1992classification}: after conducting an $E$-step
and obtaining
$\tilde{\pi}_k(\bX_i, Y_i, \bT_i; \bm{\theta}) = \tilde{\pi}_{ik} = p(Z_i = k \mid Y_i, \bX_i, \bT_i, \bm{\theta})$
for use in evaluating $Q_\beta$ and $Q_\phi$, perform a classification
or ``hard assignment''. That is, find
$k^*_i = \argmax_{k} \tilde{\pi}_{ik}$, i.e., the most probable
cluster for observation $i$ given its observed data and
$\bm{\theta}$. In the subsequent $M$-step, use a modified weight
$c_{ik} = 1$ if $k = k^*_i$ and otherwise $c_{ik} = 0$ in lieu of
$\tilde{\pi}_{ik}$.

\subsection{Degrees of Freedom}
\label{sec:app_derivations_df}

Our procedure for estimating $\check{\bm{\beta}}^{(t)}$
appears similar to the results in \cite{oelker2017uniform} where
complex regularization and non-linear models can be recast as a
(weighted) ridge regression. Using that logic, we take the trace of
the ``hat matrix'' implied by our algorithm at stationarity to
estimate our degrees of freedom. We also adjust upwards the degrees of
freedom by the number of moderator coefficients (e.g.,
\citealt{khalili2010mixture,chamroukhi2019regularized}).

Equation~\eqref{eq:df} shows our procedure where $\bm{\mathcal{R}}$
and $\check{\bm{\Omega}}$ contain expectations calculated at
convergence. $p_x$ denotes the number of moderators, i.e. the
dimensionality of $\bm{\phi}_k$. Before evaluating
Equation~\eqref{eq:df}, for any two factor levels that are
sufficiently close (e.g.,
$\sqrt{\bbeta_k^\top\bF_g \bbeta_k} < 10^{-4}$), we assume they are
fused together and consider it as an additional linear constraint on
the parameter vector $\bm{\beta}_k$.

\begin{equation}
\label{eq:df}
\begin{split}
&\mathrm{df} = \mathrm{tr}\left[\left(\check{\bm{T}}^\top \check{\bm{\Omega}} \check{\bm{T}} + \bm{\mathcal{R}}\right)^{-1} \check{\bm{T}}^\top \check{\bm{\Omega}} \check{\bm{T}}\right] + p_x \left(K - 1\right)
\end{split}
\end{equation}

From this, we can calculate a BIC criterion. We seek to find the
regularization parameter $\lambda$ that minimizes this criterion. To
avoid the problems of a naive grid-search, we use Bayesian model-based
optimization that attempts to minimize the number of function
evaluations while searching for the value of $\lambda$ that minimizes
the BIC (\texttt{mlrMBO}; \citealt{bischl2018mlrmbo}). We find that
with around fifteen model evaluations, the optimizer can usually find
a near optimal value of $\lambda$.

\subsection{Computational Improvements}
\label{sec:app_derivations_improve}

While the algorithm above provides a valid way to locate a posterior
mode, our estimation problem is complex and
high-dimensional. Furthermore, given the complex posterior implied by
mixture of experts models, we derived a number of computational
strategies to improve convergence. We use the SQUAREM algorithm
(\citealt{varadhan2008simple}). Our software provides the option to use a generalized EM algorithm to update $\bm{\beta}$ using a conjugate gradient approach and $\bm{\phi}$ using a few steps of L-BFGS.

We also outline a way to deterministically initialize the model to provide stability and, again, speed up estimation on large problems. To do this, we adapt the procedure from \cite{murphy2020init} for initializing mixture of experts: (i) initialize the groups using some (deterministic) procedure (e.g. spectral clustering on the moderators); (ii) using only the main effects, estimate an EM algorithm---possibly with hard assignment at the $E$-Step (CEM; \citealt{celeux1992classification}); (iii) iterate until the memberships have stabilized. Use those memberships to initialize the model. This has the benefit of having a deterministic initialization procedure where the group membership is based on the moderators but guided by which grouping seem to have sensible treatment effects, at least for the main effects. Given the memberships, update $\bm{\beta}$ using a ridge regression and $\bm{\phi}$ using a ridge regression and take those values as $\bm{\beta}^{(0)}$ and $\bm{\phi}^{(0)}$.

\section{Extensions to the Basic Model}
\label{sec:app_extensions}

As noted in the main text, there are five major extensions to the basic model that applied users might wish to include:

\begin{enumerate}
	\item Repeated tasks (observations) for a single individual
	\item A forced-choice conjoint experiment
	\item Survey  to weight the sample estimate to the broader population
	\item Adaptive weights for each penalty
	\item Latent overlapping groups
\end{enumerate}

All can be easily incorporated into the proposed framework above. This
section outlines the changes to the underlying model.

\subsection{Repeated Observations}
\label{sec:app_extensions_repeat}

This modification notes that in factorial and conjoint experiments it
is common for individuals to perform multiple tasks. Typically, the
number of tasks $N_i$ is similar across individuals. The updated
likelihood for a single observation $i$ is shown below; we show both
the observed and complete case. $y_{im}$ represents the choice of
person $i$ on task $m \in \{1, \cdots, N_i\}$; $p_{imk}$ is the
probability of $Y_{im} = 1$ if person $i$ was in group $k$, and $\tilde{\bm{T}}_{im}$ is the vector of treatment indicators for person $i$ on task $m$.

\begin{align}
&L\left(\{Y_{im}\}_{m=1}^{N_i}\right) = \sum_{k=1}^K \pi_{ik} \left[\prod_{m=1}^{N_i} p_{imk}^{Y_{im}} (1-p_{imk})^{1-Y_{im}} \right]; \quad p_{imk} = \frac{\exp(\psi_{imk})}{1+\exp(\psi_{imk})}; \quad \psi_{imk} = \tilde{\bm{T}}_{im}^\top\bm{\beta}_k + \mu 
\label{app_eq:repeated} \\
&L^c(\{y_{im}, \omega_{im}\} \mid Z_i) = \prod_{t=1}^{N_i}
\left[\frac{1}{2} \exp\left\{\left(Y_{im} - \frac{1}{2}\right)
\psi_{i, Z_i} - \omega_{im} \frac{\psi_{im,Z_i}^2}{2}\right\}
f_{PG}(\omega_{im} \mid 1, 0) \right]  
\end{align}

Note that because of the conditional independence of
$(y_{it}, \omega_{it})$ given $Z_i$ and the parameters, the major
modifications to the EM algorithm is that the $E$-Step must account
for all $t$ observations, i.e. the terms summed in
Equation~\eqref{app_eq:repeated}. Some additional book-keeping is
required in the code as the design of the treatments has
$\sum_{i=1}^N N_i$ rows whereas the design of the moderators has $N$
rows. Repeated observations can be easily integrated into the
uncertainty estimation procedure outlined below.

\subsection{Forced Choice Conjoint Design}
\label{sec:app_extensions_forced}

A popular design of a conjoint experiment is the forced choice design
where the respondents are required to choose between two profiles.
Therefore, the researcher does not observe an outcome for each profile
separately, but rather a single outcome is observed for each pair
indicating which is preferred. \cite{egam:imai:19} show that this can
be easily fit into the above framework with some adjustment.
Specifically, the model is modified to difference the indicators of
the treatment levels for the pair of profiles (subtracting, e.g., the
levels of the profile presented on the left from those of the profile
presented on the right). The intercept for this model can be
interpreted as a preference for picking a profile presented in a
particular location.  With this modification, estimation proceeds as
before.

\subsection{Standardization Weights}
\label{sec:app_extensions_standardization}

An additional modification to the problem is to weight the
penalty. This could be done for two reasons.  First, there is an issue
of the columns having different variances/Euclidean norms because of
the different number of factor levels $L_j$. Second, it is popular to
weight the penalty based on some consistent estimator (e.g. ridge
regression) to improve performance and, in simpler models, can be
shown to imply various oracle properties
(e.g. \citealt{zou2006adaptive}). We leave the latter to future
exploration.

Define $\xi_{gk}$ as a positive weight for the $g$-th penalty and the
$k$-th group. The kernel of the penalty is modified to include
them. 
\begin{equation}
\log p(\bm{\beta}_k \mid \lambda, \gamma, \{\bm{\phi}_k\}) \propto -\lambda \bar{\pi}_k^\gamma \sum_{g=1}^G \xi_{gk} \sqrt{\bm{\beta}_k^\top \bm{F}_g \bm{\beta}_k}
\end{equation}
This has no implication on the rank of the stacked $\bm{F}_g$ (and
thus the results in Appendix~\ref{sec:app_ssparse_prior}) as they are
all positive and thus only slightly modify the $E$-Step. 

We employ weights in all of our analyses to account for the fact that
different factors $j$ may have different number of levels $L_j$. We
use a generalization of the weights in \cite{bondell2009anova} to the
case of penalized \emph{differences}. Specifically, consider the
over-parameterized model in Appendix~\ref{sec:app_log} where the
penalty can be written entirely on the differences
$\bm{\delta}_{\mathrm{Main}}$, $\bm{\delta}_{\mathrm{Int}}$,
$\bm{\delta}_{\mathrm{Main-Copy}}$.  Note that each of those penalties
has a simple (group) LASSO form and thus we adopt the approach in
\cite{lim2015learning} of weighting by the Frobenius norm of the
associated columns in $\bm{T}_{\mathrm{LOG}}$, i.e. the
over-parameterized design matrix. At slight abuse of notation, define
$[\bm{T}_{\mathrm{LOG}}]_g$ as the columns of $\bm{T}_{\mathrm{LOG}}$
corresponding to the differences penalized in the (group) lasso $g$,
the weight can be expressed as follows:
$$\xi_{gk} = \frac{1}{\sqrt{N}}||~\left[\bm{T}_{\mathrm{LOG}}\right]_{g} ||_F$$
Ignoring the factor of $\sqrt{N}$, this exactly recovers the weight proposed in \cite{bondell2009anova} in the non-latent-overlapping non-interactive model of $(L_j + 1)^{-1} \sqrt{N^j_{l} + N^j_{l'}}$ where $N^j_l$, $N^j_{l'}$ are the number of observations for factor $j$ in level $l$ and level $l'$ that are being encouraged to fuse together by the penalty in group $g$.

\subsection{Latent Overlapping Groups}
\label{sec:app_log}

One feature of the above approach is that our groups are highly
overlapping. \cite{yan2017hierarchical} suggest that, in this setting,
a different formulation of the problem may result in superior
performance (see also \citealt{lim2015learning}). Existing work on the
topic has focused on group LASSO penalties (e.g. $\bm{F}_g = \bm{I}$)
and thus some modifications are needed for our purposes. To address
this, we note that we can again recast our model in an equivalent
fashion.  Instead of penalizing
$\sqrt{\bm{\beta}_k^\top\bm{F}_g \bm{\beta}_k}$, we can penalize the
vector of differences between levels as long as we also impose linear
constraints to ensure that the original model is maintained.

Consider a simple example with two factors each with two levels
$\{1,2\}$ and $\{A,B\}$.  The relevant differences are defined such
that $\delta^j_{1-2} = \beta^j_{1} - \beta^j_2$ and
$\delta^{jj'}_{(lm) - (l'm')} = \beta^j_{l,m} -
\beta^{j'}_{l',m'}$. The equivalent penalty can be imposed as follows:
\begin{equation}
\begin{split}
&\sqrt{\left(\delta^j_{1-2}\right)^2 + \left(\delta^{jj'}_{(1A) - (2A)}\right)^2 + \left(\delta^{jj'}_{(1B)-(2B)}\right)^2} = \sqrt{\bm{\delta}^\top \bm{\delta}}; \quad \bm{\delta} = \left(\begin{array}{c} \delta^j_{1-2} \\ \delta^{jj'}_{(1A)-(2A)} \\ \delta^{jj'}_{(1B)-(2B)} \end{array}\right) \\
&\mathrm{such~that}~\left[\begin{array}{l} \delta^j_{1-2} \\ \delta^{jj'}_{(1A)-(2A)} \\ \delta^{jj'}_{(1B)-(2B)} \end{array}\right] = \left[\begin{array}{lll} \beta^j_1 - \beta^j_2 \\ \beta^{jj'}_{1A} - \beta^{jj'}_{2A} \\ \beta^{jj'}_{1B} - \beta^{jj'}_{2B} \end{array}\right]
\end{split}
\end{equation} 
The latent overlapping group suggests a slight modification.  In
addition to the above penalization of the $\ell_2$ norm of the main
and interactive differences,\footnote{Note the related ``hierarchical
	group LASSO'' would add separate individual penalties for each of
	the interactions. It is easy to include that in our approach.} it
duplicates the main effect and penalizes it separately while ensuring
that all effects maintain the accounting identities between the
``latent'' groups and the overall effect. Specifically, it modifies
the above penalty to duplicate the column corresponding to
$\delta^j_{1-2}$ and adds a new parameter
$\delta^j_{(1-2)-\mathrm{Copy}}$.

\begin{equation}
\begin{split}
\sqrt{\bm{\delta}^\top\bm{\delta}} + |\delta^j_{(1-2)-\mathrm{Copy}}| \quad \mathrm{such~that}~\left[\begin{array}{l} \delta^j_{1-2} \\ \delta^{jj'}_{(1A)-(2A)} \\ \delta^{jj'}_{(1B)-(2B)} \end{array}\right] + \left[\begin{array}{l} \delta^j_{(1-2)-\mathrm{Copy}} \\ 0 \\ 0 \end{array}\right] = \left[\begin{array}{lll} \beta^j_1 - \beta^j_2 \\ \beta^{jj'}_{1A} - \beta^{jj'}_{2A} \\ \beta^{jj'}_{1B} - \beta^{jj'}_{2B} \end{array}\right]
\end{split}
\end{equation} 

Scoping out to the full problem, define $\bm{\delta}_{\mathrm{Main}}$
as the main effect differences, e.g. $\delta^j_{1-2}$, and
$\bm{\delta}_{\mathrm{Int}}$ as the interaction differences and
$\bm{D}_{\mathrm{Main}}$ as the matrix such that
$\bm{D}_{\mathrm{Main}} \bm{\beta} = \bm{\delta}_{\mathrm{Main}}$, and
$\bm{D}_{\mathrm{Int}}$ as the corresponding matrix to create the
vector of interactions. Define $\bm{\delta}_{\mathrm{Main}-g}$ as the
sub-vector of $\bm{\delta}_{\mathrm{Main}-g}$ that corresponds to the
(main) effect differences between levels $l$ and $l'$ of factor $j$
penalized by $\bm{F}_g$ in the original notation. Similarly define
$\bm{\delta}_{\mathrm{Int}-g}$ and
$\bm{\delta}_{\mathrm{Main-Copy}-g}$.

\begin{equation}
\begin{split} &p(\bm{\beta},\bm{\delta}_{\mathrm{Main}}, \bm{\delta}_{\mathrm{Int}}, \bm{\delta}_{\mathrm{Main-Copy}}) =  \sum_{g=1}^G \sqrt{\bm{\delta}_{\mathrm{Main}-g}^T \bm{\delta}_{\mathrm{Main}-g} +  \bm{\delta}_{\mathrm{Int}-g}^T \bm{\delta}_{\mathrm{Int}-g}} +\sum_{g'=1}^{G} \sqrt{\left[\bm{\delta}_{\mathrm{Main-Copy}-g}\right]^2}\\
&\mathrm{s.t.} \quad \left[\begin{array}{llll} \bm{C}^\top & \bm{0} & \bm{0} & \bm{0} \\ \bm{D}_{\mathrm{Main}} & -\bm{I} & \bm{0} & -\bm{I} \\ \bm{D}_{\mathrm{Int}} & \bm{0} & -\bm{I} & \bm{0} \end{array}\right] \left[\begin{array}{l} \bm{\beta} \\ \bm{\delta}_{\mathrm{Main}} \\ \bm{\delta}_{\mathrm{Int}} \\ \bm{\delta}_{\mathrm{Main-Copy}} \end{array}\right] = \bm{0} \end{split}
\end{equation} 

This also requires a modification of the design matrix $\tilde{\bm{T}}$ to ensure that (i) its dimensionality conforms with the expanded parameter vector and (ii) that for any choice of the expanded parameter that satisfies the constraints, the linear predictor for all observation (and thus the likelihood) is unchanged. Consider first the simple case without latent-overlapping groups. In this case, following \cite{bondell2009anova}, note that the expanded design can be expressed as $\tilde{\bm{T}}^\dagger = \bm{T}\tilde{\bm{M}}^\dagger$ where $\tilde{\bm{M}}^\top = [ \bm{I}, \bm{D}_{\mathrm{Main}}^\top, \bm{D}_{\mathrm{Int}}^\top]$ and $\tilde{\bm{M}}^\dagger$ is a left-inverse of $\tilde{\bm{M}}$. The latent-overlapping group formulation is a simple extension; we copy the columns of $\tilde{\bm{T}}^\dagger$ that correspond to $\bm{\delta}_{\mathrm{Main}}$ and append them to get $\bm{T}_{\mathrm{LOG}}$.

With this new design and parameterization in hand, we can again use the above results on projecting out the linear constraints to turn the problem into inference on an unconstrained vector $\bm{\beta}_k$ with a set of positive semi-definite constraints $\{\bm{F}_g\}_{g=1}^{2G}$ and inference proceeds identically to before.

\section{Estimators}\label{append:amce_acie_der}

Here we provide further details on the estimators.  In particular, we
discuss estimation of Average Marginal Component Effects (AMCEs) and
Average Marginal Interaction Effects (AMIEs) based on our model.  We
consider a traditional factorial design, where each unit receives one
treatment (profile), and a conjoint design in which each unit compares
two treatments (profiles).  We also discuss the impact of
randomization restrictions on estimators and implied changes in
interpretation of estimands.

\subsection{Factorial designs}\label{append:amce_fac}

\subsubsection{Without restrictions on randomization}\label{append:fac_unrest_rand}

For a unit in group $k$ we have
\begin{equation}
\Pr(Y_i = 1 \mid \bT_i, \bX_i)  = \zeta_{k}(\bT_i) 
\end{equation}
where $i=1,2,\ldots,N$ and for $k=1,2,\ldots,K$,
\begin{equation}
\zeta_{k}(\bT_i) \
= \ \frac{\exp(\psi_{k}(\bT_i))}{1+\exp(\psi_{k}(\bT_i))}.
\end{equation}
We model $\psi_{k}(\bT_i)$ as
\begin{equation}
\label{eq:define_lp}
\psi_{k}(\bT_i) \ =\ \mu + \sum_{j=1}^J \sum_{l =0}^{L_j-1} 
\mathbf{1}\{T_{ij} = l\} \beta^j_{kl} + \sum_{j=1}^{J-1} \sum_{j' >
	j} \sum_{l=0}^{L_j-1} \sum_{l' = 0}^{L_{j'}-1} \mathbf{1}\{T_{ij} = l,
T_{ij'} = l'\} \beta^{jj'}_{kll'}, 
\end{equation}
for each $k=1,2,\ldots,K$, with constraints
\begin{equation}
\bC^\top \bbeta_k \ = \ \bm{0} 
\end{equation}
where $\bbeta_k$ is a stacked column vector containing all coefficients
for group $k$.

We can rewrite this to aid in the interpretation of $\bm{\beta}_k$ as follows:
\begin{align*}
\text{logit}(\zeta_{k}(\bT_i))= \mu + \sum_{j=1}^J \sum_{l =0}^{L_j-1} 
\mathbf{1}\{T_{ij} = l\} \beta^j_{kl} + \sum_{j=1}^{J-1} \sum_{j' >
	j} \sum_{l=0}^{L_j-1} \sum_{l' = 0}^{L_{j'}-1} \mathbf{1}\{T_{ij} = l,
T_{ij'} = l'\} \beta^{jj'}_{kll'}.
\end{align*}
Thus, $\beta^j_{kl}-\beta^j_{kf}$ is the AMCE going from level $f$ to level $l$ of factor $j$ on the logit probability of $Y_i=1$ scale.

Let $\bm{t}$ be some combination of the $J$ factors, where $\bm{t}_j$ is the $j$th factor's level and $\bm{t}_{-j}$ is the levels for all factors except $j$.
This allows us to easily write, taking expectation over units in group $k$,
\begin{align*}
\E\left(Y_i\mid Z_i=k, {T}_{ij}=l, \bm{T}_{i,-j} = \bm{t}_{-j}\right)
&= \Pr\left(Y_i = 1|Z_i=k, \bm{T}_{i,j}=l, \bm{T}_{i,-j} = \bm{t}_{-j}\right)\\
&= \frac{\exp( \zeta_{k}(T_{ij} = l, \bT_{i,-j} = \bm{t}_{-j}))}{1+\exp(\zeta_{k}(T_{ij} = l, \bT_{i,-j} = \bm{t}_{-j}))},
\end{align*}
where $T_{ij} = l$ indicates for unit $i$ forcing factor $j$ to be assigned level $l$ and $\bT_{i,-j} = \bm{t}_{-j}$ indicates forcing the assignment on all factors except for $j$ to be assigned levels as in $\bm{t}_{-j}$.

The causal effects of interest (on the original $Y$ scale) are
defined as contrasts of these expectations.  Without additional
weighting (i.e., using traditional uniform weights for
marginalization), the AMCE for level $l$ vs $f$ of factor $j$ in
group $k$ is,
\begin{align*}
\delta^*_{jk}(l,f)=
&\frac{1}{M}\sum_{\bm{t_{-j}}}\E\left(Y_i \mid  Z_i=k, {T}_{ij}=l,
\bm{T}_{i,-j} = \bm{t}_{-j}\right) -
\E\left(Y_i\mid Z_i=k, {T}_{ij}=f, \bm{T}_{i,-j} =
\bm{t}_{-j}\right) \\ 
= &\frac{1}{M}\sum_{\bm{t_{-j}}} \frac{\exp( \zeta_{k}(T_{ij} = l, \bT_{i,-j} = \bm{t}_{-j}))}{1+\exp(\zeta_{k}(T_{ij} = l, \bT_{i,-j} = \bm{t}_{-j}))} -\frac{\exp( \zeta_{k}(T_{ij} = f, \bT_{i,-j} = \bm{t}_{-j}))}{1+\exp(\zeta_{k}(T_{ij} = f, \bT_{i,-j} = \bm{t}_{-j}))},
\end{align*}
where $M$ is the number of possible combinations of the other $J-1$ factors (e.g., if we had $J$ 2-level factors, $M=2^{J-1}$).
We can estimate this by plugging in the coefficients directly.
Note that, because of the nonlinear nature of the estimator, this approach is consistent (under model assumptions) but not unbiased.

Alternatively, instead of summing over all \textit{possible}
$\bm{t}_{-j}$, we can use the empirical distribution of $\bm{t}_{-j}$
in the sample.  This potentially changes the estimand.
Define estimators
\begin{align*}
\widehat{\psi}_{k}(\bm{t}) \ =\ \mu + \sum_{j=1}^J \sum_{l =0}^{L_j-1} 
\mathbf{1}\{t_{j} = l\} \widehat{\beta}^j_{kl} + \sum_{j=1}^{J-1} \sum_{j' >
	j} \sum_{l=0}^{L_j-1} \sum_{l' = 0}^{L_{j'}-1} \mathbf{1}\{t_{j} = l,
t_{j'} = l'\} \widehat{\beta}^{jj'}_{kll'}
\end{align*}
and 
\begin{align*}
\widehat{y}_k(\bm{t}) = \frac{\exp(   \widehat{\psi}_{k}(\bm{t}) )}{1+\exp(  \widehat{\psi}_{k}(\bm{t}) )}
\end{align*}

Then we can use the following overall estimator for the AMCE:
\begin{align*}
&\frac{1}{N}\sum_{b=1}^{N}\left(\widehat{Y}_k(T_{bj}=l,\bm{T}_{b,-j} ) -\widehat{Y}_k(T_{bj}=f,\bm{T}_{b,-j}) \right).
\end{align*}
This is a consistent estimator (under model assumptions) of
\begin{align*}
&\frac{1}{N}\sum_{b=1}^N\E\left(Y_i \mid Z_i=k, \bm{T}_{ij}=l,
\bm{T}_{i,-j} =  \bm{T}_{b,-j}\right) - \E\left(Y_i\mid
Z_i=k,  \bm{T}_{ij}=f, \bm{T}_{i,-j} = \bm{T}_{b,-j}\right)\\
= & \frac{1}{N}\sum_{b =1}^N\frac{\exp( \psi_k(T_{bj}=l,\bm{T}_{b,-j} ))}{1+\exp(\psi_k(T_{bj}=l,\bm{T}_{b,-j} ))} -\frac{\exp( \psi_k(T_{bj}=f,\bm{T}_{b,-j}))}{1+\exp(\psi_k(T_{bj}=f,\bm{T}_{b,-j} ))},
\end{align*}
conditioning on the treatments we actually observed. 

Now, we turn to examination of the AMIEs.  Without additional
weighting (i.e., using traditional uniform weights for
marginalization), the AMIE for level $l$ of factor $j$ and level $q$
of factor $s$ vs $f$ of factor $j$ and level $r$ of factor $s$ in
group $k$ is
\begin{align*}
\text{AMIE}^*_{jsk}(l,f, q, r) =\text{ACE}^*(l,f, q, r) -\delta^*_{jk}(l,f)-\delta^*_{sk}(q,r)
\end{align*}
where
\begin{align*}
&\text{ACE}^*(l,f, q, r)\\
= &\frac{1}{M^*}\sum_{\bm{t_{-(j,s)}}} \E\left(Y_i \mid Z_i=k,
T_{ij}=l,  T_{is}=q, \bm{T}_{i,-(j,s)} = \bm{t}_{-(j,s)}\right)
-\E\left(Y_i\mid Z_i=k, T_{ij}=f,  T_{is}=r, \bm{T}_{i,-(j,s)} =
\bm{t}_{-(j,s)}\right) \\  
=&\frac{1}{M^*}\sum_{\bm{t_{-(j,s)}}}\frac{\exp( \psi_k(T_{ij}=l,
	T_{is}=q, \bm{T}_{i,-(j,s)}=\bm{t}_{-(j,s)}
	))}{1+\exp(\psi_k(T_{ij}=l,
	T_{is}=q,\bm{T}_{i,-(j,s)}=\bm{t}_{-(j,s)} ))} -\frac{\exp(
	\psi_k(T_{ij}=f, T_{is}=r,\bm{T}_{i,-(j,s)}=\bm{t}_{-(j,s})
	))}{1+\exp(\psi_k(T_{ij}=f,
	T_{is}=r,\bm{T}_{i,-(j,s)}=\bm{t}_{-(j,s)} ))}, 
\end{align*}
where $M^*$ is the number of possible combinations of the other $J-2$ factors (e.g., if we had $J$ two-level factors, $M^* = 2^{J-2}$).

We can use the following overall estimator for the ACE:
\begin{align*}
\widehat{\text{ACE}}^*(l,f, q, r) =\frac{1}{N}&\sum_{b =1}^N \widehat{Y}_k(T_{bj}=l, T_{bs}=q, \bm{T}_{b,-(j,s)} ) -\widehat{Y}_k(T_{bj}=f, T_{bs}=r, \bm{T}_{b,-(j,s)}).
\end{align*}
This is then combined with the estimators for the AMCEs to get
\begin{align*}
\widehat{\text{AMIE}}^*_{jsk}(l,f, q, r) =\widehat{\text{ACE}}^*(l,f, q, r) -\widehat{\delta}^*_{jk}(l,f)-\widehat{\delta}^*_{sk}(q,r).
\end{align*}

\subsubsection{With restrictions on
	randomization}\label{append:fac_unrest_rand}

In this section we consider restricted randomization conditions.  Let
us assume that factor $j$ and factor $h$ are such that some levels of
$j$ are not well defined and hence excluded in combination with some
levels of factor $h$ under the randomization set up.  Let $\mathcal{S}(j, h) \subset \{1,\dots,L_j\}$
be the set of levels of factor $j$ that are not defined for some
levels of factor $h$.  Similarly, let
$\mathcal{S}(h, j) \subset \{1,\dots,L_h\}$ be the set of levels of
factor $h$ that are not defined for some levels of factor $j$.  In our
example, if $j$ is education and $h$ is profession, we have
$\mathcal{S}(j, h) = \{\text{No formal, 4th grade, 8th grade, High
	school}\}$ and
$\mathcal{S}(h, j) = \{\text{Financial analyst, Research scientist,
	Doctor, Computer programmer}\}$.

When estimating the AMCE for level $l$ vs $f$ of factor $J-1$ in
group $k$, using the model rather than the empirical distribution,
we consider,
\begin{align*}
&\frac{1}{M_{def(j,h)}}\sum_{\bm{t_{-j}}: \bm{t}_h \notin
	\mathcal{S}(h, j)}\E\left ( Y_i\mid Z_i=k, T_{ij}=l,
\bm{T}_{i,-j} = \bm{t}_{-j}\right) - \E\left(Y_i\mid Z_i=k, T_{ij}=f, \bm{T}_{i,-j} = \bm{t}_{-j}\right)\\
= & \frac{1}{M_{def(j,h)}}\sum_{\bm{t_{-j}}: \bm{t}_h \notin
	\mathcal{S}(h, j)}\frac{\exp( \psi_k(T_{ij}=l, \bm{T}_{i,-j}
	=\bm{t}_{-j} ))}{1+\exp( \psi_k(T_{ij}=l, \bm{T}_{i,-j}
	=\bm{t}_{-j} ))} -\frac{\exp(  \psi_k(T_{ij}=f, \bm{T}_{i,-j}
	=\bm{t}_{-j} ))}{1+\exp( \psi_k(T_{ij}=f, \bm{T}_{i,-j}
	=\bm{t}_{-j} ))}, 
\end{align*}
where $M_{def(j,h)}$ is the number of possible combinations of the
other factors, restricted such that $\bm{t}_h \notin \mathcal{S}(h,
j)$ (e.g., if we had $J$ 3-level factors, and some of the levels of
factor $j$ were not defined for one level of factor $h$, this would be
$2 \times 3^{J-2}$). 

To use empirical distribution, we need a way to deal with profiles
that are not well defined.  We can accomplish this by only aggregating
over those profiles that are sensible for all levels of factor $j$.
That is, we use the following  estimator,
\begin{align*}
\frac{1}{\sum_{i=1}^N\mathbb{I}\{T_{ih} \notin \mathcal{S}(h,
	j)\}}\sum_{b =1}^N \mathbb{I}\{T_{bh} \notin \mathcal{S}(h, j)\}\left(\widehat{Y}_k(T_{bj}=l,\bm{T}_{b,-j} ) -\widehat{Y}_k(T_{bj}=f,\bm{T}_{b,-j}) \right).
\end{align*}

Consider the case where we are estimating the AMCE for ``doctor'' vs ``gardener'' for profession.
Because of the randomization restriction between certain professions and level of education, we will remove any profiles that have ``4th grade'' as level of education.
Although ``gardener'' with ``4th grade'' education is allowable under the randomization, we must remove such profiles to have an ``apples-to-apples'' comparison with profession of doctor, which is not allowed to have ``4th grade'' education.
Note that we do this dropping of profiles even if we are comparing ``waiter'' vs ``gardener'' for profession, which are both allowed to have ``4th grade'' as level of education, to ensure that all AMCEs for profession comparable.

Similarly for the AMIEs, we restrict the profiles we marginalize over
to be only those that are defined for both factors in the
interactions.  Let factor $j$ be restricted by some other factor $h$
and let factor $s$ be restricted by some other factor $w$.  Then we
have the following estimator,
\begin{align*}
& \widehat{\text{ACE}}^*(l,f, q, r)\\
= &\sum_{b=1}^N\frac{\mathbb{I}\{T_{bh} \notin \mathcal{S}(h, j),
	T_{bw} \notin \mathcal{S}(w, s)\}}{\sum_{i=1}^N\mathbb{I}\{T_{ih}
	\notin \mathcal{S}(h, j), T_{iw} \notin \mathcal{S}(w,
	s)\}}\Big(\widehat{Y}_k(T_{bj}=l,T_{bs}=q, \bm{T}_{b,-(j,s)}
)-\widehat{Y}_k(T_{bj}=f,T_{bs}=r,\bm{T}_{b,-(j,s)})\Big). 
\end{align*}
The relevant AMCEs should be similarly restricted within the AMIE
estimator, with restrictions applied based on the restrictions for all levels
both factors in the interaction. 

\subsection{Conjoint designs}\label{append:amce_conjoint}

\subsubsection{Without restrictions on
	randomization}\label{append:con_unrest_rand}

Consider a conjoint experiment in which each unit $i$ only compares
two profiles.  The response $Y_i$ indicates a choice between two
profiles.  Let $\bm{T}_{i}^L$ be the levels for the left profile and
$\bm{T}_{i}^R$ be the levels for the right profile that unit $i$ sees.
Here, we modify how we model $\psi_k$ to
\begin{align*}
\begin{split}
\psi_k(\bT_i^L, \bT_i^R) \ = \ &  \mu + \sum_{j=1}^J \sum_{l \in L_j} \beta^j_{kl}\left(\mathbf{1}\left\{T^L_{ij}=l\right\}-\mathbf{1}\left\{T^R_{ij}=l\right\}\right)\\
& + \sum_{j=1}^{J-1} \sum_{j' > j} \sum_{l \in L_j} \sum_{l' \in L_{j'}} \beta^{jj'}_{kll'}\left(\mathbf{1}\left\{T^L_{ij}=l, T^L_{ij'}=l'\right\}-\mathbf{1}\left\{T^R_{ij}=l, T^R_{ij'}=l'\right\}\right).
\end{split}
\end{align*}
If we use $Y_i=1$ to indicate that unit $i$ picks the left profile, 
then we have,
\begin{align*}
\E\left(Y_i\mid Z_i=k, \bm{T}_{i}^L=\bm{t}^L, \bm{T}_{i}^R = \bm{t}^{R}\right)
&= \Pr\left(Y_i = 1\mid Z_i=k, \bm{T}_{i}^L=\bm{t}^L, \bm{T}_{i}^R = \bm{t}^{R}\right)\\
&= \frac{\exp(  \psi_k(\bT_i^L = \bm{t}^L, \bT_i^R=\bm{t}^R))}{1+\exp(\psi_k(\bT_i^L = \bm{t}^L, \bT_i^R=\bm{t}^R))}.
\end{align*}

We can use the symmetry assumption that choice order does not affect
the appeal of individual attributes.  That is, there may be some
overall preference for left or right accounted for by $\mu$, but this
preference is not affected by profile attributes.  Then, we can
define our effects, on the original $Y$ scale, as contrasts of these
expectations.  Without additional weighting, the AMCE for level $l$
vs $l^\prime$ of factor $j$ in group $k$ is,
\begin{align*}
\delta_{jk}(l, l^\prime) & \ = \ \frac{1}{2}\E \left[
\left\{\Pr\left(Y_i = 1 \mid Z_i=k, T^L_{ij}=l,
\bT^L_{i,-j}, \bT^R_{i}\right) -
\Pr\left(Y_i = 1 \mid Z_i=k, T^L_{ij}=l^\prime,
\bT^L_{i,-j}, \bT^R_{i}\right)
\right\}\right. \\
& \hspace{.75in} \left.
+\left\{\Pr\left(Y_i = 0 \mid Z_i=k, T^R_{ij}=l,
\bT^R_{i,-j}, \bT^L_{i}\right) -
\Pr\left(Y_i = 0 \mid Z_i=k, T^R_{ij}=l^\prime,
\bT^R_{i,-j}, \bT^L_{i}\right)
\right\}\right]. 
\end{align*}
To save space, the outer expectation is over the random assignment, which corresponds to the expectation over the $\tilde{M}$ possible combinations of the two profiles on the other $J-1$ factors (e.g., if we had $J$ two-level factors, this would be $4^{J-1}$).
We can again estimate this by plugging in our coefficient estimates directly.

Alternatively, instead of summing over all \textit{possible} $\bm{t}^L_{-j}$ and $\bm{t}^R_{-j}$, we can use the empirical distribution of $\bm{t}^L_{-j}$ and $\bm{t}^R_{-j}$ in the sample.
Define
\begin{align*}
\widehat{Y}_k(\bm{t}^L, \bm{t}^R)  =& \frac{\exp( \widehat{\psi}(\bm{t}^L, \bm{t}^R))}{1+\exp( \widehat{\psi}(\bm{t}^L, \bm{t}^R))}.
\end{align*}
Then we can use the estimator
\begin{align*}
\widehat{\delta}_{jk}(l, l^\prime)= 
&\frac{1}{2N}\sum_{i=1}^N\Bigg[\left\{ \widehat{Y}_k(T^L_{ij}=l, \bT^L_{i,-j}, \bT^R_{i}) - \widehat{Y}_k(T^L_{ij}=l^\prime, \bT^L_{i,-j}, \bT^R_{i}) \right\}\\
&\qquad \qquad -\left\{ \widehat{Y}_k(T^R_{ij}=l,
\bT^R_{i,-j}, \bT^L_{i}) - \widehat{Y}_k(T^R_{ij}=l^\prime,
\bT^R_{i,-j}, \bT^L_{i}) \right\}\Bigg].
\end{align*}

Now we turn to examination of the AMIEs.  Without additional weighting
(i.e., using traditional uniform weights for marginalization), the
AMIE for level $l$ of factor $j$ and level $q$ of factor $s$ vs $m$ of
factor $j$ and level $r$ of factor $s$ in group $k$ is
\begin{align*}
\text{AMIE}_{jsk}(l,f, q, r) =\text{ACE}(l,f, q, r) -\delta_{jk}(l,f)-\delta_{sk}(q,r)
\end{align*}
Here we can use the estimator
\begin{align*}
\widehat{ \text{ACE}}(l,f, q, r)=
&\frac{1}{2N}\sum_{i=1}^N\left[(\widehat{Y}_k(T^L_{ij}=l, T^L_{is}=q, \bm{T}^L_{i,-(j,s)}, \bT^R_{i} ) - \widehat{Y}_k(T^L_{ij}=f, T^L_{is}=r, \bm{T}^L_{i,-(j,s)}, \bT^R_{i} )\right)\\
&-\frac{1}{2N}\sum_{i=1}^N\left(\widehat{Y}_k(T^R_{ij}=l, T^R_{is}=q, \bm{T}^R_{i,-(j,s)}, \bT^L_{i} )  -\widehat{Y}_k(T^R_{ij}=f, T^R_{is}=r, \bm{T}^R_{i,-(j,s)}, \bT^L_{i} )\right).
\end{align*}
This gives us
\begin{align*}
\widehat{\text{AMIE}}_{jsk}(l,f, q, r) =\widehat{\text{ACE}}(l,f, q, r) -\widehat{\delta}_{jk}(l,f)-\widehat{\delta}_{sk}(q,r).
\end{align*}

\subsubsection{With restrictions on randomization}
\label{append:con_rest_rand}

Similar to Appendix~\ref{append:fac_unrest_rand}, adjustments to
estimation need to be made when we have restricted randomizations.  We
again will do this by dropping profiles that have levels of factors
not allowable for all levels of the factor(s) whose effects we are
estimating (e.g., profiles with ``4th grade'' for education when
estimating an effect for profession).  However, now we estimate the
effect for the right profile and the effect for the left profile, and
then average the two (they should be equal under symmetry).  When
estimating the effect for the right profile, therefore, we will only
drop pairings if the \textit{right} profile has a level that is not
allowed for some level of the factor we are estimating an effect of.
For example, dropping pairings where the right profile has ``4th
grade'' as level of education when estimating main effects of
profession because ``doctor'' cannot have level ``4th grade.''  Again,
this will drop more profiles than those that are not allowed under
randomization to ensure an ``apples-to-apples'' comparison across
levels of profession.

In this calculation, we use the empirical distribution for the levels
of the left profile (which represents the ``opponent'').  Thus, the
distribution of other factors for the profile we are calculating the
effect of may differ than that distribution for its opponents.
Similarly, when estimating the effect for the left profile, we only
drop pairings in which the left profile has a restricted level for
some level of the factor of interest.  Estimation for the AMIE under
randomization restrictions follows similarly.

\section{Quantification of Uncertainty}
\label{sec:app_uncertainty}

We quantify uncertainty in our parameter estimates by inverting the negative Hessian of the log-posterior at the estimates $\hat{\bm{\theta}}$, i.e. $\left[-\frac{\partial}{\partial \bm{\theta}\bm{\theta}^T} \log p(\bm{\theta} | Y_i)\right]_{\bm{\theta} = \hat{\bm{\theta}}}$ or $\mathcal{I}(\hat{\bm{\theta}})$. This can be stably and easily computed using terms from the AECM algorithm following \cite{loui:82}'s method. Specifically, consider the model from the main text augmented with $Z_i$, i.e. the group memberships. Recall that $z_{ik} = \mathbf{1}\{Z_i = k\}$ for notational simplicity.

\begin{equation}
\begin{aligned}
L^c(\bm{\theta}) = &\sum_{i=1}^N \left[\sum_{k=1}^K z_{ik}
\log(\pi_{ik}) + z_{ik} \log L(Y_i \mid \bm{\beta}_k)\right] + \\
&\sum_{k=1}^K m \log(\lambda) + m \gamma \log(\bar{\pi}_k) - \lambda
\bar{\pi}^\gamma_k \left[\sum_{g=1}^G \xi_{gk} \sqrt{\bm{\beta}^\top_k
	\bm{F}_{g} \bm{\beta}_k}\right] + \log p(\{\bm{\phi}_k\}).
\end{aligned}
\end{equation}

\cite{loui:82} notes that equation can be used to compute $\mathcal{I}_L(\hat{\bm{\theta}})$, where the subscript $L$ denotes its computation via this method.

\begin{equation}
\mathcal{I}_{L}(\hat{\bm{\theta}}) = E_{p\left(\{Z_i\}_{i=1}^N \mid \{Y_i, \bX_i, \bT_i\}_{i=1}^N,\hat{\bm{\theta}}\right)}\left[-\frac{\partial L^c(\bm{\theta})}{\partial \bm{\theta}\bm{\theta}^\top}\right] - \mathrm{Var}_{p\left(\{Z_i\}_{i=1}^N \mid \{Y_i, \bX_i, \bT_i\}_{i=1}^N, \hat{\bm{\theta}}\right)}\left[\frac{\partial L^c(\bm{\theta})}{\partial \bm{\theta}}\right]
\end{equation}

To address the issue with the non-differentiability of the penalty on $\bm{\beta}$ (and thus $L^c(\bm{\theta})$), we follow the existing research in two ways. First, for restrictions that
are sufficiently close to binding, we assume them to bind and estimate the uncertainty \emph{given} those restrictions. That is, we identify the binding restrictions such that $\sqrt{\bm{\beta}_k^\top\bm{F}_g\bm{\beta}_k}$ is sufficiently small (say $10^{-4}$) and note that if these are binding, we can use the null space projection technique to transform $\bm{\beta}_k$ such that it lies in an unconstrained space. 

To further ensure stability, we modify the penalty with a small positive constant $\epsilon \approx 10^{-4}$ to ensure that the entire objective is (twice) differentiable. For notational simplicity, we derive the results below assuming $\bm{\beta}_k$ represent the parameter vector after projecting into a space with no linear constraints. The approximated log-posterior is shown below and denoted with a tilde. We thus evaluate $\mathcal{I}_L(\hat{\bm{\theta}})$ using $\tilde{L}^c$ in place of $L^c$.

\begin{equation}
\begin{aligned}
\tilde{L}^c(\bm{\theta}) = &\sum_{i=1}^N \left[\sum_{k=1}^K z_{ik} \log(\pi_{ik}) + z_{ik} \log L(y_i | \bm{\beta}_k)\right] + \\ &\sum_{k=1}^K m \log(\lambda) + m \gamma \log(\bar{\pi}_k) - \lambda \bar{\pi}^\gamma_k \left[\sum_{g=1}^G \xi_{gk} \sqrt{\bm{\beta}^\top_k \bm{F}_{g} \bm{\beta}_k + \epsilon}\right] + \log p(\{\bm{\phi}_k\})
\end{aligned}
\end{equation}

This procedure has some pleasing properties that mirror existing results on approximate standard errors after sparse estimation; consider a simple three-level case: $\beta^j_1, \beta^j_2, \beta^j_3$. If $\beta^j_1$ and $\beta^j_2$ are fused, then their approximate point estimates and standard errors will be identical but \emph{crucially} not zero. This is because while their difference is zero and assumed to bind with no uncertainty, this does not imply that the effects, themselves, have no uncertainty: $\beta^j_1 - \beta^j_2$ will have a standard error of zero in our method. This thus mirrors the results from \cite{fan2001variable} where effects that are shrunken to zero by the LASSO are not estimated with any uncertainty. One might relax this with fully Bayesian approaches in future research.

Second, note that if all levels are fused together, i.e. $\beta^j_1 = \beta^j_2 = \beta^j_3$, then all point estimates must be zero by the ANOVA sum-to-zero constraint \emph{and} all will have an uncertainty of zero. Thus, when an entire factor is removed from the model, the approximate standard errors return a result consist with existing research.

\subsection{Derivation of Hessian}\label{append:hess}
\allowdisplaybreaks
To calculate the above terms, the score and gradient of $\tilde{L}^c$ are required. They are reported below: 
\begin{align*} 
\tilde{S}^c(\mu) &= \sum_{i=1}^N \left[\sum_{k=1}^K z_{ik} (Y_i - p_{ik})\right] \\
\tilde{S}^c(\bm{\beta}_k) &= \sum_{i=1}^N z_{ik} \cdot (Y_i - p_{ik}) \tilde{\bm{T}}_i - \lambda \bar{\pi}^\gamma_k \sum_{g=1}^G \xi_{gk} (\bm{\beta}_k^\top\bm{F}_{g}\bm{\beta}_k)^{-1/2} \cdot \bm{F}_{g} \bm{\beta}_k \\
\begin{split}\tilde{S}^c(\bm{\phi}_k) &= \sum_{i=1}^N \left[z_{ik} - \pi_{ik}\right] \bm{X}_i + \frac{\partial \log p(\{\bm{\phi}_k\})}{\partial \bm{\phi}_k} + \\ &\sum_{k'=1}^K m \gamma \frac{\partial
	\log(\bar{\pi}_{k'})}{\partial \bm{\phi}_k} - \lambda \gamma
\bar{\pi}_{k'}^{\gamma-1} \cdot \frac{\partial
	\bar{\pi}_{k'}}{\partial \bm{\phi}_k} \cdot \left[\sum_{g=1}^G
\xi_{g,k'}
\sqrt{\bm{\beta}_{k'}^\top\bm{F}_{g,k'}\bm{\beta}_{k'}}\right]\end{split} \\
H^c(\mu, \mu) & = \sum_{i=1}^N \left[-\sum_{k=1}^K z_{ik} p_{ik} (1-p_{ik})\right]\\
H^c(\mu, \bm{\beta}_k) & = -\left[\sum_{i=1}^N z_{ik} p_{ik}
(1-p_{ik}) \tilde{\bm{T}}_i\right] \\
H^c(\bm{\beta}_k, \bm{\beta}_k) &=-\left[\sum_{i=1}^N z_{ik} \cdot
p_{ik} (1-p_{ik})
\tilde{\bm{T}}_i\tilde{\bm{T}}_i^\top\right] -
\lambda \bar{\pi}^\gamma_k
\sum_{g=1}^G \xi_{gk} \bm{D}_{gk}  
\end{align*}
where
$\left[\bm{D}_{gk}\right]_{a,b}
=-\left(\bm{\beta}_k^\top\bm{F}_{g}\bm{\beta}_k\right)^{-3/2}
\bm{\beta}_k^\top\left[\bm{F}_{g}\right]_a
\bm{\beta}_k^\top\left[\bm{F}_{g}\right]_b +
\left(\bm{\beta}_k^\top\bm{F}_{g}\bm{\beta}_k\right)^{-1/2}
\left[\bm{F}_{g}\right]_{a,b}$. 
\begin{align*}
H^c(\left[\bm{\beta}_k\right]_i, \bm{\phi}_\ell) & = -\lambda \gamma \bar{\pi}^{\gamma-1}_k  \left[\sum_{g=1}^G \xi_{gk} (\bm{\beta}_k^\top\bm{F}_{g}\bm{\beta}_k)^{-1/2} \cdot  \bm{\beta}_k^\top \left[\bm{F}_{g}\right]_i \right] \frac{\partial \bar{\pi}_k}{\partial \bm{\phi}_\ell}  \\
H^c(\bm{\phi}_k, \bm{\phi}_\ell) & =
\sum_{i=1}^N -\left[\left(I[k=\ell] -
\pi_{ik}\right)\pi_{i\ell}\right]\bm{X}_i\bm{X}_i^\top +
\frac{\partial^2 \log p(\{\bm{\phi}_k\})}{\partial
	\bm{\phi}_k\bm{\phi}_\ell^\top} + \sum_{k'=1}^K m \gamma
\frac{\partial \log(\bar{\pi}_{k'})}{\partial
	\bm{\phi}_k\bm{\phi}_\ell^\top} + \\
\sum_{k'=1}^K -\lambda\gamma&\left[\sum_{g=1}^G \xi_{g,k'} \sqrt{\bm{\beta}_{k'}^\top\bm{F}_{g,k'}\bm{\beta}_{k'}}\right]\left[ I(\gamma \notin \{0,1\}) \cdot (\gamma-1) \bar{\pi}_{k'}^{\gamma-2} \cdot \textcolor{black}{\left[\frac{\partial \bar{\pi}_{k'}}{\partial \bm{\phi}_k}\right]}\left[\frac{\partial \bar{\pi}_{k'}}{\partial \bm{\phi}_\ell}\right]^\top  + \bar{\pi}_{k'}^{\gamma-1} \frac{\partial \bar{\pi}_{k'}}{\partial \bm{\phi}_k\bm{\phi}_\ell^\top} \right]
\end{align*}
The above results use the following intermediate derivations:
\begin{align*}
\frac{\partial \bar{\pi}_{k'}}{\partial \bm{\phi}_k} &= \frac{1}{N} \sum_{i=1}^N \pi_{i,k'}\left[I(k=k') - \pi_{ik}\right] \bm{X}_i \\
\frac{\partial \bar{\pi}_{k'}}{\partial \bm{\phi}_k\bm{\phi}_\ell^\top} &=\frac{1}{N}\sum_{i=1}^N\left[\pi_{i,k'}\left(I(k'=\ell) - \pi_{i\ell}\right)\left(I(k=k') - \pi_{ik}\right) -\pi_{i,k'}\pi_{ik}\left(I(k=\ell)- \pi_{i\ell}\right)\right] \bm{X}_i\bm{X}_i^\top
\\
\frac{\partial \log(\bar{\pi}_{k'})}{\partial \bm{\phi}_k} &= \frac{1}{\bar{\pi}_{k'}} \cdot \frac{\partial \bar{\pi}_{k'}}{\partial \bm{\phi}_k} \\
\frac{\partial \log(\bar{\pi}_{k'})}{\partial \bm{\phi}_k\bm{\phi}_\ell^\top} &= -\frac{1}{\bar{\pi}_{k'}^{2}} \left[\frac{\partial \bar{\pi}_{k'}}{\partial \bm{\phi}_k}\right]\left[\frac{\partial \bar{\pi}_{k'}}{\partial \bm{\phi}_\ell}\right]^\top + \frac{1}{\bar{\pi}_{k'}} \cdot \frac{\partial \bar{\pi}_{k'}}{\partial \bm{\phi}_k\bm{\phi}_\ell^\top}
\end{align*}
Second, the variance of $\tilde{S}^c(\bm{\theta})$ over $p(\{z_{ik}\} \mid \bm{\theta})$. This is derived blockwise below.
\begin{align*}
\mathrm{Cov}\left[\tilde{S}^c(\bm{\beta}_k), \tilde{S}^c(\bm{\beta}_{\ell})\right] &= \sum_{i=1}^N (Y_i - p_{ik}) \cdot (Y_i - p_{i\ell}) \cdot \E(z_{ik}) \left(I(k =\ell) - \E(z_{i\ell})\right) \tilde{\bm{T}}_i\tilde{\bm{T}}_i^\top \\
\mathrm{Cov}\left[\tilde{S}^c(\bm{\beta}_k), \tilde{S}^c(\bm{\phi}_{\ell})\right] &= \sum_{i=1}^N (Y_i - p_{ik}) \cdot \E(z_{ik}) \left(I(k =\ell) - \E(z_{i\ell})\right) \tilde{\bm{T}}_i \bm{X}_i^\top \\
\mathrm{Cov}\left[\tilde{S}^c(\bm{\phi}_k), \tilde{S}^c(\bm{\phi}_{\ell})\right] &= \sum_{i=1}^N \E(z_{ik}) \left(I(k =\ell) - \E(z_{i\ell})\right) \bm{X}_i\bm{X}_i^\top \\
\mathrm{Cov}\left[\tilde{S}^c(\mu), \tilde{S}^c(\mu)\right] &= \sum_{i=1}^N \left[\sum_{k=1}^K \sum_{k'=1}^K \E(z_{ik}) \left(I(k=k') - \E(z_{ik'})\right)(Y_{i} - p_{ik}) (Y_i - p_{ik'})\right] \\
\mathrm{Cov}\left[\tilde{S}^c(\bm{\phi}_k), \tilde{S}^c(\mu)\right] &= \sum_{i=1}^N \left[\sum_{k'=1}^K \E(z_{ik}) \left(I(k=k') - \E(z_{ik'})\right) (Y_i - p_{ik'}) \bm{X}_i\right] \\
\mathrm{Cov}\left[\tilde{S}^c(\bm{\beta}_k), \tilde{S}^c(\mu)\right] &= \sum_{i=1}^N \left[\sum_{k'=1}^K \E(z_{ik}) \left(I(k=k') - \E(z_{ik'})\right)(Y_{i} - p_{ik}) (Y_i - p_{ik'}) \tilde{\bm{T}}_i\right] 
\end{align*}

This provides all terms needed to compute $\mathcal{I}_L(\hat{\bm{\theta}})$.

\subsection{Repeated Observations}\label{append:hes_rep_obs}

Now consider the case of repeated observations per individual $i$. In
this scenario, each individual $i$ performs $N_i$ tasks. Note, after augmentation, the score has exactly the same form and thus the
complete Score $\tilde{S}^c$ and Hessian $\tilde{H}^c$ are identical
where the sum merely now runs over $\sum_{i=1}^N
\sum_{m=1}^{N_i}$. The average for $\bar{\pi}_k$ is similarly a
weighted average by $N_i$, although note that often each respondent
answers an identical number of tasks so it is, effectively, the same
as before. The covariance of $\tilde{S}^c$ is adjusted as shown
below.

\begin{align*}
\mathrm{Cov}\left[\tilde{S}^c(\bm{\beta}_k), \tilde{S}^c(\bm{\beta}_\ell)\right] &= \sum_{i=1}^N \E(z_{ik})\left(I[k=\ell] - \E(z_{i\ell})\right) \left[\sum_{m=1}^{N_i} (Y_{im} - p_{imk})\tilde{\bm{T}}_{im}\right]\left[\sum_{m'=1}^{N_i} (Y_{im} - p_{im\ell})\tilde{\bm{T}}_{im'}^\top\right] \\
\mathrm{Cov}\left[\tilde{S}^c(\bm{\beta}_k), \tilde{S}^c(\bm{\phi}_\ell)\right] &= \sum_{i=1}^N \E(z_{ik})\left(I[k=\ell] - \E(z_{i\ell})\right) \left[\sum_{m=1}^{N_i} (Y_{im} - p_{imk}) \tilde{\bm{T}}_{im}\right]\bm{X}_i^\top \\
\mathrm{Cov}\left[\tilde{S}^c(\mu), \tilde{S}^c(\mu)\right] &= \\ \sum_{i=1}^N \Bigg[ \sum_{k=1}^K \sum_{k'=1}^K \E(z_{ik})&\left(I[k=k'] - \E(z_{ik'})\right) \left[\sum_{m=1}^{N_i} (Y_{im} - p_{imk})\tilde{\bm{T}}_{im}\right]\left[\sum_{m=1}^{N_i} (Y_{im} - p_{imk})\tilde{\bm{T}}_{im}^\top\right]\Bigg] \\
\mathrm{Cov}\left[\tilde{S}^c(\bm{\beta}_k), \tilde{S}^c(\mu)\right] &= \sum_{i=1}^N \left[\sum_{k'=1}^K \E(z_{ik})\left(I[k=k'] - \E(z_{ik'})\right) \left[\sum_{m=1}^{N_i} (Y_{im} - p_{imk})\tilde{\bm{T}}_{im}\right]\left[\sum_{m=1}^{N_i} (Y_{im} - p_{imk'})\right]\right]
\end{align*}

\subsection{Standard Errors on Other Quantities of Interest}\label{append:uncert_delta}

Given the above results, we derive an approximate covariance matrix on $\hat{\bm{\theta}}$. We calculate uncertainty on other quantities of interest, e.g. AMCE and marginal effects, using the multivariate delta method. As almost all of our quantities of interest can be expressed as (weighted) sums or averages over individuals $i \in \{1, \cdots, N\}$, calculating the requisite gradient for the multivariate delta method simply requires calculating the relevant derivative for each observation. For example, all derivatives needed in the AMCE are of the following form; see Appendix~\ref{append:amce_acie_der} for more details.
\begin{equation*}
\frac{\partial}{\partial \bm{\theta}}\left[ \frac{\exp(\psi_{ik})}{1+\exp(\psi_{ik})}\right]
\end{equation*}

\section{Simulations}
\label{sec:app_simulations}

We detail our simulations and provide additional results in this
section.

\subsection{Setup}

We generate the $\bm{\beta}_k$ used in our simulations following Equation~\ref{eq:basic_models} and calibrating their implied AMCEs to be roughly comparable to the magnitude found in our empirical example, i.e. ranging between around $-0.30$ and $0.30$. The $\bm{\beta}_k$ and $\{\bm{\phi}_k\}_{k=2}^3$ used in all simulations are determined using one draw from the following procedure:

\begin{itemize}
	\item[] Simulating $\bm{\beta}_k$:
	\begin{enumerate}
		\item For each factor $j$ and group $k$, draw the number of unique levels $u$ with equal probability from $\{1, 2, 3\}$. 
		\item Draw $u$ normal random variables independently from $N(0, 1/3)$; call these $b^j_{ku}$.
		\item For $u = 1$, set $\beta^j_{kl} = 0$
		\item For $u = 3$, de-mean $\{b^j_{ku}\}_{u=1}^3$ drawn in (2) and set all $\beta^j_{kl}$ equal to the corresponding value.
		\item For $u = 2$, assign $b^j_{k3}$ equal to one of the two $b^j_{ku}$ with equal probability. De-mean the $\{b^j_{ku}\}_{u=1}^3$ and set $\beta^j_{kl}$ equal to the corresponding values.
	\end{enumerate}
	\item[] Simulating $\bm{\phi}_k$: $\{\bm{\phi}_k\}_{k=1}^K \sim N(\bm{0}, 2 \cdot \bm{I})$
\end{itemize}

To evaluate our method, we calculate the AMCEs in each group simulations using Monte Carlo simulation where we sample 1,000,000 pairs of treatment profiles for the other attributes to marginalize over the other factors. The distribution of the
$\bm{\beta}_k$ and average marginal component effects (with a baseline level of `1') used in the simulations are shown below:
\begin{figure}[!h]
	\includegraphics[width=\textwidth]{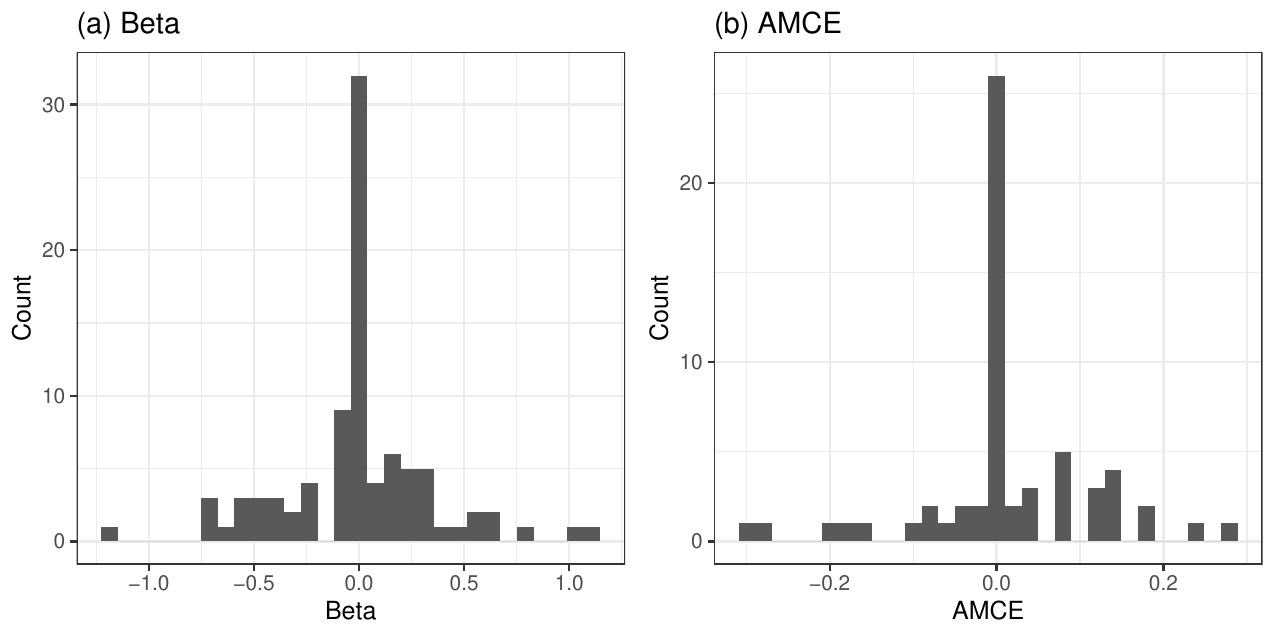}
	\caption{The distribution of parameters and AMCEs used in the simulation.}
\end{figure}

For each simulation, we draw $N$ individuals who rate $T$ profiles
where $(N, T) \in \{(1000,5), (2000, 10)\}$. For each individual $i$,
we draw its moderators $\bm{x}_{i}$ from a correlated multivariate
normal where $\bm{x}_{i} \sim N(\bm{0}_5, \bm{\Sigma})$ with
$\bm{\Sigma}_{ij} = 0.25^{|i-j|}$ for $i,j \in \{1, \cdots, 5\}$. The
distribution of group assignment probabilities $\pi_{ik}$ is shown
below from one million Monte Carlo simulation draws of
$[1, \bm{x}_{i}^\top]$.

\begin{figure}[!htbp]
	\caption{Group Membership Probabilities}
	\includegraphics[width=\textwidth]{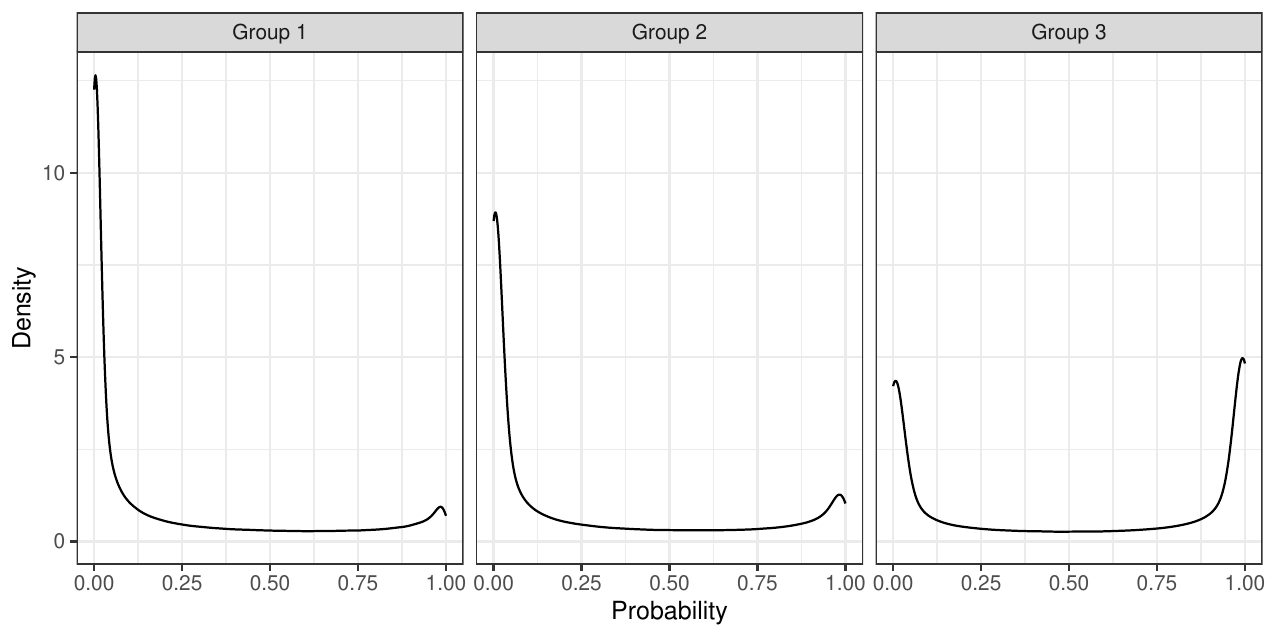}
\end{figure}

We see that the members are well-separated; the groups are somewhat
unbalanced, i.e. $\bar{\bm{\pi}} = [0.217, 0.261, 0.522]$. If we
consider the maximum probability for each person $i$, i.e.
$\pi^*_i = \max_{k \in \{1, 2, 3\}} \pi_{ik}$, this distribution has a
median of 0.93, a 25th percentile of 0.75 and a 75th percentile of
0.99.

In terms of simulating the treatment profiles and outcome, for each
individual $i$, we draw a group membership $Z_i$ using
$\bm{\pi}_{i}$ generating using $\bX_i$, $\bm{\phi}$ and Equation~\ref{eq:basic_models}. For each task $t$, we then randomly draw a pair of
treatments and then, given $Z_i$, draw the outcome $Y_i$ given their observed treatments using the model in the main text.

After estimating our model with $K = 3$, we resolve the problem of
label switching by permuting our estimate group labels to minimize
the absolute error between the estimated posterior membership
probabilities $\{E[z_{ik} | \bm{\theta}]\}_{k=1}^K$ and $\bm{z}_i$
(the one-hot assignment of group membership).

\subsection{Additional Results}
\label{sec:app_simulations_coverage}

We provide additional simulation results to complement those presented
in the main text. Figure~\ref{fig:app_simulation_beta} presents the
results for the simulations in the main text when considering the
$\bm{\beta}_k$ (instead of the AMCE). It shows a similar pattern of
some bias even at the larger sample size.

\begin{figure}[!ht]
	\centering \spacingset{1}
	\includegraphics[width=\textwidth]{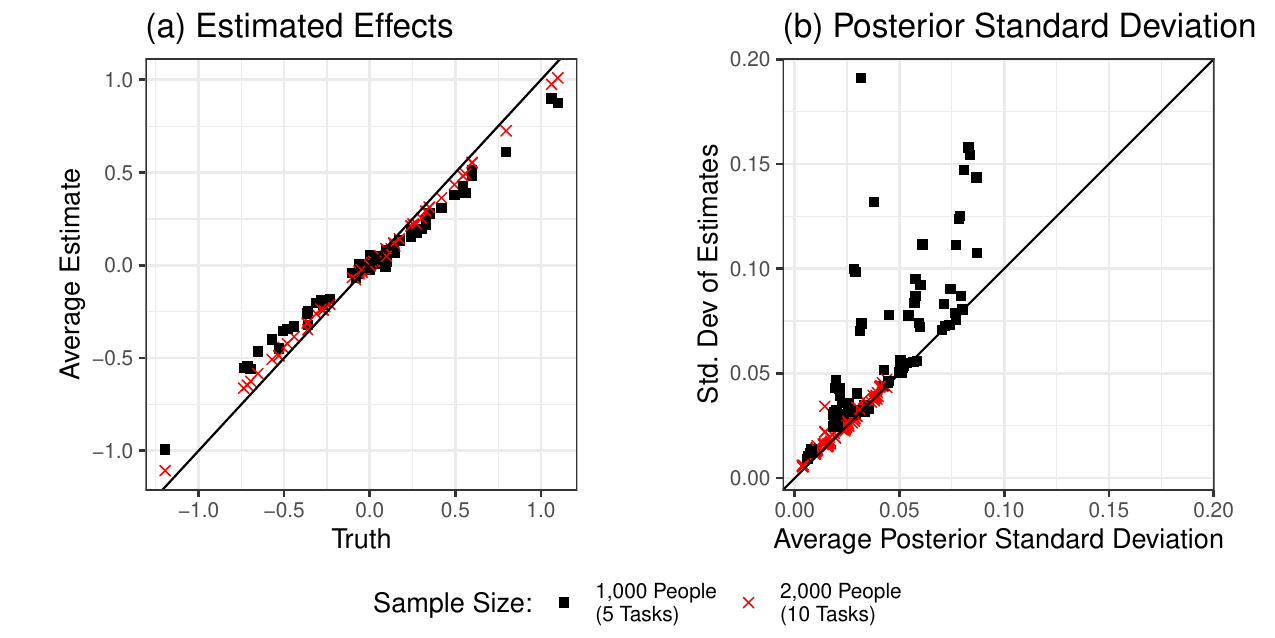}
	\caption{The empirical performance of the proposed estimator on
		simulated data. The black squares indicate the effects estimated in each group
		with the smaller sample size (1,000 people completing 5 tasks);
		the red crosses indicate effects estimated with the larger sample
		size (2,000 people completing 10 tasks).}
	\label{fig:app_simulation_beta}
\end{figure}

To address this issue, we consider an alternative procedure based on
sample splitting. We fit the model using half of the data (selected at
random) and then refit the model. To refit the model, we hold fixed
the sparsity pattern estimated in the original estimation hold (i.e.,
which levels are fused together) using a tolerance of $10^{-3}$. We
also fix the estimated moderator relationship, i.e. $\pi_{k}(\bX_i)$,
and only estimate the treatment effect coefficients after
fusion. Algorithm~\ref{alg:refit} states the procedure. To calculate
the average marginal effects, as noted in
Appendix~\ref{append:amce_acie_der}, we use the empirical distribution
of treatments to marginalize over other factors. In this split
version, we also use the distribution from the full dataset.

\begin{algorithm}[!ht]\spacingset{1.25}
	\caption{Refitting Procedure}
	\label{alg:refit}
	\begin{algorithmic}
		\State{1. Randomly split the observations $i \in \{1, \cdots, N\}$ into two groups indexed by $\mathcal{I}_1$ and $\mathcal{I}_2$} 
		\State{2. Using the data $i \in \mathcal{I}_1$, estimate the parameters of the model using Algorithm~\ref{alg:main} in the main text. Define the resulting parameters from this as $\tilde{\bm{\theta}}$: $\{\tilde{\bm{\beta}}_k\}_{k=1}^K$, $\{\tilde{\bm{\phi}}_k\}_{k=2}^K,~\tilde{\mu}$}
		\State{3. Fuse levels $l$ and $l'$ of factor $j$ for group $k$ where the following condition holds for tolerance $\epsilon$
			
			$$\max \left\{\left|\tilde{\beta}^j_{kl} - \tilde{\beta}^j_{kl'}\right|\right\} \bigcup \left\{\bigcup_{j' \neq j} \bigcup_{m=0}^{L_{j'}-1} \left| \tilde{\beta}^{jj'}_{klm} - \tilde{\beta}^{jj'}_{kl'm}\right|\right\} \leq \epsilon$$
			
			For each combination where this is satisfied, construct matrices $\bm{R}_k$ that contain the required equality constraints, i.e. where $\bm{R}_k^T\tilde{\bm{\beta}}_k$ ensures that $\tilde{\beta}^j_{kl} = \tilde{\beta}^j_{kl'} = 0$ and/or $\tilde{\beta}^{jj'}_{klm} - \tilde{\beta}^{jj'}_{kl'm} = 0$. 
			
			Define $\tilde{\pi}_k(\bX_i)$ as follows:
			
			$$\tilde{\pi}_k(\bX_i) = \frac{\exp(\bX_i^\top \tilde{\bphi}_k)}{\sum_{k'=1}^K
				\exp(\bX_i^\top \tilde{\bphi}_{k'})}$$	
		}
		\State{4. Using the other half of the data $i \in \mathcal{I}_2$, estimate the refit parameters for the treatment effects, where $\bC$ contains the original sum-to-zero constraints discussed in the main text.
			\begin{equation*}
			\{\hat{\bm{\beta}}^{\mathrm{refit}}_k\}_{k=1}^K, \hat{\mu}^{\mathrm{refit}} = \argmax_{\{\bm{\beta}_k\}_{k=1}^K, ~\mu}~\sum_{i \in \mathcal{I}_2} \log \left(\sum_{k=1}^K \tilde{\pi}_k(\bm{X}_i) \zeta_k(\bm{T}_i)^{Y_i} \{1 - \zeta_k(\bT_i)\}^{1-Y_i}\right) \quad \mathrm{s.t.} \quad \bC^T \bm{\beta}_k = \bm{0},~\bm{R}_k^T\bm{\beta}_k = \bm{0}
			\end{equation*}
		}
	\end{algorithmic}
\end{algorithm}

Figure~\ref{fig:app_dist_simulations} compares the estimators from the
split sample and full data (``Full Sample'', i.e. the methods shown in
the main text) approaches. It shows the distribution of the root
mean-squared error (RMSE), bias, and coverage across the estimated
AMCE and coefficients. We split the results by whether the true
underlying effect is zero to compare differences across those
cases. We also consider one even larger sample size (4,000 respondents
with 10 tasks) to examine a scenario where the split sample method has
the same amount of data as the full sample method for the second step
in the estimation process.

\begin{figure}
	
	\begin{subfigure}{\textwidth}	
		\caption{Results for AMCE}
		\includegraphics[width=\textwidth]{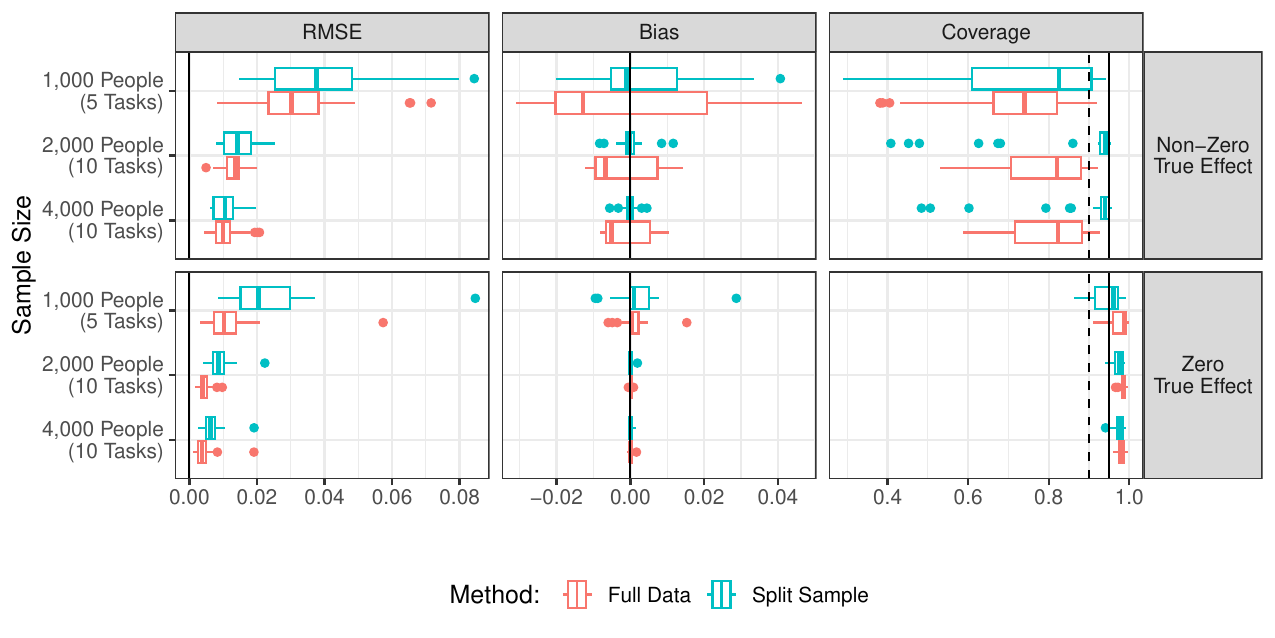}
	\end{subfigure}
	
	\begin{subfigure}{\textwidth}	
		\caption{Results for $\bm{\beta}_k$}
		\includegraphics[width=\textwidth]{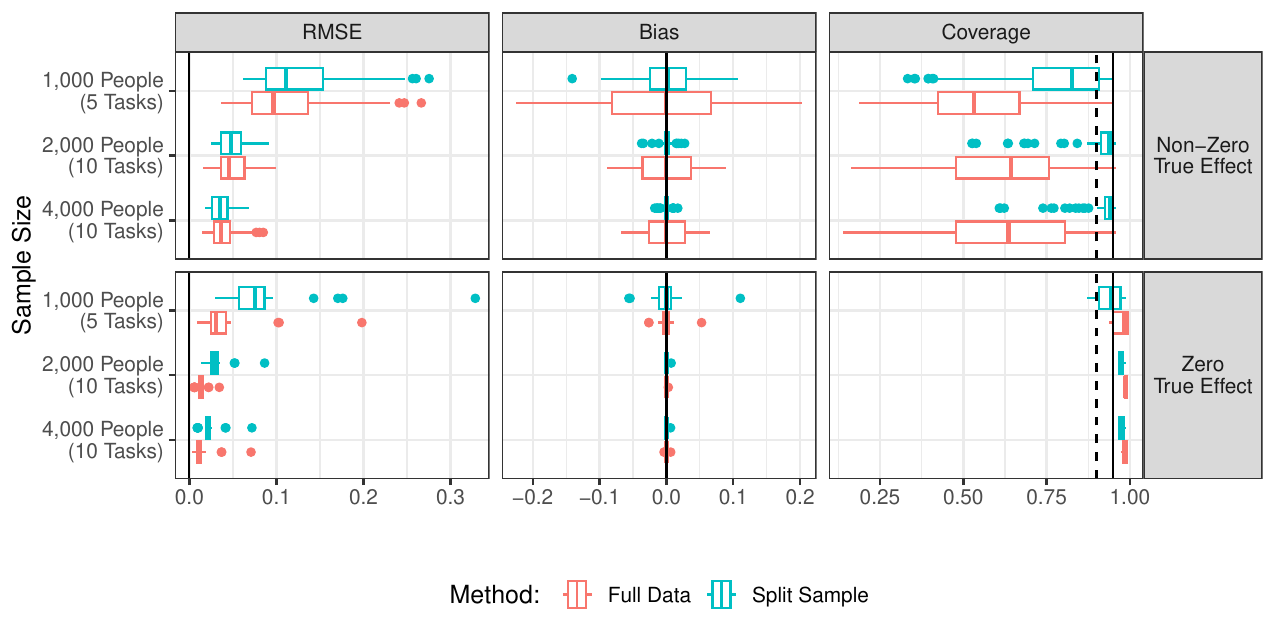}
	\end{subfigure}
	
	\caption{The distribution of performance for each estimator across sample sizes. The top figure shows results for the AMCE; the lower figure shows results for the coefficients $\bm{\beta}_k$. Inside each figure, results are split by whether the true effect is zero (``Zero True Effect'') or not (``Non-Zero True Effect''). The boxplot shows the distribution across all effects for each group. For the plots on RMSE and bias, the solid vertical line indicates zero. For coverage, the solid line indicates 95\% coverage and the dashed line indicates 90\%.}\label{fig:app_dist_simulations}
\end{figure}

The figure corroborates the initial results.  Specifically, the full
data method has non-trivial bias that decreases slowly even at the
largest sample sizes. By contrast, the bias is small in the split
sample method. As the panel on coverage shows, this results in
considerably better coverage---especially for quantities with a
non-zero true effect. At the two larger sample sizes, the median
frequentist coverage of the split sample method is close to the
nominal 95\%, with a few outliers that have low coverage. In terms of
RMSE, the methods perform similarly.

\subsection{Robustness to Misspecification}\label{sec:missp}

As noted in the main text, our methodology is not predicated on the
assumption that the true data generating process is a mixture model.
Rather, fitting a mixture model or a mixture of experts model is
equivalent to finding maximally heterogeneous groups. Nevertheless, we consider a simulation setting in which the true data generating process is a mixture model.  Under this assumption, we explore how the
specification of different parts of the model (e.g., $K$ and the
choice of moderators) affects performance.  Specifically, we explore different choices of $K$ and misspecification of the moderator model $\pi_k(\bX_i)$ from the ones used to generate the data.

\subsubsection{Data-Driven Choice of $K$}\label{sec:app_sim_choose_k}

First, as noted in the main text, a common approach to choosing $K$ can be information criterion. We use the BIC to calibrate our choice of $\lambda$, i.e. pick the $\lambda$ that minimizes the BIC. In our simulations, we compare the BIC across $K \in \{1,2,3,4\}$ to see which it would suggest choosing. Table~\ref{tab:empk_BIC} reports the probability of each $K$ being chosen across 1,000 simulations. It shows that, even for the smallest data size, the BIC correctly identifies $K=3$. The probability of correct selection rises as the sample size grows. However, as we note in the main text, this simulation example has relatively well separated clusters, and correctly specified likelihoods, and thus the information criterion approach is expected to perform well.

\begin{table}[!htbp]
	\begin{centering}
		\begin{tabular}{lcccc}
			\hline\hline
			Sample Size & $K=1$ & $K=2$ & $K=3$ & $K=4$ \\
			\hline
			1,000 People
(5 Tasks) & 0 & 0.01 & 0.941 & 0.049 \\
2,000 People
(10 Tasks) & 0 & 0.00 & 0.999 & 0.001 \\
4,000 People
(10 Tasks) & 0 & 0.00 & 0.994 & 0.006 \\ \hline\hline

		\end{tabular}
		\caption{Probability of $K$ being chosen using smallest BIC}\label{tab:empk_BIC}
	\end{centering}
\end{table}

Other criterion based on cross-validation---e.g., splitting the sample and taking the model with the highest out-of-sample predictive likelihood or lowest RMSE---also show a high probability of choosing $K=3$ (84\% for the smallest sample size and 97-98\% for the larger sample sizes).

\subsubsection{Effect of Choice of $K$ on Estimates}\label{sec:app_sim_vary_k}

We first consider how different choices of $K$ impact our
results in the simulation study. To do this, we focus on the CAMCE
discussed in the main text (Section~\ref{subsec:cjbart}) as this
quantity is comparable across models with different $K$. For each
individual $i$, we calculate our estimate of CAMCE using their
moderators $\bm{X}_i$ and compare this against the true value, which
can be calculated by plugging in the true values of $\pi_k(\bX_i)$ and
$\delta_{jk}(l,l')$ into Equation~\eqref{eq:camce}. We run models with
$K \in \{2,3,4\}$ with both split-sample and full data methods
discussed above.

Figure~\ref{fig:binscatter_K} shows a binned scatterplot of the true
CAMCEs against the estimated CAMCEs for each individual $i$, i.e., for
all true CAMCE in a bin, what is the average estimated CAMCE? As
above, it shows that for the correct choice of $K=3$, the estimates
track the truth well. Interestingly, $K=4$ also shows good performance
but $K=2$ shows some weaker performance, especially for certain ranges
of the true CAMCE.

\begin{figure}[t!]
	\includegraphics[width=\textwidth]{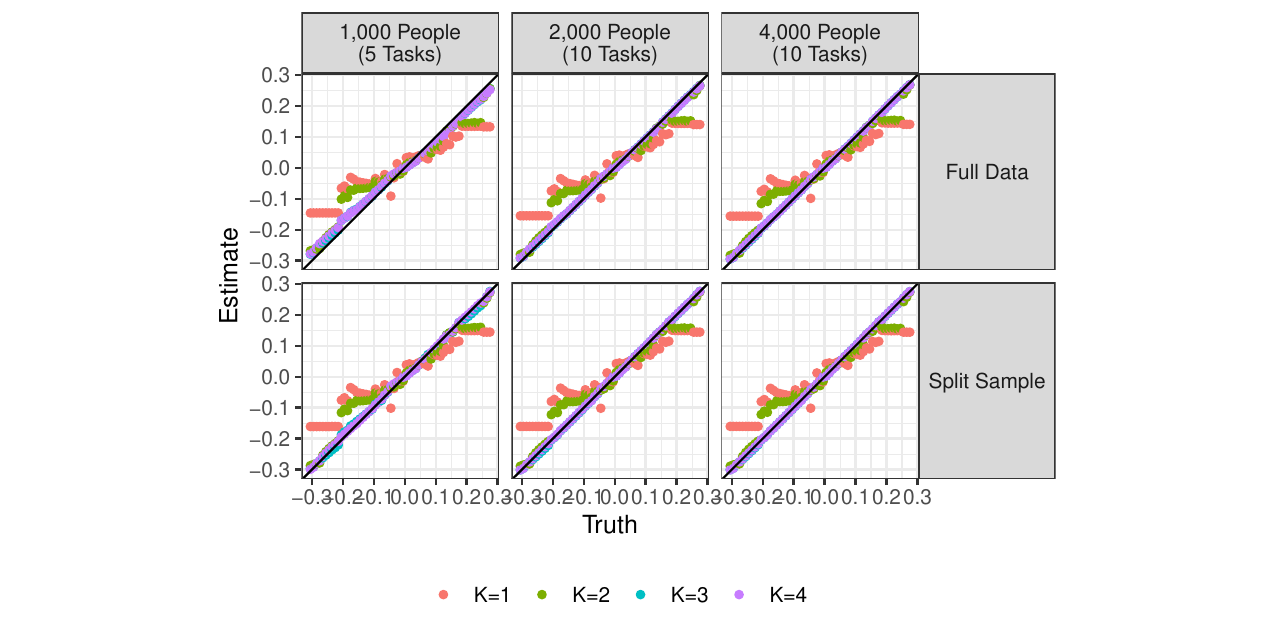}
	\caption{The binned scatterplot of the true CAMCEs versus the estimated CAMCEs. Results are shown for different sample sizes and estimation method (e.g., full data versus split sample). The color of the dot indicates the number of groups $K$.}\label{fig:binscatter_K}
\end{figure}

We also compute the marginalized error (i.e., the error in the
estimated CAMCE vs the true CAMCE, averaged across all people and
CAMCEs estimated in a simulation) and RMSE of the estimated
CAMCEs. Figure~\ref{fig:wrong_Ksims} plots the distribution of RMSE
and marginalized error across the 1000 simulations. Consistent with
our earlier results, the figure shows that the full sample method for
all choice of $K$ has some non-vanishing bias while the split-sample
method exhibits a considerably smaller error. Further, while the
estimated error looks similar for $K \in \{2, 3, 4\}$, the correct
choice ($K=3$) has lower RMSE than either $K=2$ or $K=4$.  The results
for $K=4$ are comparable to those for $K=3$, but the case of $K=2$
sees a considerably worse performance.

\begin{figure}[!htbp]
	\includegraphics[width=\textwidth]{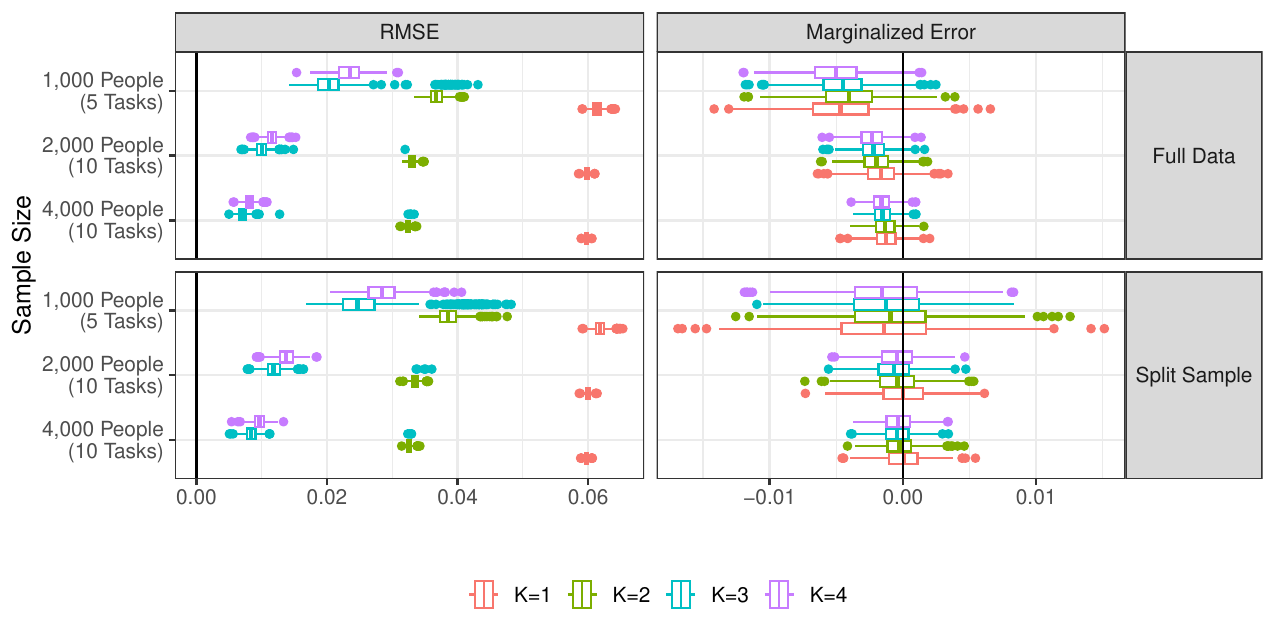}
	\caption{The distribution of performance across simulations. The top panel shows the performance in terms of RMSE and marginalized error, across all individuals and CAMCEs, for the model fit on the entire dataset. The bottom panel shows the results for a method estimated using the split sample method. The color of the boxplot indicates the number of groups $K$.}\label{fig:wrong_Ksims} 
\end{figure}

Next, we consider how different choices of $K$ affect the ability to
recover the average marginal effect.  To do this, we average the CAMCE
across all individuals used to fit the model and compare that AMCE in
the population. Figure~\ref{fig:app_recons_AME} plots the bias of the
estimated AMCE by aggregating the individual-level effects; it is
largely unaffected by the choice of $K$, corroborating
Figure~\ref{fig:wrong_Ksims}. As expected, there is regularization
bias for the full data method that using the split sample approach
eliminates.

\begin{figure}[!htbp]
	\includegraphics[width=\textwidth]{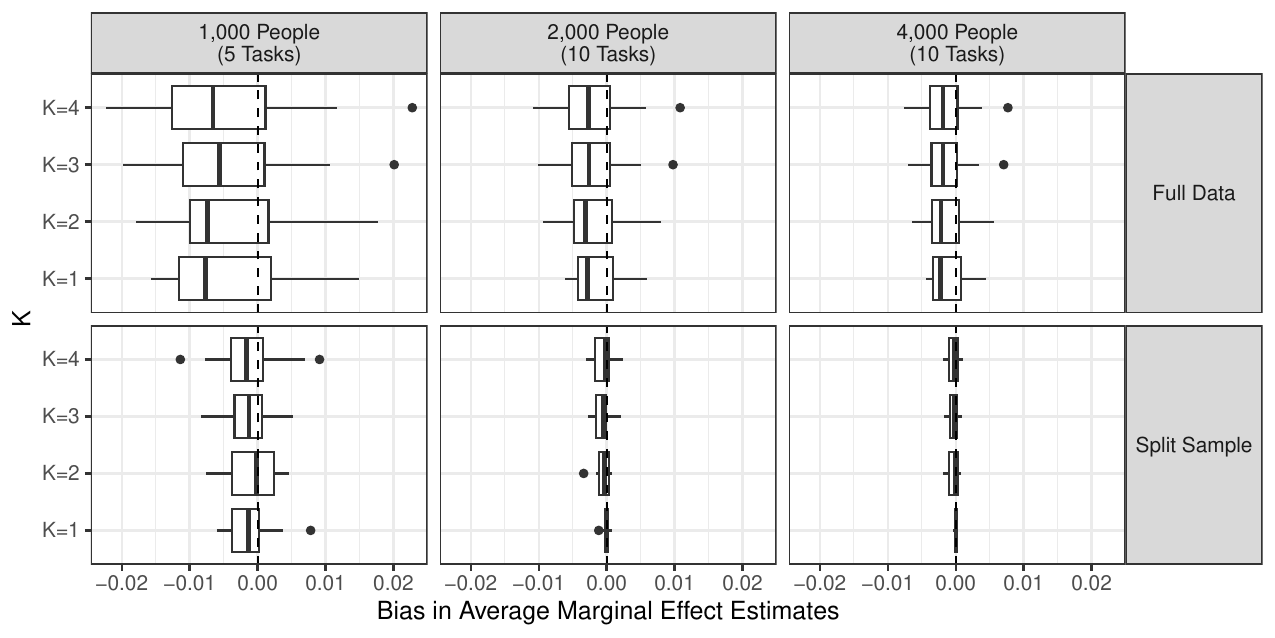}
	\caption{The distribution of bias in AMCEs by averaging CAMCEs by different $K$}\label{fig:app_recons_AME}
\end{figure}

As a final illustration on the choice of $K$, we also examine how much
variability in the \emph{true} CAMCE is explained by the estimated
groups, inspired by how one might assess the quality of clustering in
$k$-means. We compute this as follows: For each observation $i$,
obtain its estimated group membership probabilities
$\hat{\pi}_k(\bX_i)$ for $k \in \{1, \cdots, K\}$. Using its true
CAMCE, i.e. $\mathrm{CAMCE}^*_{j}(l, l'; \bX_i)$, compute the total
variability in CAMCE across the $N$ units and the between-group
variability using $\hat{\pi}_k$ as group weights. Formally, we compute
$B_{K}$ and the total variability $T$.
\begin{align*}
B_{K} &= \sum_{k=1}^{K} \sum_{j=1}^J \sum_{l'_j=1}^{L_j - 1} N_{k} \left[\overline{\mathrm{CAMCE}}^*_{k,j}(l_j, l'_j) - \overline{\mathrm{CAMCE}}^*_{j}(l_j, l'_j)\right]^2; \quad N_k = \sum_{i=1}^N \hat{\pi}_k(\bX_i); \\
T  &= \sum_{j=1}^J \sum_{l'_j=1}^{L_j - 1} \sum_{i=1}^N \left[\mathrm{CAMCE}^*_{j}(l_j, l'_j; \bX_i) - \overline{\mathrm{CAMCE}}^*_{j}(l_j, l'_j)\right]^2 \\
\overline{\mathrm{CAMCE}}^*_{k,j} &= \frac{1}{N_k}\sum_{i=1}^N \hat{\pi}_k(\bX_i) \cdot \mathrm{CAMCE}^*_{j}(l_j, l_j'; \bX_i); \quad \overline{\mathrm{CAMCE}}^*_{j} = \frac{1}{N}\sum_{i=1}^N \mathrm{CAMCE}^*_j(l_j, l_j'; \bX_i)
\end{align*}

Figure~\ref{fig:between_tot_K} reports the ratio of the between-group
variability over the total variability across the 1,000 simulations
for $K \in \{2, 3, 4\}$. With $K=2$, we already able to explain around 50\%
of the variability in the data. As expected, $K=2$ shows considerably
lower $B_{K}/T$ than higher $K$'s, suggesting its groups are less
distinct---or, equivalently, more internally heterogeneous---than
$K \in \{3,4\}$. There is limited improvement in quality with $K=4$,
which is consistent with the earlier results that the correct choice
($K=3$) adequately summarizes the variability in the data.

\begin{figure}[!htbp]
	\includegraphics[width=\textwidth]{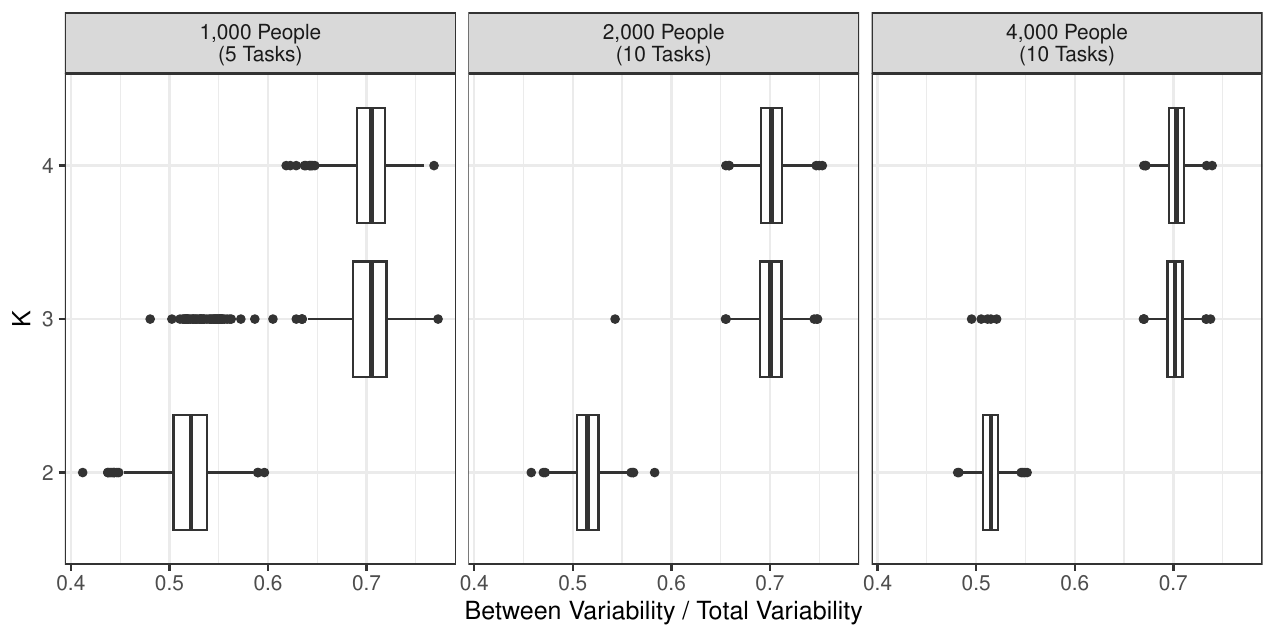}
	\caption{The distribution of $B_{K}/T$ across simulations. The
		top panel shows results for the model fit on the entire
		dataset. The bottom panel shows the results for a method
		estimated using the split sample
		method.}\label{fig:between_tot_K}
\end{figure}

\subsubsection{Misspecified Moderators}\label{sec:app_sim_vary_moderator}

We next consider how misspecifying the model for the moderators
$\pi_k(\bX_i)$ affects our simulated results. We show this in two
ways; first, we fit a model with no moderators, that is, $\bX_i =
1$. While this model has a number of limitations---e.g., for
classifying and predicting heterogeneous effects for new individuals,
it is a useful benchmark. Second, instead of using the true moderators
(e.g., $\bX_i$), we assume the researcher only has available the
following non-linear transformations of the moderators (following
\citealt{kang2007dr}) and uses those instead:
\begin{align*}
\bm{A}_{i,1} &= \sqrt{3} \exp(\bX_{i,1}/2) - 2 \\
\bm{A}_{i,2} &= \sqrt{3} \bX_{i,2}/\left[1+\exp(\bX_{i,1})\right] \\ 
\bm{A}_{i,3} &= 1/19 \left[\bX_{i,1} + \bX_{i,3} + 0.6\right]^3 \\
\bm{A}_{i,4} &= 1/3 \left[\bX_{i,2} + \bX_{i,4}\right]^2 - 1 \\ 
\bm{A}_{i,5} &= 2.5 \sqrt{|\bX_{i,5} + \bX_{i,1}|} - 2.5.
\end{align*}

We rescale the moderators $\{\bm{A}_i\}_{i=1}^N$ to have zero mean and unit variance in each simulated dataset.

Figure~\ref{fig:app_nomod_ame} replicates Figure~\ref{fig:app_dist_simulations} on the performance on estimating the AMCE where we show results with all moderators (i.e., in Figure~\ref{fig:app_dist_simulations}) and with both types of mis-specification (``No Moderators'' and ``Non-Linear Transf.'' when $\bm{A}_i$ are used).

\begin{figure}[!htbp]
	\includegraphics[width=\textwidth]{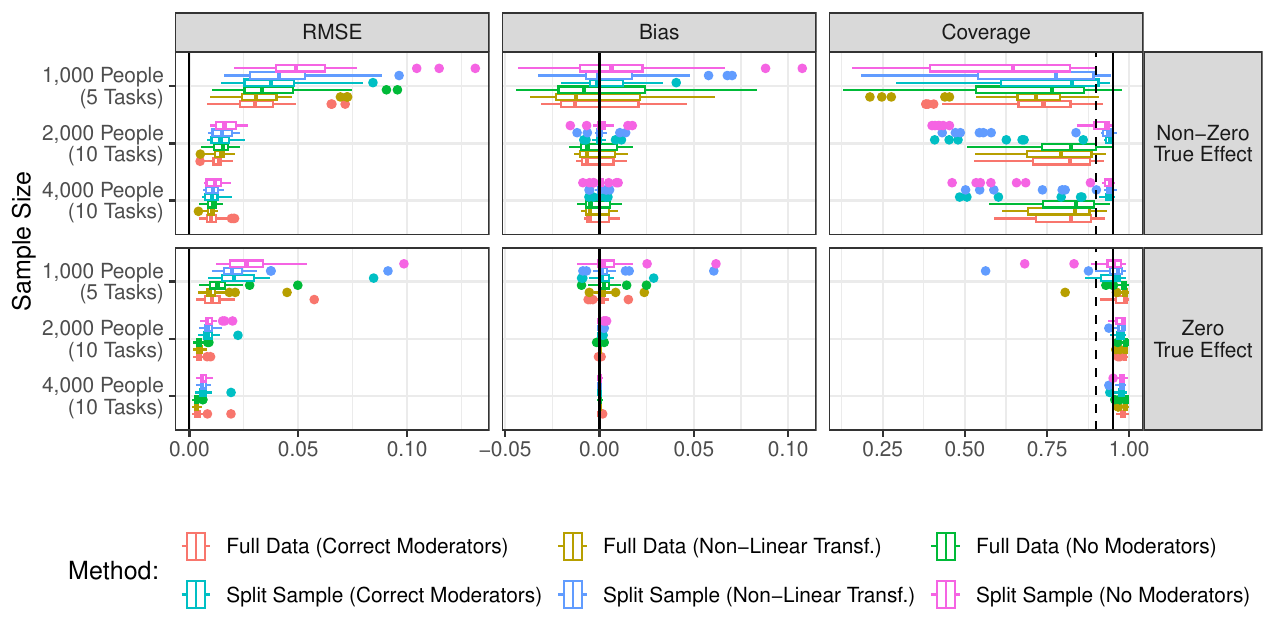}
	\caption{The distribution of performance for each estimator across sample sizes, with and without moderators. Inside each figure, results are split by whether the true effect is zero (``Zero True Effect'') or not (``Non-Zero True Effect''). The boxplot shows the distribution across all effects for each group. For the plots on RMSE and bias, the solid vertical line indicates zero. For coverage, the solid line indicates 95\% coverage and the dashed line indicates 90\%.}\label{fig:app_nomod_ame}
\end{figure}

It shows that, for the smallest sample size, the no-moderator model incurs a penalty in terms of the RMSE of the estimated AMCEs, although it does not have considerably larger bias. At larger sample sizes, the difference between the moderator and no-moderator models decreases. With moderators that are included but mis-specified using some non-linear transformation, the performance is rather close to the one that uses the correct moderators.

To further illustrate the impact of excluding moderators, Figure~\ref{fig:app_nomod_posterior} plots the estimated average posterior and posterior predictive probability (i.e., $\hat{\pi}_k(\bX_i)$) in the group corresponding to the individual's sampled $Z_i$ for all observations in the estimation data. It shows, as expected, that using the correctly specified moderators results in a considerably higher probability of each individual being assigned to group that corresponds to their sampled $Z_i$. The model with included but mis-specified moderators (``Non-Linear Transf.'') is somewhere between the model without moderators and the correctly specified one.

\begin{figure}[!htbp]
	\includegraphics[width=\textwidth]{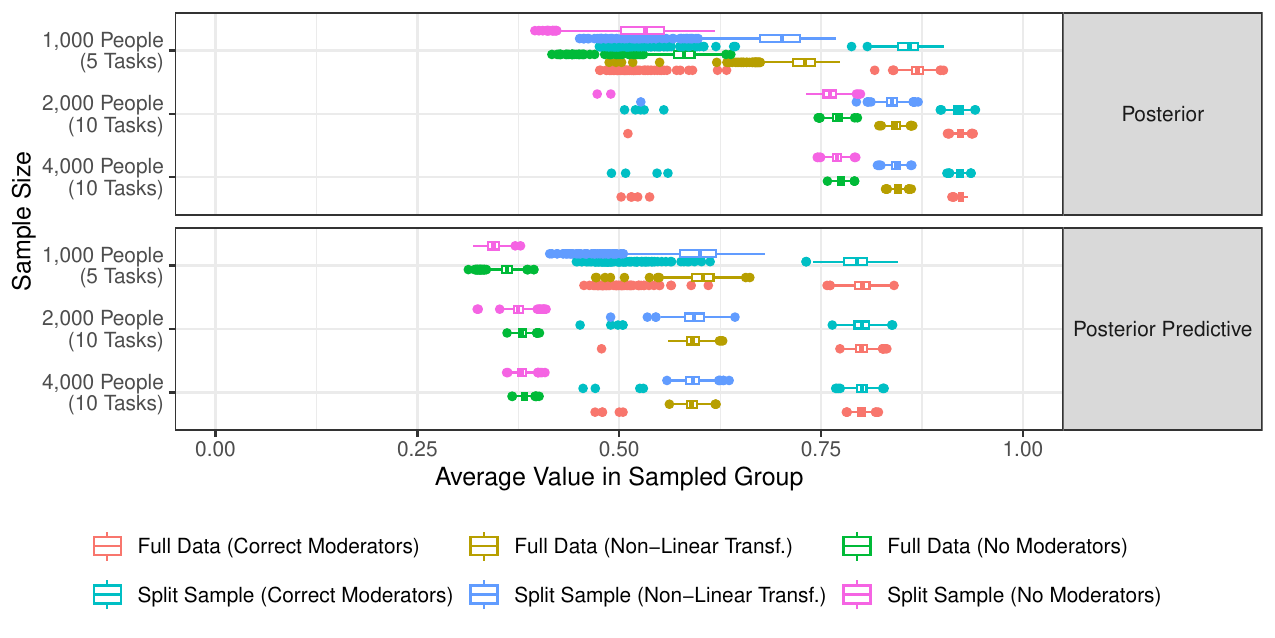}
	\caption{The average probability that is assigned to the group corresponding to an individual's sampled $Z_i$, showing the distributions across simulations.}\label{fig:app_nomod_posterior}
\end{figure}

\section{Additional Results for Immigration Conjoint Experiment}
\label{app:emp_marg_means}

We provide some additional results for our main empirical analysis. First, focusing on the three-group model, we report a different quantity of interest. We use an analogue to the ``marginal means'' estimator in \cite{leeper2020measuring}. We compute the probability of a profile being chosen \emph{without} specifying a baseline category. The equation is shown below for the forced choice case; note it consists of two of the terms used for the AMCE. 

\begin{align}
\mathrm{MM}_{jk}(l) & \ = \ \frac{1}{2}\E \left[
\left\{\Pr\left(Y_i = 1 \mid Z_i=k, T^L_{ij}=l,
\bT^L_{i,-j}, \bT^R_{i}\right) + \Pr\left(Y_i = 0 \mid Z_i=k, T^R_{ij}=l,
\bT^R_{i,-j}, \bT^L_{i}\right)
\right\}\right]. 
\end{align}

The below plot ignores randomization restrictions when estimating this quantity to center the estimate around 0.50 as in \cite{leeper2020measuring}. The results are substantively similar to the analysis in shown in the main paper using AMCEs.

\begin{figure}[t!]
	\centering \spacingset{1}
	\includegraphics[width=\textwidth]{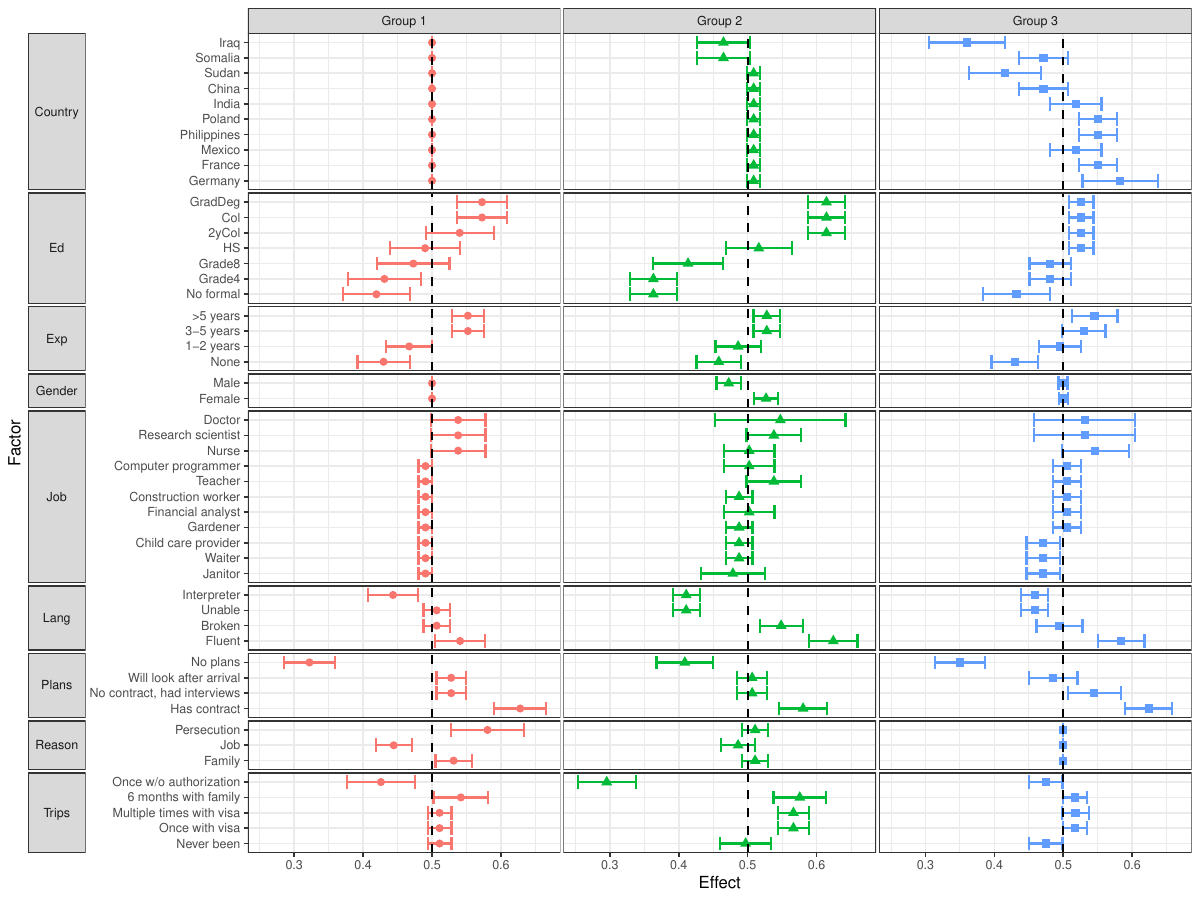}
	\caption{Estimated average marginal means using a
		three-group (right) analysis. The point
		estimates and 95\% Bayesian credible intervals are shown.} \label{fig:marg_means}
\end{figure}

Second, as noted in the main text, we found that sample splitting and refitting the model (see Appendix~\ref{sec:app_simulations_coverage}) was somewhat unstable given different splits of the data. To illustrate this point, Figure~\ref{fig:perm_AMCE} shows the 25th-75th percentile (and median) of the AMCEs estimated across twenty repetitions of splitting the data into halves and then using the refitting procedure described above. We address the problem of label switching using a permutation of labels that minimizes the average mean absolute error between all pairs of estimates; we find a permutation by randomly permuting the labels for a randomly chosen set of estimates and repeat this repeatedly until the average mean absolute error stabilizes. 

While Figure~\ref{fig:perm_AMCE} shows instability in some of the estimated AMCE, it broadly shows a similar result to that in the main text. For example, one group (Group 2 when $K=2$; Group 3 when $K=3$) shows a clear effect of country across most splits whereas one group (Group 1 when $K=2$ and Groups 1 and 2 when $K=2$) generally shows a large penalty for immigrants who entered without legal authorization. 

\begin{figure}[!htbp]
	\centering \spacingset{1}
	\includegraphics[width=\textwidth]{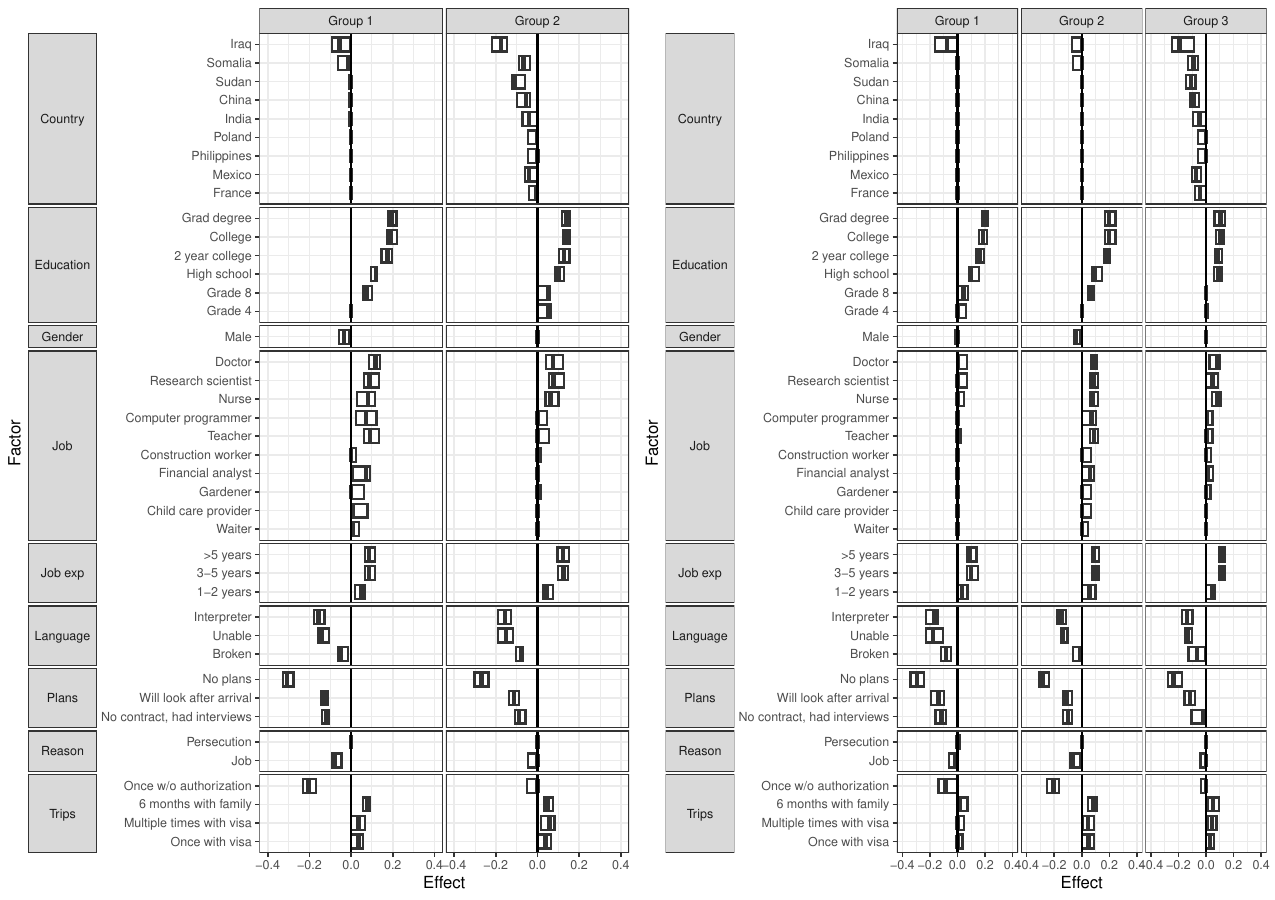}
	\caption{The distribution of AMCE from a two-group and three- model with twenty random splits of the data. The interquartile range and median are shown.} \label{fig:perm_AMCE}
\end{figure}

Third, Figure~\ref{fig:mod} in the main text reports the average effect of changing some moderator from $x_0$ to $x_1$ on $\pi_k$, i.e.,

\begin{equation}
\label{eq:app_mfx_moderator}
\mathbb{E}\left[\pi_k(X_{ij} = x_1, \bm{X}_{i,-j}) - \pi_{k}(X_{ij} =
x_0, \bm{X}_{i,-j})\right].
\end{equation}

Figure~\ref{fig:abs_mfx} considers the impact on the average \emph{absolute} distance, i.e. 

\begin{equation}\label{eq:app_mfx_abs}
\mathbb{E}\left[\left|\pi_k(X_{ij} = x_1, \bm{X}_{i,-j}) - \pi_{k}(X_{ij} =
x_0, \bm{X}_{i,-j})\right|\right],
\end{equation}
to prevent positive and negative changes from canceling each other out. To interpret this quantity, Figure~\ref{fig:abs_mfx} also the absolute value of the difference reported in the main text, i.e., the absolute value of Equation~\ref{eq:app_mfx_moderator} in a red $*$. Uncertainty is computed by drawing samples from the estimated asymptotic distribution of $\hat{\bm{\phi}}$, evaluating Equation~\ref{eq:app_mfx_abs} over those samples, and reporting the mean and $[0.025, 0.975]$ percentile interval. Figure~\ref{fig:abs_mfx} shows that, for certain groups, some covariates show a small average effect but a larger average of absolute effects (e.g., with $K=3$, Group 2 and ``Not Strong Republican'' versus the baseline of ``Strong Republican'') .

\begin{figure}[!htbp]
	\centering \spacingset{1}
	\includegraphics[width=\textwidth]{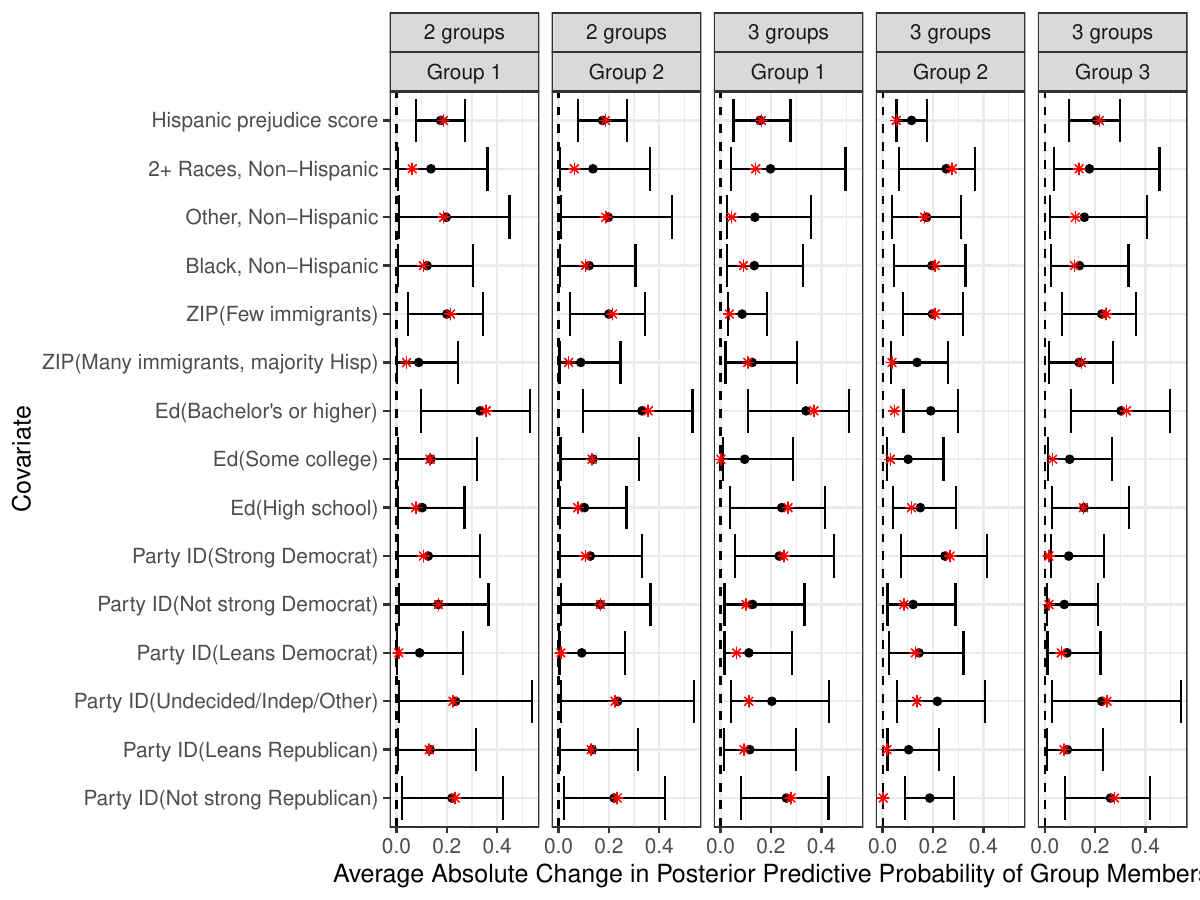}
	\caption{The average absolute effect of changing a moderator. The 2.5\% to 97.5\% percentile interval is shown.} \label{fig:abs_mfx}
\end{figure}

Next, we discuss the two-factor interactions. The largest average marginal interaction effect (AMIE) was found between education and job in the three group analysis. This is visualized in Figure~\ref{fig:int}. The largest AMIE occurs between the levels of Teacher and High School and has magnitude of 0.0021.

Compared in magnitude to the AME, which for education was on average 0.111 and for job was on average 0.0237, this is clearly negligible. Given this, we have little hope of finding substantial higher-order interactions in this example.

If higher-order interactions were of interest, a pre-processing step to do some basic screening \citep[see, e.g.,][]{shi2023forward} might be implemented on the full dataset to a priori reduce the number of interactions considered. The sparsity inducing penalties of our method would then impose additional regularization.

\begin{figure}[!htbp]
	\centering \spacingset{1}
	\includegraphics[width=\textwidth]{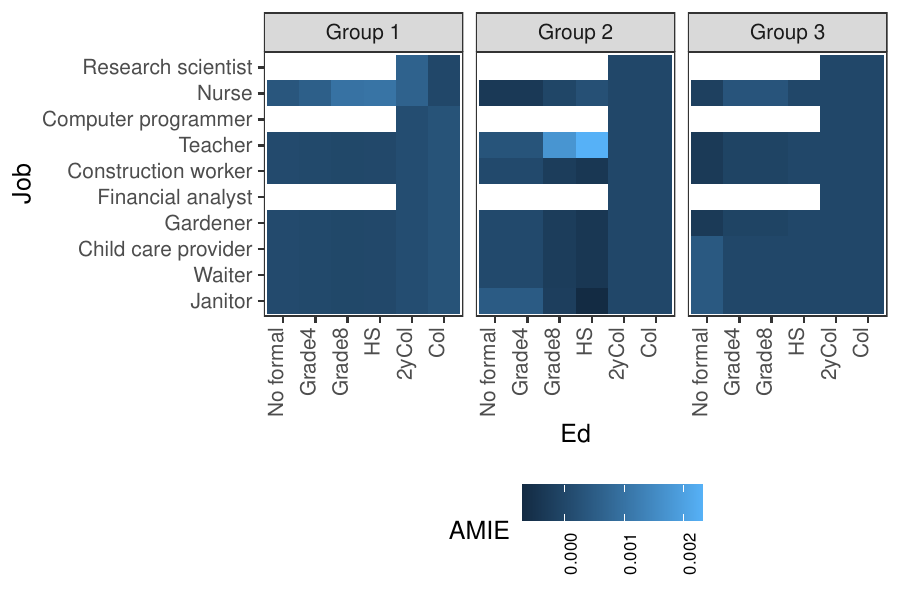}
	\caption{The average marginal interaction effect between education and job.} \label{fig:int}
\end{figure}

Finally, we briefly remark upon choosing $K$ using an information criterion. While this works well in the simulated example (see Appendix~\ref{sec:app_sim_choose_k}), we find less clear results on the full data. Table~\ref{tab:ic_hhy} the results of optimizing the BIC over $\lambda$ for $K \in \{1, 2, 3, 4\}$ as well as optimizing the AIC over $\lambda$. It shows that, if one uses the BIC, this suggests $K=1$. However, if one uses the AIC, this suggests $K=4$.
\begin{table}[htbp]
	\begin{center}
		\begin{tabular}{llll}
			\hline\hline
			\multicolumn{4}{c}{Optimizing BIC over $\lambda$} \\
			$K=1$ & $K=2$ & $K=3$ & $K=4$ \\
6125 & 6270 & 6391 & 6529\\
\hline
\multicolumn{4}{c}{Optimizing AIC over $\lambda$} \\
$K=1$ & $K=2$ & $K=3$ & $K=4$ \\
5968 & 5902 & 5871 & 5833\\ \hline\hline

		\end{tabular}
	\end{center}
	\caption{Information criterion for different $K$}\label{tab:ic_hhy}
\end{table}

	\end{set11}
\end{appendices}

\end{document}